\documentclass[prb,aps,floats,twocolumn]{revtex4}

\usepackage{graphicx}
\usepackage{amsfonts}      
\usepackage{amssymb}      
\usepackage{amsbsy}      
\usepackage{amsmath}
\usepackage{xcolor}
\usepackage{hyperref}
\usepackage{titlesec}
\usepackage{bm}
\usepackage{ulem}

\makeatletter

\newcommand{\pair}[1]{\langle{#1}\rangle}

\newcommand{\normal}[1]{\,:\!{#1}\!:\,}

\newcommand{\bk}{{\bf k}}

\DeclareMathOperator{\tr}{Tr}

\DeclareMathOperator{\re}{\rm Re}

\makeatother

\begin{document}

\title{Dynamical structure factor of the triangular antiferromagnet: the Schwinger boson theory beyond the mean field approach}

\author{E. A. Ghioldi$^{1}$, M. G. Gonzalez$^{1}$, Shang-Shun Zhang$^{2}$, Yoshitomo Kamiya$^{3}$, L. O. Manuel$^{1}$, A. E. Trumper$^{1}$  and C. D. Batista$^{2,4}$}

\affiliation{$^1$ Instituto de F{\'i}sica Rosario (CONICET) and Universidad Nacional de Rosario, Boulevard 27 de Febrero 210 bis, (2000) Rosario, Argentina}
\affiliation{$^2$ Department of Physics and Astronomy, University of Tennessee, Knoxville, Tennessee 37996-1200, USA}
\affiliation{$^3$ Condensed Matter Theory Laboratory, RIKEN, Wako, Saitama 351-0198, Japan}
\affiliation{$^4$ Quantum Condensed Matter Division and Shull-Wollan Center, Oak Ridge National Laboratory, Oak Ridge, Tennessee 37831, USA}

\begin{abstract}
We compute the zero temperature dynamical structure factor $S({\bm q},\omega)$ of the triangular lattice Heisenberg model (TLHM) using a Schwinger boson approach that includes the Gaussian fluctuations ($1/N$ corrections) of the saddle point solution. While the ground state of this model exhibits a 
well-known 120$^{\circ}$ magnetic ordering, experimental observations have revealed a  strong quantum character of the excitation spectrum. 
We conjecture that this phenomenon arises from the  proximity of the ground state of the TLHM to the quantum melting point separating the  magnetically 
ordered and spin liquid states. Within this scenario,  magnons are described as collective modes (two spinon-bound states) of a spinon condensate 
(Higgs phase) that spontaneously breaks the SU(2) symmetry of the TLHM.  Crucial to our results is the proper account of this  spontaneous  symmetry breaking. 
The main qualitative difference relative to  semi-classical treatments ($1/S$ expansion) is the presence of a high-energy spinon continuum extending  up to 
about three times the single-magnon bandwidth. In addition,  the magnitude of the ordered moment ($m=0.224$) agrees very well with numerical results and  the 
low energy part of the  single-magnon dispersion is in very good agreement with  series expansions.  Our results indicate that the Schwinger boson approach is an adequate 
starting point for describing the excitation spectrum of some magnetically ordered compounds that are near the quantum melting point separating this Higgs phase 
from the {\it deconfined} spin liquid state.   
\end{abstract}

\maketitle

\section{Introduction}
Novel quantum states in strongly interacting electron systems are boosting a new era of quantum materials.~\cite{Tokura2017}
Understanding their basic constituents is necessary to predict their behavior under different conditions and to derive low-energy theories that can describe 
the interplay between charge and spin degrees of freedom in doped magnets. 
It is then imperative to develop new approaches beyond the  conventional paradigms.  
More specifically, the  increasingly refined spectra produced by recent advances in inelastic neutron 
scattering~\cite{Han2012,Zhou2012,Banerjee2016a,Ma2016,Paddison2017,Ito2017} are demanding new theories that can account for  multiple anomalies observed in 
dynamical spin structure factor $S({\bm q}, \omega)$ of frustrated  quantum  antiferromagnets.\\ 

In the conventional paradigm,~\cite{Landau1937} magnetic order develops at low enough temperatures via spontaneous symmetry breaking.~\cite{Anderson1997}
The elementary low-energy quasi-particles  are  spin one modes known as magnons. In the  new paradigm~\cite{Anderson1973}, zero-point or quantum fluctuations 
enhanced by  magnetic frustration and/or low dimensionality may preclude conventional symmetry breaking, leading to a quantum spin liquid phase 
at $T=0.$~\cite{Lacroix2010}
Topologically ordered quantum spin liquids  are different from simple quantum paramagnets because they cannot be adiabatically connected with any product state 
and they can support excitations with fractional quantum numbers.~\cite{Wen2002,Sachdev2008,Normand2009,Balents2010,Savary2017,Zhou2017}
The first proposal of a topologically ordered quantum spin liquid was the resonant valence bond (RVB) state introduced by P. W. Anderson to describe the 
ground state of the TLHM.~\cite{Anderson1973}
The RVB state is a linear superposition of different configurations of short range singlet pairs, whose resonant character leads to the decay of spin one 
modes into pairs of free $S=1/2$ spinons.

The nature of the ground state of the triangular Heisenberg antiferromagnet was a controversial topic for a long time.~\cite{Zheng2006} Finally, a sequence 
of numerical works~\cite{Huse1988,Bernu1992,Singh1992,Capriotti1999,White2007} provided enough evidence in favor of  long range  N\'eel magnetic order
($120^{\circ}$ ordering) with a relatively small ordered moment ($41\%$ of the full moment).~\cite{Capriotti1999,Zheng2006,White2007} 
This sizable reduction  of the ordered moment is indicative  of strong quantum fluctuations and of the proximity to a quantum spin liquid phase.  
In a semi-classical  treatment of the problem ($1/S$ expansion), the presence of strong quantum fluctuations manifests via a large $1/S$ correction of 
the magnon bandwidth along with single to two magnon decay in a large  region of the Brillouin zone.~\cite{Starykh06,Chernyshev2006,Chernyshev2009,Zhitomirsky2013}

\begin{figure}[ht]
\vspace*{0.cm}
\includegraphics*[width=0.4\textwidth,angle=0]{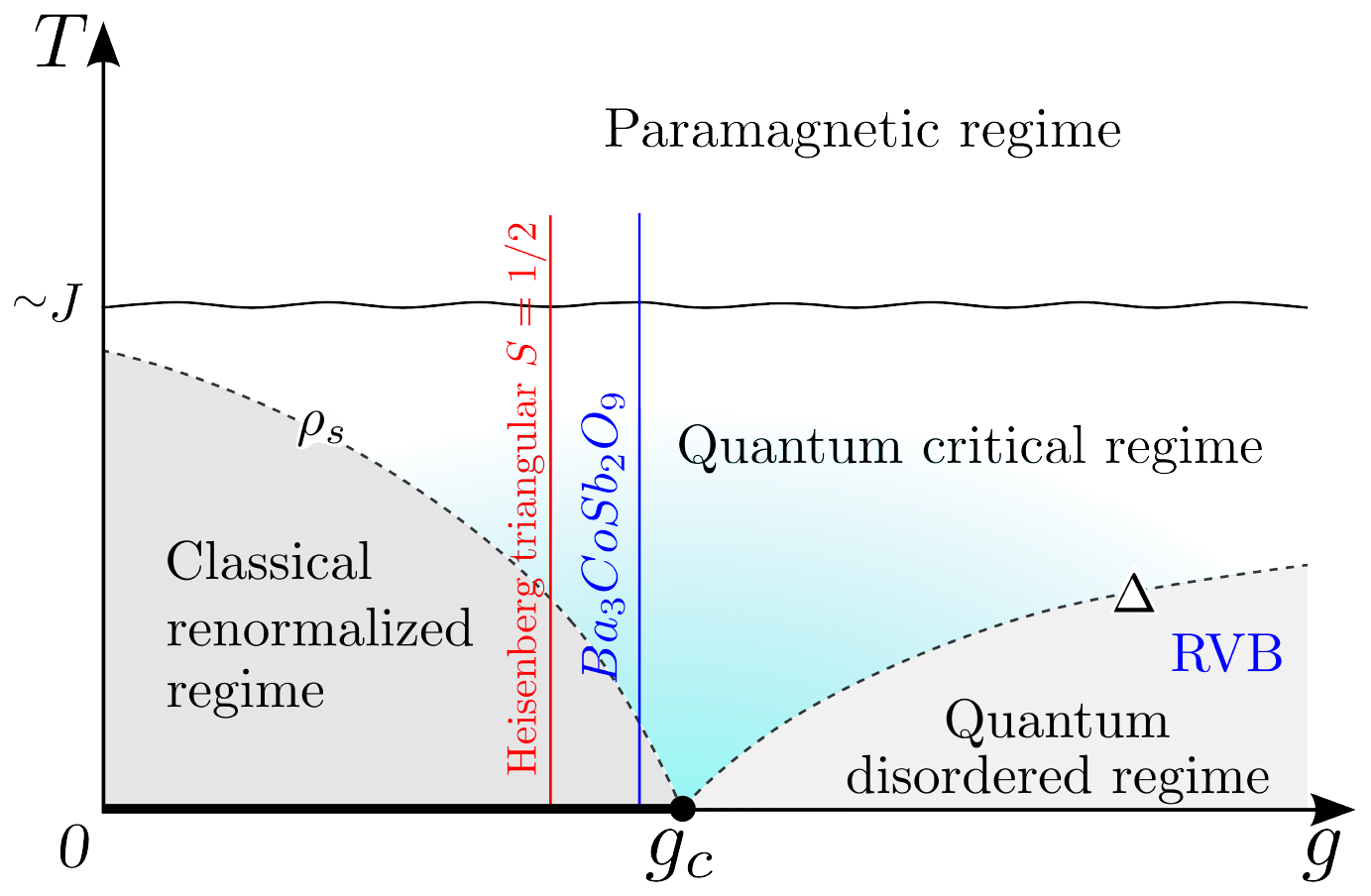}
\caption{Schematic finite temperature phase diagram for $2D$ frustrated antiferromagnets.~\cite{Coldea2003} $g$ is a generic measure of the zero point quantum 
fluctuations and $g_c$ connects continually a spiral magnetic state to quantum spin liquid state. The dashed lines indicate the crossover from the classical 
renormalized and the quantum disordered regimes to the quantum critical regime. The corresponding energy scales are the spin stiffness $\rho_s$ and the triplet excitation $\Delta$, respectively. The conjectured location of the spin-$\frac{1}{2}$ triangular Heisenberg model 
and the compound Ba$_3$CoSb$_2$O$_9$ are indicated by the vertical lines.}
\label{fig1:diagram}
\end{figure}

Early studies of the low temperature properties based on an effective quantum field theory  suggested the need of adopting alternative descriptions to the 
semiclassical approach.~\cite{Read1991,Sachdev1991,Chubukov1994} In particular, Chubukov {\textit et al.}~\cite{Chubukov1994} proposed that  the AF triangular Heisenberg 
model is in the crossover region between a classical renormalized  and a quantum critical regime of deconfined spinons at temperatures $T\sim 0.4 J$.  
As shown in Fig.~\ref{fig1:diagram},  this observation is consistent with the  proximity of the $120^{\circ}$ N\'eel order  to a zero temperature  quantum melting point 
(QMP). If the quantum phase  transition between the mangetically  ordered state and the spin liquid phase turns out to be continuous (or quasi-continuous), 
the magnon modes should be described as weakly bounded two-spinon bound states in the proximity of the QMP. In other words,  the two-spinon confinement 
length $\xi_{\rm conf}$ should become significantly larger than the lattice spacing ($\xi_{\rm conf} \gg a$) near the QMP. Indeed, it is known that 
the $J_1-J_2$  (nearest and next-nearest exchange coupling) triangular Heisenberg model exhibits a transition into a spin liquid state 
at $J_2/J_1 \simeq 0.06$.~\cite{Manuel1999,Mishmash13,Kaneko14,Li15,Zhu15,Hu15,Iqbal16,Saadatmand16,Bauer17,Gong17,Zhu18} 
Recent numerical studies~\cite{Zhu18} indicate that this is a  {\it continuous} quantum phase transition between the $120^{\circ}$ N\'eel ordered state and 
the spin liquid phase.

The above described picture  is analogous to color confinement in quantum chromodynamics (QCD): hadrons are described as composite  states of quarks, although 
quarks cannot be directly observed because they are confined by the gluon field that creates some kind of ``string" between them. The analogy with a two-body 
problem with a linear interaction potential is an oversimplification because it does not account for the quantum nature of the gluon field:
excited bound states of the linear potential (heavy hadron particles) are unstable and decay into lighter ones.~\cite{Gell-Mann95,Greiner94}
A similar situation is expected for the above described quantum magnet:  high-energy two-spinon bound states, corresponding to longitudinal modes, are expected 
to decay into multiple pairs of two-spinon bound states transforming the two-body problem of confinement into a many-body one. These processes should leave their 
fingerprint in the high-energy continuum of the dynamical spin susceptibility, which can be measured with inelastic neutron scattering (INS). Unlike other 
experimental techniques, INS can reveal the internal structure of the magnon modes. Identifying these signatures is then crucial to determine if a given compound 
provides a realization of this strong quantum mechanical effect.
 
Identifying condensed matter  analogues of confined fractional particles  is  important for multiple reasons. In the first place, we can connect the original 
lattice or microscopic (high-energy) model with the effective (low-energy) field theory that is obtained in the long wavelength limit.~\cite{Chubukov1994} 
Consequently, we can relate the parameters of the microscopic theory to the properties of the particles (such as hadron masses in the context of QCD) that emerge 
at low energies. The physics of spin ladders provides a simple 1D analog of this physics,~\cite{Shelton1996,Lake2009} where the role of quarks is played by 
spinons, although there are also some obvious differences because the interaction between spinons is not usually attributed to gauge fields.  In the second 
place,  the emergence of gauge fields and fractionalized excitations in dimension higher than one could shed light  on the unusual behavior of different 
classes of correlated electron materials in the proximity of a quantum critical point.~\cite{Starykh1994,Powell2011,Chubukov1995,Lee2006,Sachdev2008,Vojta2018}

Recent inelastic neutron scattering measurements performed in Ba$_3$CoSb$_2$O$_9$~\cite{Ma2016,Ito2017,Kamiya2018} --an experimental realization of a quasi-2D 
triangular $S=1/2$ antiferromagnet (AF)--  are indeed suggesting that semi-classical approaches (large-$S$ expansion) do not reproduce several aspects of the 
dynamical structure factor $S({\bm q},\omega)$, despite the existence of magnetic long-range order. 
In particular, the magnon bandwidth $W$, the observed line broadening~\cite{Ma2016} and, more importantly, the very unusual dispersive continuum extending up 
to $6W$~\cite{Ito2017} are the most salient features, which cannot be reproduced by linear spin wave theory (LSW) plus $1/S$ (LSW+$1/S$) 
corrections.~\cite{Chernyshev2009,Mourigal2013a}  It is important to note that  the easy-plane anisotropy and the finite inter-layer exchange of 
Ba$_3$CoSb$_2$O$_9$ preclude spontaneous single to two-magnon decay at the  LSW+$1/S$ level,~\footnote{The easy-plane anisotropy gaps out the Goldstone mode 
at the $K$ point. Consequently, unlike the isotropic case, the kinematic conditions prevent single Goldstone modes at the $\Gamma$-point from decaying  into 
pairs of Goldstone modes at the $K$-point. } which is obtained  for the {\it isotropic} 2D Heisenberg model.~\cite{Chernyshev2006,Chernyshev2009} 
In other words, a low order $1/S$ expansion for the $S=1/2$ model relevant to  Ba$_3$CoSb$_2$O$_9$ does not even anticipate strong quantum effects in 
this material. Then, as for the case of ${\mathrm{Cs}}_{2}\mathrm{Cu}{\mathrm{Cl}}_{4}$,~\cite{Coldea2001,Coldea2003,Isakov05} it is natural to ask if the  anomalies observed 
in the dynamical structure factor of Ba$_3$CoSb$_2$O$_9$ can be attributed to a long confinement length $\xi_{\rm conf}  \gg a$ of spinons, as hypotesized 
in Fig.~\ref{fig1:diagram}, and if a $1/N$ expansion ($N$ is the number of flavors of the fractional particles) can account for the  observed anomalies.

The Schwinger boson (SB) theory --originally developed by Arovas and Auerbach~\cite{Arovas1988}-- is an adequate technique to answer this question.  
The control parameter $N$ can be naturally introduced in the SB theory by increasing the number of flavors of the Schwinger 
bosons.~\cite{Arovas1988,Read1991,Sachdev1991,Timm1998,Flint2009} The saddle point result  becomes exact in the $N \to \infty$ limit. 
At this level, the system is described as a gas of non-interacting spinons  and long range magnetic ordering manifests via a Bose condensation of 
the SBs.~\cite{Hirsch1989,Sarker1989,Chandra1990} The resulting dynamical spin structure factor, $S({\bm q},\omega)$,  only includes a two-spinon 
continuum, which misses the true collective modes (magnons) of a magnetically ordered 
state.~\cite{Auerbach1988, Chandra1990,Mila1991,Yoshioka1991,Sachdev1992,Lefmann1994,Mattsson1995,Mezio2011} 
As we demonstrate in this work, magnons already arise at the Gaussian fluctuation level ($1/N$ corrections) as a result of the interaction with  
fluctuations of the emergent gauge fields. The crucial difference relative to previous formulations of this problem~\cite{Arovas1988} is that we 
compute $S({\bm q},\omega)$ on top of the spinon condensate (Higgs phase) that spontaneously breaks the SU(2) symmetry of the TLHM (broken symmetry 
ground state).

The broken symmetry spinon condensate is selected by adding an infinitesimal symmetry breaking field, $h$, which is sent to zero after taking the 
thermodynamic limit. The resulting local magnetization of the $120^{\circ}$ N\'eel ordering is $m=0.224$ which quantitatively agrees with quantum 
Monte Carlo (QMC)~\cite{Capriotti1999} and density matrix renormalization group (DMRG)~\cite{White2007} predictions. 
On the other hand,  the excitation spectrum, revealed by $S(\bm q,\omega)$, has a strong quantum character, which is not captured by low-order $1/S$ expansions. 
The low-energy magnons consist of two-spinon bound states confined by gauge fluctuations of the auxiliary fields. The  good agreement with the relation 
dispersion predicted by series expansions~\cite{Zheng2006} indicates that  magnons may indeed have the composite nature predicted by the SB theory. 
Moreover, the resulting high-energy two-spinon continuum, which extends up to about three times the single-magnon bandwidth, may account for 
the first high-energy peak that is observed in Ba$_3$CoSb$_2$O$_9$.~\cite{Ito2017} Furthermore, as we show in the next sections, the inclusion of 
Gaussian corrections removes other problems of the saddle point approximation, such as the spurious modes arising from unphysical density fluctuations 
of the bosonic field.\\     

The article is organized as follows: 
in Sec. II we briefly review the Schwinger boson approach for treating AF Heisenberg models. 
Using the saddle point expansion, we derive the saddle point solution consisting of a spinon condensate that spontaneously breaks the SU(2) 
symmetry of the TLHM (Higgs phase) and the effect of Gaussian fluctuations of the auxiliary (gauge) fields on the ground state energy and the 
dynamical susceptibility. In Sec. III we show the main consequences of properly accounting for the spontaneous $SU(2)$ symmetry breaking. 
Sec. IV contains the results obtained for the TLHM, including the magnitude of the ordered moment, the dynamical structure factor, and the 
magnon dispersion relation along with a detailed analysis of the long wavelength limit. The physical implications of these results are discussed in Sec.~V.

\section{Schwinger boson theory}
In this section we present the path integral formulation of the Schwinger boson theory specialized for isotropic frustrated AF models whose  ground 
states break the $SU(2)$ symmetry of the Heisenberg Hamiltonian. In particular, we consider the $S=1/2$ AF TLHM,
\begin{eqnarray}
{\cal H} = J \sum_{\pair{ij}} \bm S_i \cdot \bm S_j,
\label{Heis}
\end{eqnarray}
whose ground state is known to exhibit 120$^{\circ}$ magnetic ordering.~\cite{Bernu1992,Huse1988,Capriotti1999,White2007} 
We introduce the  Schwinger boson representation of the spin $\bm S_i$ in terms of spin$-\frac{1}{2}$ boson operators $b_{i\sigma},$ 
through the relation
\begin{equation}
 \bm S_{i} = \frac{1}{2}\bm b^\dagger_i \cdot \bm{\sigma}\cdot \bm b_i,
 \label{spins}
\end{equation}
where $\bm b^\dagger_i = \left(b^\dagger_{i\uparrow},b^\dagger_{i\downarrow}\right),$ 
$\bm \sigma =\left(\sigma^x, \sigma^y, \sigma^z\right)$ is the vector of Pauli matrices, 
and the bosons are subject to the number constraint, 
\begin{equation}
\sum_{\sigma} b_{j \sigma}^{\dagger} b_{j\sigma}^{\;} = 2S = 1.
\label{constr}
\end{equation} 
The Heisenberg interaction can be expressed as~\cite{Ceccatto1993}
\begin{equation}
{\bm S}_{i} \cdot {\bm S}_{j} = \normal{{B}_{ij}^{\dagger} {B}_{ij}} - {A}_{ij}^{\dagger} {A}_{ij},
\label{decoup}
\end{equation}
in terms of  the SU(2) invariant  bond operators
$$ {A}_{ij}  = \frac{1}{2} \left( b_{i \uparrow} b_{j \downarrow} - b_{i\downarrow} b_{j\uparrow}\right), \;\;
B_{ij}  = \frac{1}{2} \left( b_{i\uparrow} b_{j \uparrow}^{\dagger} + b_{i \downarrow} b_{j\downarrow}^{\dagger}\right).$$
$A^{\dagger}_{ij}$ creates singlet states, while $B^{\dagger}_{ij}$ makes them resonate. These are the two key ingredients of the 
RVB theory proposed by P. W. Anderson.~\cite{Anderson1973} 
By using the  operator identity $:\!{B}^{\dagger}_{ij}{B}_{ij}\!:\! +\! A^{\dagger}_{ij}{A}_{ij} = S^2$, the Heisenberg interaction was 
originally expressed in terms of the $A_{ij}$ operators only.~\cite{Arovas1988,Read1991,Sachdev1992} 
However, keeping the $A_{ij}$ and $B_{ij}$ operators in Eq.~\eqref{decoup} for the saddle point approximation has two important advantages. 
It better accounts for non-collinear magnetic orderings, like the 120$^{\circ}$ structure, that typically appear in frustrated 
magnets,~\cite{Gazza1993,Manuel1994,Mezio2011,Mezio2012} and it enables a proper  extension from SU(2) $\simeq$ Sp(2) to Sp($N$), which is formally required to take the 
large $N$ limit with generators of the Lie algebra that are odd under time reversal.~\cite{Flint2009} This two-singlet bond structure is 
currently used to classify quantum spin liquids based on the projective symmetry group.~\cite{Wang2006,Messio2013}\\

The partition function for ${\cal H}$  is expressed in terms of the functional integral over coherent states:~\cite{Auerbach1994}
\begin{multline}
 \mathcal{Z}[j] = \int D [\overline b,b]D[\lambda] \ e^{\! -\int_{0}^{\beta} d\tau \! \left[ \sum\limits_{i \sigma} \overline b_{i \sigma}^{\tau} \partial_{\tau} b_{i \sigma}^{\tau} + \ \mathcal{H}(\overline b,b) \ + \mathcal{J}_{s} + \mathcal{J}_{b}    \right] } \\
 \times e^{ -\int_{0}^{\beta} d\tau \  i \sum\limits_{i} \lambda_{i}^{\tau} \big(\sum\limits_{\sigma} \overline b_{i \sigma}^{\tau} b_{i \sigma}^{\tau} - 2S \big)  },
\label{partition}
\end{multline}
where 
\begin{equation}
\mathcal{J}_{s} = \sum_{i}  j_{i}^{\tau  \mu } \ \bm b_i^{\tau \dagger}  \cdot \sigma^{\mu} \cdot \bm b_i^\tau,  
\end{equation}
represents the Zeeman coupling to a space ($i$) and  time ($\tau$) dependent external field $j_{i}^{\tau\mu}$ ($\mu = x,y,z$), while   
\begin{equation}
\mathcal{J}_{b} = \sum_i  h_{i}^{\mu} \; \bm b_i^{\tau \dagger} \cdot \sigma^{\mu} \cdot \bm b_i^\tau
\end{equation}
represents a linear coupling between the order parameter and  a finite {\it static symmetry breaking field}  
$\bm h_j = \left(h\cos(\bm Q \cdot \bm r_j ), h\sin(\bm Q \cdot \bm r_j ),0\right)$ with $\bm Q=(\frac{2\pi}{3},\frac{2\pi}{\sqrt{3}})$, corresponding to 
the $120^\circ$ magnetic structure. 
The integration over the time and space dependent auxiliary field $\lambda_{i}^{\tau}$ accounts for
the local constraint \eqref{constr}. The integration measures are $ D[\overline b,b] = \prod_{i \tau \sigma} \frac{d\bar b _{i\sigma}^{\tau} db_{i\sigma}^{\tau}}{2 \pi i}$, 
and $ D[\lambda] = \prod_{i \tau} \frac{d\lambda_{i}^{\tau}}{2 \pi}$.\\

\noindent The Hamiltonian 
\begin{equation}
\mathcal{H}=\frac{1}{2} \sum_{\langle i j \rangle} J_{ij} (\overline A_{ij}^{\tau } A_{ij}^{\tau} - \overline B_{ij}^{\tau} B_{ij}^{\tau} )
\end{equation}
is quartic in the complex numbers $\overline b$ and $b$. This terms can be decoupled into quadratic terms using a Hubbard-Stratonovich (HS) transformation that introduces auxiliary fields $\overline W^{A}$, $W^{A}$ and $\overline W^{B}$, $W^{B}$  to decouple the $\overline AA$ and $\overline B B$ terms, respectively: 
\begin{eqnarray}\label{HSAA}
e^{J_{ij} \overline {A}_{ij}^{\tau} A_{ij}^{\tau} } & = &  \int D[\overline W^{A}\!,\!W^{A}] \ e^{ -J_{ij} \overline W_{ij}^{A\tau} W_{ij}^{A\tau}} \times   \\ & & \;\;\;\;\;\;\;\;\;\;\;\;\times \; e^{J_{ij} \left( \overline W_{ij}^{A \tau} A_{ij}^{\tau} + W_{ij}^{A\tau} \overline A_{ij}^{\tau} \right) } \nonumber  ,
\end{eqnarray}
\noindent and
\begin{eqnarray}\label{HSBB}
e^{-J_{ij} \overline B_{ij}^{\tau} B_{ij}^{\tau} } & = & \int D[\overline W^B\!,\!W^B] \ e^{ -J_{ij} \overline W_{ij}^{B\tau} W_{ij}^{B\tau}} \times  \\ & & \;\;\;\;\;\;\;\;\;\;\;\;\times \; e^ {J_{ij} \left( -\overline W_{ij}^{B\tau} B_{ij}^{\tau} + W_{ij}^{B\tau} \overline B_{ij}^{\tau} \right) }\nonumber,
\end{eqnarray} 
with integration measure $D[\overline W^{r}\!,\!W^r] =  \prod_{ij\tau} \frac{ d\overline {W}_{ij}^{r\tau} d W_{ij}^{r\tau}}{2 \pi i/ J_{ij}} $, and $r=A,B$.
After replacing Eqs.~\eqref{HSAA} and \eqref{HSBB} in Eq.~\eqref{partition}, the Gaussian integrals over $\overline{b}$ and $b$ can be formally carried out. 
The resulting partition function becomes
\begin{equation}
 \mathcal{Z}[j] = \int D[\overline W,W]D[\lambda] \ e^{-S_{\rm eff}(\overline W, W, \lambda, j)},
\label{generatriz}
\end{equation}
where the effective action can be split into two terms,
\begin{equation}
 S_{\rm eff}(\overline W, W, \lambda,j) = S_{0}(\overline W, W, \lambda) + S_{\rm bos}(\overline W, W, \lambda,j),
\label{effective}
\end{equation}
with
\begin{equation}
S_{0}(\overline W, W, \lambda)\! = \!\int_{0}^{\beta} d\tau \sum\limits_{ijr} J_{ij} \overline W_{ij}^{r\tau} W_{ij}^{r\tau}\!-\!2Si \sum_{i} \lambda_{i}^{\tau}, 
\label{S0}
\end{equation}
and
\begin{eqnarray}
 S_{\rm bos}(\overline W, W, \lambda,j) & \!= \!& \frac{1}{2}  \tr \ln \left[ \mathcal{G}^{-1}(\overline W, W, \lambda, j) \right]=\\ \nonumber
                                         &\!=\!& - \frac{1}{2} \ln \ \mathcal{Z}_{\rm bos}(\overline W, W, \lambda,j).
\end{eqnarray}\\
\noindent $\overline W$ and $ W $ are the HS fields  $W_{ij}^{r  \tau}$, $\overline W_{ij}^{r  \tau}$ ($r=A,B$) and  
$\mathcal{G}^{-1}=\mathcal{M}$ is the bosonic dynamical matrix with the trace taken over space, time, and  boson flavor indices. 
The bosonic partition function $\mathcal{Z}_{\rm bos}$ can be formally integrated out to get 
\begin{eqnarray}
 \mathcal{Z}_{\rm bos}(\overline W, W, \lambda,j)\!\!&=&\!\!\! \int\! D [\overline b,b] 
 e^{- \vec b^\dagger \cdot \mathcal{G}^{-1}(\overline W, W, \lambda,j) \cdot \vec b}=
 \nonumber \\ 
 &= & \det \left[ \mathcal{G}(\overline W, W, \lambda,j)\right],\nonumber 
\end{eqnarray}
where $\vec b$ is a vector containing all the variables $b_{i\sigma}^{\tau}$. 

The effective action \eqref{effective} is invariant under the $U(1)$ gauge transformation
$b_{i\sigma}^{\tau}\rightarrow b_{i\sigma}^{\tau} e^{i \theta^{\tau}_{i}}$ if the auxiliary fields transform as
\begin{eqnarray}
 W_{ij}^{r  \tau} &\rightarrow& W_{ij}^{r  \tau} e^{i(\theta^{\tau}_{i}\pm \theta^{\tau}_{j})}, \nonumber \\
 \overline{W}_{ij}^{r  \tau} &\rightarrow& \overline{W}_{ij}^{r  \tau} e^{-i(\theta^{\tau}_{i}\pm \theta^{\tau}_{j})}, \label{transf} \\
 \lambda_{\bm i}^{\tau} &\rightarrow& \lambda_{\bm{i}}^{ \tau} - \partial_{\tau}\theta^{\tau}_{i} \nonumber,
\end{eqnarray}
where the $+$ and $-$ signs hold for the $A$ and $B$ fields respectively.
In other words, the phase fluctuations of the auxiliary fields represent the emergent gauge fluctuations of the SB theory. 

The Fourier transformation to Matsubara frequency and momentum space is done using 
\begin{equation}
 f^\tau_i=\frac{1}{\sqrt{N_s\beta}} \sum_{\bm k, i\omega_n} f_{\bm k}^{i\omega_n} e^{-i({\bm k}\cdot \bm r_i- \omega_n \tau)}
\end{equation}
for any field $f^\tau_i$, where $i\omega_n= 2\pi in/\beta$ are the bosonic Matsubara frequencies  and $\bm k$ the momenta. For convenience, in what follows, we denote  $f_{\bm k}^{i\omega_n}$ as $f_{\bm k}^{\omega}$.
For simplicity,  we perform a rotation to a local reference frame such that the magnetic ordering, and consequently ${\bm h}$, become spatially uniform. In this case the bosonic variables transform as $b_{\bm k \uparrow}^{\omega} \rightarrow b_{\bm k+\frac{\bm Q}{2} \uparrow}^{\omega}$ and $b_{\bm k \downarrow}^{\omega} 
\rightarrow b_{\bm k - \frac{\bm Q}{2} \downarrow}^{\omega}$. After introducing  the representation $ \vec b_{\bm k,\omega}^{\dagger} = \left( \bar b_{\bm k \uparrow}^{\omega}, b_{-\bm k \downarrow}^{-\omega}, 
\overline b_{\bm k \downarrow}^{\omega}, b_{-\bm k \uparrow}^{-\omega} \right)$,  $\mathcal{G}^{-1}$ is 
\begin{widetext}
\begin{equation}
 \mathcal{G}_{\bm{k,k'}}^{-1 \ \omega,\omega'} = 
 \left(
\begin{array}{cccc}
 F_{\bm Q}^{B}(\bm{k,k'},i\omega)     &   F_{\bm Q}^{A}(\bm{k,k'})           &   -\frac{h}{2} \delta_{\bm{k,k'}}             &      0     \\ \\
 \overline F_{\bm Q}^{A}(\bm{k',k})  &   F_{\bm Q}^{B}(\bm{k',k},-i\omega)    &            0            & -\frac{h}{2} \delta_{\bm{k,k'}}  \\ \\
 -\frac{h}{2} \delta_{\bm{k,k'}}    &  0      &  F_{-\bm Q}^{B}(\bm{k,k'},i\omega)  &  -F_{-\bm Q}^{A}(\bm{k,k'}) \\  \\
 0  &   -\frac{h}{2} \delta_{\bm{k,k'}}       & -\overline F_{-\bm Q}^{A}(\bm{k',k})         &   F_{-\bm Q}^{B}(\bm{k',k},-i\omega)
\end{array} \right), 
\end{equation}
\noindent with matrix elements  
\begin{equation}
 F_{\bm Q}^{B}(\bm{k,k'},i\omega) = \frac{1}{2} i \omega \ \delta_{\bm{k,k'}} 
 \delta_{\omega,\omega'} + \frac{i \lambda_{k-k'}^{\omega-\omega'} }{2\sqrt{N_s\beta}} - \sum_{\delta>0} \frac{ J_{\delta}}{4\sqrt{N_s \beta}} 
 \left( W_{\bm{k-k'},\delta}^{B\ \omega-\omega'} e^{-i(\bm k'+ \frac{\bm Q}{2})\cdot \bm{\delta} } -  \overline W_{\bm{k'-k}, \delta}^{B\ \omega'-\omega} e^{i(\bm k + \frac{\bm Q}{2})\cdot \bm{\delta}} \right) \ ,
\end{equation}
\noindent and
\begin{equation}
 F_{\bm Q}^{A}(\bm{k,k'}) = \sum_{\delta>0}  \frac{J_{\delta}}{4\sqrt{N_s \beta}} W_{\bm{k-k'}, \delta}^{A\ \omega-\omega'} 
 \left( e^{i(\bm k + \frac{\bm Q}{2})\cdot \bm{\delta}} - e^{-i(\bm k'+ \frac{\bm Q}{2})\cdot \bm{\delta} }  \right) \ ,
\end{equation}
\end{widetext}
where $\bm \delta$ represents (half of) the vectors connecting the nearest neighours of the triangular lattice.

\subsection{The saddle point expansion} 

To evaluate Eq.~\eqref{generatriz} we  expand the effective action $S_{\rm eff}$ about its saddle point,~\cite{Auerbach1994} 
defined by the set of saddle point equations
\begin{equation} \label{spcond}
\frac{\partial S_{\rm eff}}{\partial \phi_{\alpha}} \bigg|_{\rm sp} = 
\frac{\partial \  S_0}{\partial \phi_{\alpha}}\bigg|_{\rm sp} + \frac{1}{2} \tr \bigg[\mathcal{G}^{\rm sp} \frac{\partial \mathcal{G}^{-1}}{\partial \phi_{\alpha}}\bigg] = 0,
\end{equation}
where   
$\phi_{\alpha}$ denotes the fields $\left\{ \overline W_{\bm{k},\delta}^{r  \omega},\ W_{\bm{k},\delta}^{r\omega},\  \lambda_{\bm k}^{\omega} \right\}$ 
($\alpha$ includes field, momentum, and frequency indices). The expansion of the effective action  becomes

\begin{equation}\label{Seffexp}
 S_{\rm eff} = S^{\rm sp}_{\rm eff} + \sum_{\alpha_1 \alpha_2} S_{\alpha_1 \alpha_2}^{(2)} \ \Delta \phi_{\alpha_1} \Delta \phi_{\alpha_2}  + S_{int},
\end{equation}
where $S^{\rm sp}_{\rm eff}$ corresponds to the effective action evaluated at the saddle point fields $\phi_{\alpha}^{\rm sp}$, the second term takes into account the auxiliary field fluctuations $\Delta \phi_{\alpha}^{} = \phi_{\alpha} - \phi_{\alpha}^{\rm sp}$ at the Gaussian level,  and the last term $S_{int}=\sum_{n=3}^{\infty} \sum_{\alpha_1 \cdots \alpha_n} S_{\alpha_1 \cdots \alpha_n}^{(n)} \ \Delta \phi_{\alpha_1} \cdots \Delta \phi_{\alpha_n}$ includes auxiliary field fluctuations of third and higher orders. The coefficients are defined as 
\begin{equation}\label{derivada}
S_{\alpha_1 \cdots \alpha_n}^{(n)}= \frac{1}{n!}\left.\frac{\partial^{n} S_{\rm eff}}{\partial \phi_{\alpha_1} \cdots \ \partial \phi_{\alpha_{n}}}\right|_{\rm sp},
\end{equation}
for $n\geq 2$. The corresponding partition function \eqref{generatriz} is
\begin{equation}
\mathcal{Z}[j] = e^{-S^{\rm sp}_{\rm eff}{(\overline{W}_{\rm sp},{W}_{\rm sp},\lambda_{\rm sp},j)}}\times\int D[\bar \phi,\phi] \ 
e^{-\Delta \vec \phi_{}^{\dagger}\cdot S_{}^{(2)} \cdot \Delta \vec \phi_{}+S_{int}}.\nonumber
\end{equation}
The first factor represents the partition function within the saddle point approximation, while the second one is the contribution from the 
fluctuations of the auxiliary fields: $\Delta \vec{\phi}^{\dagger} = \vec{\phi}^{\dagger} - \vec{\phi}^{{\rm sp} \dagger} $  with 
$\vec{\phi}^{\dagger} = \left( \overline W_{\bm k,\delta}^{r \omega},\ W_{-\bm k,\delta}^{r  -\omega},\  \lambda_{-\bm k}^{-\omega} \right)$.

\subsubsection{Saddle Point Approximation}

For a static and homogeneous saddle point solution, the Fourier transformed fields satisfy 
 $\left.\phi_{\bm{k}}^{\omega}\right|_{\rm sp} = \sqrt{N_s \beta} \; \phi \; 
 \delta_{\bm{k,0}} \delta_{\omega,0}$, for $\phi=\overline W_{\delta}^{r}, W_{\delta}^{r}, \lambda$. 
 We consider the ansatz 
\begin{eqnarray}
 W_{\delta}^{A}\big|_{\rm sp} = iA_{\delta}, & \ & \  \overline W_{\delta}^{A}\big|_{\rm sp} = -iA_{\delta}, \nonumber\\   
 W_{\delta}^{B}\big|_{\rm sp} = -B_{\delta}, & \ & \  \overline W_{\delta}^{B}\big|_{\rm sp} = B_{\delta} \ , \label{ansat} \\  
 \lambda|_{\rm sp} &=& i \lambda, \nonumber
\end{eqnarray}
with $A_{\delta}$, $B_{\delta}$, and $\lambda$ real, which is consistent with magnetic ordering in the $xy$ plane.~\cite{Ghioldi2015}
While $\overline {W}_{\delta}^{A}|_{\rm sp} = (W_{\delta}^{A}|_{\rm sp})^*$, it turns out that  
$(W_{\delta}^{B}|_{\rm sp})^* = - \overline W_{\delta}^{B}|_{\rm sp}$.~\cite{Trumper1997,Flint2009}
This corresponds to a \textit{distorted} saddle point solution as a consequence of the sign difference between the  
$\overline{W}^{A}A$ and $\overline{W}^{B}B$ terms in the HS decouplings of Eqs.~(\ref{HSAA}) and ~(\ref{HSBB}). 
Furthermore, the real field $\lambda^{\tau}_i$ takes an imaginary value at the SP. 
To reach this distorted saddle point, it is necessary to perform an analytical continuation for computing the partition function.~\cite{Auerbach1994} 
The  resulting effective saddle point action describes a system of non-interacting bosons coupled with the static and homogeneous SP auxiliary fields: 

\begin{widetext}

\begin{equation}\label{Seff-sp}
S_{\rm eff}^{\rm sp} (\overline W_{\rm sp}, W_{\rm sp},\lambda_{\rm sp}) = -N_s\beta \sum\limits_{\delta> 0}  J_{\delta} \left( B_{\delta}^{2} - A_{\delta}^{2} \right)  - 2SN_s\beta \lambda- \ln \mathcal{Z}_{\rm bos}^{\rm sp},  
\end{equation}
where $ \mathcal{Z}_{\rm bos}^{\rm sp} = \int d [\bar b,b] \ e^{-S_{\rm bos}^{\rm sp}} $, $ S_{\rm bos}^{\rm sp} = 
\sum_{\bm k, \omega} \vec b_{\bm k, \omega}^{\dagger} \cdot \mathcal{M}_{\bm k, \omega}^{\rm sp} \cdot \vec b_{\bm k, \omega}$ and 
$\mathcal{M}^{\rm sp}$ is the  dynamical matrix of the bosons evaluated at the saddle point solution,

\begin{equation}\label{dynmat}
 \mathcal{M}_{\bm{k}, \omega}^{\rm sp} = {
 \left(
\begin{array}{cccc}
 (i \omega  + \lambda + \gamma_{\bm{k+\frac{Q}{2}}}^{B}) e^{-i\omega \eta}  &   -\gamma_{\bm{k +\frac{Q}{2}}}^{A}  &   -\frac{h}{2} &  0  \\ \\
 -\gamma_{\bm{k+\frac{Q}{2}}}^{A}  &  (-i \omega + \lambda + \gamma_{\bm{k+\frac{Q}{2}}}^{B} ) e^{i\omega \eta}    &  0  & -\frac{h}{2}  \\  \\
 -\frac{h}{2} &  0  &  (i \omega  + \lambda + \gamma_{\bm{-k + \frac{Q}{2}}}^{B} ) e^{-i\omega \eta}  &  -\gamma_{\bm{-k+\frac{Q}{2}}}^{A} \\  \\
 0  &   -\frac{h}{2}  &  -\gamma_{\bm{-k+\frac{Q}{2}}}^{A}  &  (-i \omega +  \lambda +  \gamma_{\bm{-k+\frac{Q}{2}}}^{B} ) e^{i\omega \eta}
\end{array} \right)} \ ,
\end{equation} \\
\end{widetext}
with
\begin{eqnarray}
 \gamma_{\bm k}^{A} &=& \sum\limits_{\delta>0} J_{\delta} A_{\delta} \sin \left(\bm{k\cdot\delta}\right), \\   
 \gamma_{\bm k}^{B} &=& \sum\limits_{\delta>0} J_{\delta} B_{\delta} \cos \left(\bm{k\cdot\delta}\right). 
 \end{eqnarray}
The convergence factors $e^{\pm i\omega \eta}$ with $\eta=0^{+}$ arise from the time ordering of the functional integral and they are crucial for the Matsubara frequency sum in Eq.~\eqref{spcond} to be well defined.~\cite{Timm1998}

\noindent The single spinon Green's function is obtained by computing $\mathcal{G}^{\rm sp} = (\mathcal{M}^{\rm sp})^{-1}$,
\begin{equation}\label{green}
 \mathcal{G}^{\rm sp}(\bm k, i\omega)= \sum_{\sigma} \frac{g_{\bm k}^{- \sigma}}{i\omega - \varepsilon_{\bm k \sigma}}\!  + \! 
 \frac{g_{\bm k}^{+\sigma}}{i\omega + \varepsilon_{\bm k \sigma}}\ ,
\end{equation}
($\sigma = \pm$) where the two band spinon relation dispersion are
\begin{equation}
\varepsilon_{\bm k \sigma}  = \sqrt{ \frac{1}{2} \left[ \left( \alpha_{\bm{k+\frac{Q}{2}}}^{2} + \alpha_{\bm{-k+\frac{Q}{2}}}^{2} + \frac{h^{2}}{2} \right) +
\sigma \Delta_{\bm k}^{2} \right] } \ ,    
\end{equation}
with  $\alpha_{\bm{k}}^{2} = \left(\lambda + \gamma_{\bm{k}}^{B} \right)^{2} - \left( \gamma_{\bm{k}}^{A} \right)^{2},$ and
\begin{widetext}
\begin{equation}
 \Delta_{\bm k}^{2} = \sqrt{ \Big(\alpha_{\bm{k+\frac{Q}{2}}}^{2} - \alpha_{\bm{-k+\frac{Q}{2}}}^{2} \Big)^{2}   + \bigg[\Big((\lambda+\gamma_{\bm{k+\frac{Q}{2}}}^{B})+(\lambda + \gamma_{\bm{-k+\frac{Q}{2}}}^{B}) \Big)^{2} - \Big( \gamma_{\bm{k+\frac{Q}{2}}}^{A} - \gamma_{\bm{-k+\frac{Q}{2}}}^{A} \Big)^{2} \bigg] \; h^{2} } .
\end{equation}

\noindent The $4\times4$ matrices $g_{\bm k}^{\pm \; \sigma}$ are 
 \begin{equation}
  g_{\bm k}^{+\;\sigma} = \left(
  \begin{array}{cccc}
   E_{\bm k \sigma}  & C_{\bm k \sigma}  & F_{\bm k \sigma}   & D_{-\bm k \sigma} \\
   C_{\bm k \sigma}  & A_{\bm k \sigma}  & D_{\bm k \sigma}   & B_{\bm k \sigma}  \\
   F_{\bm k \sigma}  & D_{\bm k \sigma}  & E_{-\bm k \sigma}  & C_{-\bm k \sigma} \\
   D_{-\bm k \sigma} & B_{\bm k \sigma}  & C_{-\bm k \sigma}  & A_{-\bm k \sigma} \\
  \end{array} \right) \ \quad
  g_{\bm k}^{-\; \sigma} = -\left(
  \begin{array}{cccc}
   A_{\bm k \sigma} & C_{\bm k \sigma}  & B_{\bm k \sigma}   & D_{\bm k \sigma} \\
   C_{\bm k \sigma} & E_{\bm k \sigma}  & D_{-\bm k \sigma}  & F_{\bm k \sigma} \\
   B_{\bm k \sigma} & D_{-\bm k \sigma} & A_{-\bm k \sigma}  & C_{-\bm k \sigma} \\
   D_{\bm k \sigma} & F_{\bm k \sigma}  & C_{-\bm k \sigma}  & E_{-\bm k \sigma} \\
  \end{array} \right),
 \end{equation}
with matrix elements 
\begin{equation}
 A_{\bm k \sigma} = v_{\bm k \sigma}^{2} \frac{\Delta_{-\bm k \sigma}^{2}}{\Delta_{\bm k}^{2}} +\sigma u_{-{\bm k\sigma}}^{2} \frac{h^{2}}{4\Delta_{\bm k}^{2}},\qquad 
 E_{\bm k \sigma} = u_{\bm k \sigma}^{2} \frac{\Delta_{-\bm k \sigma}^{2}}{\Delta_{\bm k}^{2}} +\sigma v_{-\bm k \sigma}^{2} \frac{h^{2}}{4\Delta_{\bm k}^{2}},\qquad 
 C_{\bm k \sigma} = z_{\bm k \sigma} \frac{\Delta_{-\bm k \sigma}^{2}}{\Delta_{\bm k}^{2}} -\sigma z_{-\bm k \sigma} \frac{h^{2}}{4\Delta_{\bm k}^{2}},
 \end{equation}
 \begin{equation}
  B_{\bm k \sigma} = -\sigma \varepsilon_{\bm k \sigma} \left[ v_{\bm k \sigma}^{2} v_{-\bm k \sigma}^{2} + z_{\bm k \sigma} z_{-\bm k \sigma}- 
  \left(\frac{h}{4\varepsilon_{\bm k \sigma}} \right)^{2} \right] \frac{h}{\Delta_{\bm k}^{2}} \ , \quad
  F_{\bm k \sigma} = -\sigma \varepsilon_{\bm k \sigma} \left[ u_{\bm k \sigma}^2 u_{-\bm k \sigma}^{2} + z_{\bm k \sigma} z_{-\bm k \sigma} - 
  \left(\frac{h}{4\varepsilon_{\bm k \sigma}} \right)^{2} \right] \frac{h}{\Delta_{\bm k}^{2}} \ , \quad
 \end{equation}
\end{widetext}
 and
 \begin{equation}
  D_{\bm k \sigma} = -\sigma \varepsilon_{\bm k \sigma} \left[ v_{\bm k \sigma}^2 z_{-\bm k \sigma} + 
  u_{-\bm k \sigma}^2 z_{\bm k \sigma} \right] \frac{h}{2\Delta_{\bm k}^{2}} \ ,
 \end{equation}
where
 \begin{equation}
  u_{\bm k \sigma}^{2} =  \frac{ \lambda + \gamma_{\scriptscriptstyle{\bm{k+\frac{Q}{2}}}}^{B} }{2 \varepsilon_{\bm k \sigma}} + \frac{1}{2}  \ ,  \qquad
  v_{\bm k \sigma}^{2} =  \frac{ \lambda + \gamma_{\scriptscriptstyle{\bm{k+\frac{Q}{2}}}}^{B} }{2 \varepsilon_{\bm k \sigma} } - \frac{1}{2}  \ ,
  \end{equation}
  \begin{equation}
  z_{\bm k \sigma} = \frac{ \gamma_{\scriptscriptstyle{\bm{k+\frac{Q}{2}}}}^{A} }{ 2\varepsilon_{\bm k \sigma}} \ , \qquad   
  \Delta_{\bm k \sigma}^{2} = \sigma \left(\varepsilon_{\bm k \sigma}^{2} - \alpha_{\bm{k+\frac{Q}{2}}}^{{2}} \right) \ .
 \end{equation}
\noindent By replacing the Green's functions and the dynamical matrix given in Eqs.~\eqref{green} and \eqref{dynmat}, respectively, the saddle point condition Eq.~\eqref{spcond} at $T=0$, yields the following self-consistent equations  for the mean field parameters $A_{\delta}$, $B_{\delta}$, and $\lambda$,
 \begin{eqnarray}\label{selfeqs}
  A_{\delta} & = & \frac{1}{N_s} \sum_{\bm k \sigma} C_{\bm k \sigma} \sin\left(\bm k+\frac{\bm Q}{2}\right)\cdot \bm{\delta} \nonumber, \\
  B_{\delta} & = & \frac{1}{N_s} \sum_{\bm k \sigma} A_{\bm k \sigma} \cos\left(\bm k+\frac{\bm Q}{2}\right)\cdot \bm{\delta},  \\
  S \ & = & \frac{1}{N_s} \sum_{\bm k \sigma} A_{\bk \sigma}\nonumber .  \qquad \qquad \quad \quad  
 \end{eqnarray}

\noindent These equations coincide with the Schwinger boson mean field theory (SBMFT). In particular, the usual system of equations for a singlet ground state is recovered for $h=0$,~\cite{Mezio2011}
\begin{eqnarray}
 A_{\delta} &=& \frac{1}{N_s} \sum_{\bm k} \frac{\gamma_{\bm k}^{A}}{2 \alpha_{\bm k}} \sin(\bm{k\cdot\delta}), \nonumber \\
 B_{\delta} &=& \frac{1}{N_s} \sum_{\bm k} \frac{\lambda + \gamma_{\bm k}^{B}}{2 \alpha_{\bm k}} \cos(\bm{k\cdot\delta}), \\
 S + \frac{1}{2} &=& \frac{1}{N_s} \sum_{\bm k} \frac{\lambda + \gamma_{\bm k}^{B}}{2 \alpha_{\bm k}}, \nonumber
\end{eqnarray}
where $\alpha_{\bm k}$ is the spinon dispersion relation in a global reference frame for $h=0$.

\subsubsection{Gaussian fluctuation approximation}

The $S_{int}$ term of Eq.~\eqref{Seffexp} is neglected within the Gaussian fluctuation approximation, so as to keep the field fluctuations in the effective action up to quadratic order,
\begin{equation}\label{Seff-fl}
 S_{\rm eff} \simeq S^{\rm sp}_{\rm eff} + \sum_{\alpha_1 \alpha_2} \Delta \vec \phi_{\alpha_1}^{\dagger} \cdot S_{\alpha_1 \alpha_2}^{(2)} \cdot \Delta \vec \phi_{\alpha_2} \ . 
\end{equation}
\noindent The coefficients of the quadratic terms $S^{(2)}$ define the fluctuations matrix 
\begin{eqnarray}\label{flucmatrix}
 S_{\alpha_1 \alpha_2}^{(2)}\!\!\! &=& \frac{1}{2} \ \frac{\partial^2 S_{\rm eff}}{\partial \phi_{\alpha_1} \partial \phi_{\alpha_2} } \bigg|_{\rm sp} = \\
                             &=&\frac{1}{2} \ \Bigg\{ \frac{\partial^2 S_{0}}{\partial \phi_{\alpha_1} \partial \phi_{\alpha_2} } - \frac{1}{2} \tr \Big[ \mathcal{G}^{\rm sp} \ v_{\phi_{\alpha_1}} \     \mathcal{G}^{\rm sp}\ v_{ \phi_{\alpha_2}}    \Big] \Bigg\} \equiv \nonumber \\ 
                             &\equiv&  \Big( \Pi_{0} - \Pi \Big)_{\alpha_1 \alpha_2} \nonumber .
\end{eqnarray}
$\Pi_{0}$ is a diagonal matrix containing the coupling constant $J_{\bm \delta}$ in the diagonal, and a zero in the $\lambda\lambda$ element, $\Pi$ is the so-called polarization matrix, and $v_{\phi} = \frac{\partial \mathcal{G}^{-1}}{\partial \phi}$ are the internal vertices, i. e., the derivatives  of the bosonic dynamical matrix with respect to the auxiliary fields.~\cite{Auerbach1994}\\ 

The distorted character of the SP has two important consequences. The first one is the need to perform an analytic continuation of the real and the imaginary parts of the auxiliary fields in the Gaussian integral 

\begin{equation}\label{intgauss}
\int D[\bar \phi, \phi] \ e^{-\Delta \vec \phi^{\dagger}  \cdot S^{(2)} \cdot \Delta  \vec \phi}, 
\end{equation}
in order to pass through the SP. The second one is the non-Hermiticity of the fluctuation matrix.

Regarding the first issue, the analytical continuation leads to a $\Delta \vec \phi^{\dagger}=\vec\phi^{\dagger}-\vec\phi^{{\rm sp}\dagger}$ in Eq.~\eqref{intgauss} that is not necessarily the Hermitian conjugated of $\Delta \vec \phi$. One way to circumvent this problem in the evaluation of the Gaussian integral is to make a simple change of integration variables, $\Delta\vec\phi\to\vec\phi$, consisting of the rigid shifts of the real and the imaginary parts of the auxiliary field axes, so that the SP is shifted to the origin of the integration domain and, hence, $\vec\phi^{\dagger}$ and $\vec\phi$ become conjugate variables. In fact, as the integrand is an analytical function of the real and imaginary parts of the auxiliary fields, this change of variables can be seen as a ``rectangular'' deformation of the real and imaginary parts of the auxiliary field axes.

On the other hand, due to the non-Hermiticity of the fluctuation matrix, the stability of the Gaussian fluctuations about the SP must be analized carefully. In  particular, the convergence of the Gaussian integral
$\int D [\bar \phi, \phi] e^{- \vec \phi^{\dagger} \cdot S^{(2)} \cdot \vec\phi},$
is given by the {\it positive-stable} condition of the fluctuation matrix (see Appendix \ref{appendix_gaussian}). 
A matrix is positive-stable if all of its eigenvalues have a positive real part. 
This condition is less restrictive than the usual requirement of positive-definiteness of the Hermitian part of the matrix.~\cite{Negele1998} 
In fact, in our case we have found that for the TLHM the fluctuation matrix is always positive-stable while its Hermitian part is indefinite.\\

It is worth mentioning that, if the positive-stable condition is satisfied, the Gaussian integral (\ref{intgauss}) can be evaluated alternatively through the steepest descent method.~\cite{Auerbach1994}
In this method, the real and imaginary parts of the auxiliary field axes are deformed such that the integrand, along the deformed path, has a constant phase close to the SP. Mathematically, the expected optimal direction of the path is obtained from the singular value decomposition (SVD) of the fluctuation matrix:
\begin{eqnarray}\label{svd}
 S^{(2)} = U \cdot \Sigma \cdot V^{\dagger}, 
\end{eqnarray}
where $U,V$ are complex unitary matrices and $\Sigma = \text{diag}[s_{1},...,s_{n}]$ is a semi-positive definite diagonal matrix with $n\times n$ being the dimension of the fluctuation matrix.
The steepest descent path is then given by

\begin{eqnarray}\label{path}
\vec \phi^{\dagger} = \vec \phi^{H}\cdot V \cdot U^{\dagger},
\end{eqnarray}

\noindent where the superscript $H$ denotes the Hermitian conjugate of $\vec \phi$.
Notice that along the steepest descent path $\vec \phi^{\dagger}$ and $\vec \phi$ may not be conjugate variables. Substituting Eq.~\eqref{path} into the Gaussian effective action yields

\begin{equation}\label{Seff-flh}
 S_{\rm eff} \simeq S^{\rm sp}_{\rm eff} + \sum_{\alpha_1 \alpha_2} \Delta \vec \phi_{\alpha_1}^{H} \cdot \tilde{S}_{\alpha_1 \alpha_2}^{(2)} \cdot  \Delta \vec \phi_{\alpha_2} \ . 
\end{equation}
with $\tilde{S}^{(2)} = V\cdot \Sigma\cdot V^{\dagger},$ which is Hermitian and semi-positive definite. So, this reformulation allows us to work with, both, a pair of conjugate variables, $\Delta\vec\phi^{H}$ and $\Delta\vec\phi$, and a Hermitian fluctuation matrix. However, it should be stressed that the stability of the Gaussian fluctuation approximation is given by the positive-stable condition of the original $S^{(2)}$ fluctuation matrix. In our case of the TLHM we have checked 
that the two procedures mentioned above, using non-Hermitian and Hermitian fluctuation matrices, give always the same quantitative results.

\subsection{Ground state energy}
In this subsection we calculate the ground state energy within the Gaussian fluctuation approximation aiming to the evaluation of the local magnetization. For this purpose we must compute the partition function at the Gaussian level with a finite symmetry breaking field $h$ (and $\mathcal{J}_s$ switched off). The partition function is given by

\begin{equation}
 \mathcal{Z}^{(2)}(h) = e^{-S_{\rm eff}^{\rm sp}} \times \int D[\vec \phi^{\dagger}, \vec \phi] \ e^{-\vec \phi_{}^{\dagger}\cdot S_{}^{(2)} \cdot \vec \phi_{}}.
 \label{zgauss}
\end{equation}

Within the SP approximation, the ground state energy is obtained by taking the limit $T\to 0$ of $F^{\rm sp}=S^{\rm sp}_{\rm eff}/\beta$. However, to get the correct zero point energy one should mantain the discrete character of the imaginary time $\tau$ (a very tedious procedure) to respect the correct equal-time ordering in the path integral. Alternatively, one can keep track of this time ordering by transforming the conjugate bosonic field as $\overline{b}_{\bm k}^{\omega}\to \overline{b}_{\bm k}^{\omega} e^{-i\omega \eta}$, with the convergence factor $\eta$  mentioned previously. In this case the ground state energy ({\it per} site) turns out to be
\begin{eqnarray}
E^{\rm sp}_{\rm gs}(h) &  =  & \frac{1}{2 N_s} \sum_{\bm k \sigma} \varepsilon_{\bm k \sigma}(h) - \label{egsp} \\ 
& - & \sum_{\delta >0} J_\delta \left( B^2_\delta -A^2_\delta \right)-  \lambda \left(2S+1\right),\nonumber
\end{eqnarray}
which is identical to the ground state energy derived within the canonical  mean field approximation.~\cite{Auerbach1994,Mezio2011}\\

Hereafter, we proceed to compute the Gaussian correction to the ground state energy. We use the original non-Hermitian fluctuation matrix $S^{(2)}$ and the  Gaussian integral in Eq.~\eqref{zgauss} yields, for a positive-stable matrix (see Appendix \ref{appendix_gaussian}), 
\begin{equation}
\mathcal{Z}^{\rm fl}(h) \equiv\int D[\vec \phi^{\dagger}, \vec\phi] \ e^{-\vec \phi^{\dagger}\cdot S^{(2)} \cdot \vec \phi} =\frac{1}{{\rm det} \;S_{}^{(2)}}.
\label{Gauss}
\end{equation} 
\noindent  Because of the redundant local $U(1)$ gauge degree of freedom introduced by the Schwinger boson representation [see Eq.~\eqref{transf}], the fluctuation matrix contains infinite zero gauge modes (one for each $\omega$ and ${\bm k}$  value),  corresponding to  the null space of $S^{(2)}$.  The artificial infinities arising from these zero gauge modes are avoided by integrating only over the genuine physical fluctuations (fluctuations with restoring force). This is formally done following the Faddeev-Popov prescription.~\cite{Trumper1997}\\

\noindent By applying an infinitesimal gauge transformation to the auxiliary fields, 
$\vec\phi \to \vec \phi_{}(\theta)$, the gauge fluctuations around the saddle point solution, $\vec\phi^{\rm sp}_{} (\theta)=\vec\phi^{\rm sp}_{}+ \delta \vec \phi^{\rm sp}_{}(\theta)$, take the form
\begin{equation}
    \delta \vec\phi^{\rm sp}_{}(\theta) =  \left(
  \begin{array}{c}
   \delta W_{\bm k, \delta}^{r\ \omega}(\theta)     \\
     \delta \overline W_{-\bm k, \delta}^{r \ -\omega}(\theta)  \\
      \delta \lambda_{\bm k}^{\omega}(\theta)  \\
  \end{array} \right) ,   
 \end{equation}
\noindent where
\begin{eqnarray}
 \delta W_{\bm k, \delta}^{r \;\omega}(\theta) & = & i (1\pm e^{i \bm k  \cdot \bm \delta}) \left.{W_{\delta}}^{r}\right|_{\rm sp} 
 \theta_{{\bm k}\omega}, \nonumber \\
  \delta \overline{W}_{-\bm k, \delta}^{r \; -\omega}(\theta) & = & -i (1 \pm e^{i \bm k \cdot \bm \delta}) \left. \overline W_{\delta}^{r}\right|_{\rm sp} \theta_{{\bm k}\omega}, \\
 \delta \lambda_{\bm k}^{\omega}(\theta) &=&  -i \omega_n \theta_{{\bm k}\omega} \nonumber.
\end{eqnarray}

\noindent As the fluctuation matrix is non-Hermitian, the right (R) and left (L) zero gauge modes of $S^{(2)}$, corresponding to each $(\bm k, i\omega)$, are not necessarily Hermitian conjugate. In particular, the right zero mode is computed as $\vec\phi^R_{0 }={\partial \vec\phi^{\rm sp}_{}(\theta)}/{\partial  \theta_{{\bm k} \omega}}$; while the left zero mode is defined by $\vec\phi^L_{0 }={\partial \vec\phi^{sp \dagger}_{}(\overline\theta)}/{\partial  \overline\theta_{{\bm k} \omega}}$:
\begin{equation}
 \vec\phi^R_0(\bk,i\omega_n) = i\left(
 \begin{array}{c}
 \left(1+e^{i\bk\cdot \bm \delta}\right)\left.W^A_{\delta}\right|_{\rm sp} \\
- \left(1+e^{i\bk\cdot \bm \delta}\right) \left.\bar W^A_{\delta}\right|_{\rm sp} \\
   \left(1-e^{i\bk\cdot \bm \delta}\right)\left.W^B_{\delta}\right|_{\rm sp} \\
- \left(1-e^{i\bk\cdot \bm \delta}\right)\left.\bar W^B_{\delta}\right|_{\rm sp} \\
\omega_n
 \end{array}
 \right), \label{rightzero}
\end{equation}
and 
\begin{widetext} 
\begin{equation}
\vec\phi^L_{0}(\bk,i\omega_n)  =  -i\Big(\left(1+e^{-i\bk\cdot \bm \delta}\right)\left.\bar W^{A}_{\delta}\right|_{\rm sp}, 
- \left(1+e^{-i\bk\cdot \bm \delta}\right) \left. W^{A}_{\delta}\right|_{\rm sp},
- \left(1-e^{-i\bk\cdot \bm \delta}\right) \left.\bar W^{B}_{\delta}\right|_{\rm sp},
\left(1-e^{-i\bk\cdot \bm \delta}\right) \left. W^{B}_{\delta}\right|_{\rm sp}, \omega_n\Big)\label{leftzero}
\end{equation}
\end{widetext}

To get rid of the redundant gauge fluctuations of the auxiliary fields
we impose the {\it natural} gauge conditions~\cite{Trumper1997}
\begin{equation} 
g(\overline\theta)=\vec \phi^{\dagger}(\overline\theta) \cdot \vec \phi^R_0=0,\;\;\;\;\; 
h(\theta)=\vec \phi_0^{L} \cdot \vec \phi(\theta)=0,
\end{equation}
by means of the Faddeev-Popov trick, that consists of expressing the identity as
\begin{equation} 
1= \Delta_{\rm FP} (\vec\phi^{\dagger},\vec\phi) \times \int \frac{d\overline \theta d\theta}{2\pi i} \delta(g(\overline\theta), h(\theta) ).
\label{identity}
\end{equation}
Here the Dirac delta function has been generalized to the complex plane (see Appendix \ref{appendix_fp}) and $\Delta_{\rm FP} (\vec\phi^\dagger, \vec\phi)$ is the so-called Faddeev-Popov determinant which, at SP level, is given by 
\begin{equation}
 \Delta_{\rm FP}={\vec \phi^{L}_0} \cdot \vec \phi^R_0.
\end{equation}
Using Eqs. \eqref{rightzero} and \eqref{leftzero}, the explicit expression of the Faddeev-Popov determinant is 
\begin{eqnarray}
\Delta_{\rm FP}({\bm k},i \omega_n)\!\!&=&\!\! 4\sum_{\bm\delta} \left[(1+\cos {\bm k} \cdot {\bm \delta}) A^2_{\bm \delta}\right. \nonumber \\
&-& \left. (1-\cos {\bm k} \cdot {\bm \delta}) B^2_{\bm \delta}\right] + \omega_n^2. \label{detFP}
\end{eqnarray}

\noindent After introducing the identity \eqref{identity} in the Gaussian integral \eqref{Gauss} and drawing upon the gauge invariance of the action and the measure, we obtain (see Appendix \ref{appendix_fp}) 
\begin{equation}
\mathcal{Z}^{\rm fl}(h)= \prod_{{\bm k},i\omega_n>0}  \frac{\Delta_{\rm FP}({\bm k},i \omega_n)}{{\rm det} \; S_{\perp}^{(2)}\!({\bm k}, i\omega_n)} ,
\end{equation} 

\noindent where $S_{\perp}^{(2)}\!({\bm k}, i \omega_n)$ 
is the projection of the fluctuation matrix $S^{(2)}({\bm k}, i\omega_n)$ onto the subspace orthogonal to the zero gauge modes for $({\bm k},i\omega_n)$. The determinant
of  $S^{(2)}_{\perp}$ is simply the product of all the non-zero (complex) eigenvalues of the fluctuation matrix $S^{(2)}$.\\

\noindent The Gaussian correction to the free energy can then be expressed as
\begin{eqnarray}
F^{\rm fl}(h) &=& -\frac{1}{\beta}\ln \left(\mathcal{Z}^{\rm fl}(h)/\mathcal{Z}_0\right)= \nonumber\\ &=&-\frac{1}{2\beta} \sum_{{\bm k},i\omega_n} \ln\left[ \frac{\Delta_{\rm FP}({\bm k},i \omega_n)}{\omega^2_n {\;\rm det} \; S_{\perp}^{(2)}\!({\bm k}, i\omega_n)} \right]. 
\end{eqnarray}
In this expression $\mathcal{Z}_0$ corresponds to the Gaussian correction of the partition function when all the Hamiltonian parameters (exchange interactions, external magnetic fields) are set to zero. This normalization by $\mathcal{Z}_0$ 
fixes the zero reference  level of the free energy (see Appendix \ref{appendix_fp}).\\  
  
\noindent  At $T=0$, we get the following Gaussian correction to the ground state energy 

\begin{equation}
E_{\rm gs}^{\rm fl}(h) \!= \!-\frac{1}{4\pi N_s} \int^{\infty}_{-\infty}\!\!\! d\omega \sum_{\bm k} \ln\left[ \frac{\Delta_{\rm FP}({\bm k},\omega)}{\omega^2 \; {\rm det} \; S_{\perp}^{(2)}\!({\bm k}, \omega)} \right].
\label{EGS}
\end{equation}
Within the Gaussian fluctuation approximation the ground state energy is $E_{\rm gs}(h)=E_{\rm gs}^{\rm sp}(h)+E_{\rm gs}^{\rm fl}(h) $. It can be shown that this expression, in a $1/N$ expansion, includes all the terms up to $1/N$ order.~\cite{Auerbach1994} It has been shown that the Gaussian fluctuation approximation yields a ground state energy that compares quantitatively very well with numerical predictions for several Heisenberg models.~\cite{Trumper1997,Manuel1998,Manuel1999, Gonzalez2017} \\

Alternatively, one can avoid the Faddeev-Popov prescription by fixing the gauge phase, $\theta^{\tau}_i$, in Eq.~\eqref{transf} such that the transformed $\lambda$ field becomes 
$\tau-$independent. In this case, the $\lambda$ fluctuations are only restricted to the $\omega=0$ subspace. As $\omega=0$ is a single point in the logarithmically convergent integral (\ref{EGS}) we can discard the fluctuations coming from this subspace. On the other hand, for $\omega \neq 0$ the fluctuation matrix $S^{(2)}$ must be truncated by eliminating the $\lambda-$column and row. We call this matrix the
{\it truncated} fluctuation matrix $S^{(2)}_{\rm tr}$. Owing to the gauge fixing,  $S^{(2)}_{\rm tr}$ has no zero gauge modes anymore. Hence, one can evaluate directly the  Gaussian integral (\ref{Gauss}) and the Gaussian correction to the ground state energy results,~\cite{Trumper1997}

\begin{equation}\label{EGStrunc}
E_{\rm gs}^{\rm fl}(h) = -\frac{1}{4\pi N_s} \int^{\infty}_{-\infty} d\omega \sum_{\bm k} \ln\left[ \frac{1}{ {\rm det} \; S_{\rm tr}^{(2)}\!({\bm k}, \omega)} \right]. 
\end{equation}

\noindent In Appendix \ref{appendix_determinants} we analytically show that both expressions for $E^{\rm fl}_{\rm gs}(h)$, Eqs. (\ref{EGS}) and (\ref{EGStrunc}), are identical. Furthermore, we  have numerically checked that this identity is fulfilled. \\

\subsection{Dynamical Spin Susceptibility}
\noindent The dynamical spin susceptibility in the frequency and momentum space,~\cite{Auerbach1994,Shindou2013}
\begin{equation}
 \chi_{\mu\nu}(\bm q,i\omega) = \frac{\partial^{2} ln \mathcal{Z}[j]}{\partial j_{\bm q,i\omega}^{\ \mu} \ \partial j_{-\bm q,-i\omega}^{\ \nu}} \bigg|_{j=0} \ .
\end{equation}
\noindent  can be separated in two contributions: 
\begin{equation}
 \chi = \chi_{_I} + \chi_{_{II}} 
\end{equation}

\begin{widetext}
\begin{equation}\label{chi1}
 \chi_{_{I} \mu\nu}(\bm q,i\omega) = \frac{1}{2\mathcal{Z}[j=0]} \int D[\overline \phi, \phi]\ \tr \Big[ \mathcal{G} \ u^{\mu}(\bm q,i\omega) \mathcal{G} \ u^{\nu}(-\bm q,-i\omega) \Big] \ \ e^{-S_{\rm eff}(\overline \phi, \phi,j=0)}   
\end{equation}
and
\begin{equation}\label{chi2}
 \chi_{_{II} \mu\nu}(\bm q,i\omega) = \frac{1}{4\mathcal{Z}[j=0]} \int D[\overline \phi,\phi] \ \tr \Big[ \mathcal{G} \ u^{\mu}(\bm q,i\omega) \Big] \tr \Big[ \mathcal{G} \ u^{\nu}(-\bm q,-i\omega) \Big] \ \ e^{-S_{\rm eff}(\overline \phi, \phi,j=0)}  \ ,
\end{equation}
\end{widetext}
\noindent where $ u^{\mu}(\bm q,i\omega) = \partial \mathcal{G}^{^{-1}} \! / \partial j_{\bm q,i\omega}^{\; \mu} $ is the external vertex, with $\mu=x,y,z$. The traces go over momentum, Matsubara frequency, and boson indices. The analytic continuation $ i\omega \rightarrow \omega + i 0^{+}$ yields the real and imaginary parts of the dynamical spin susceptibility. At $T=0$, the latter is related with the dynamical structure factor as,
\begin{equation}
 \mathcal{S}(\bm q, \omega) = -\frac{1}{\pi} {\rm Im} \Big[\; \chi(\bm q, \omega)\Big]
\end{equation}

\begin{figure}[!t]
  \includegraphics[width=\columnwidth,bb=0 0 481 318]{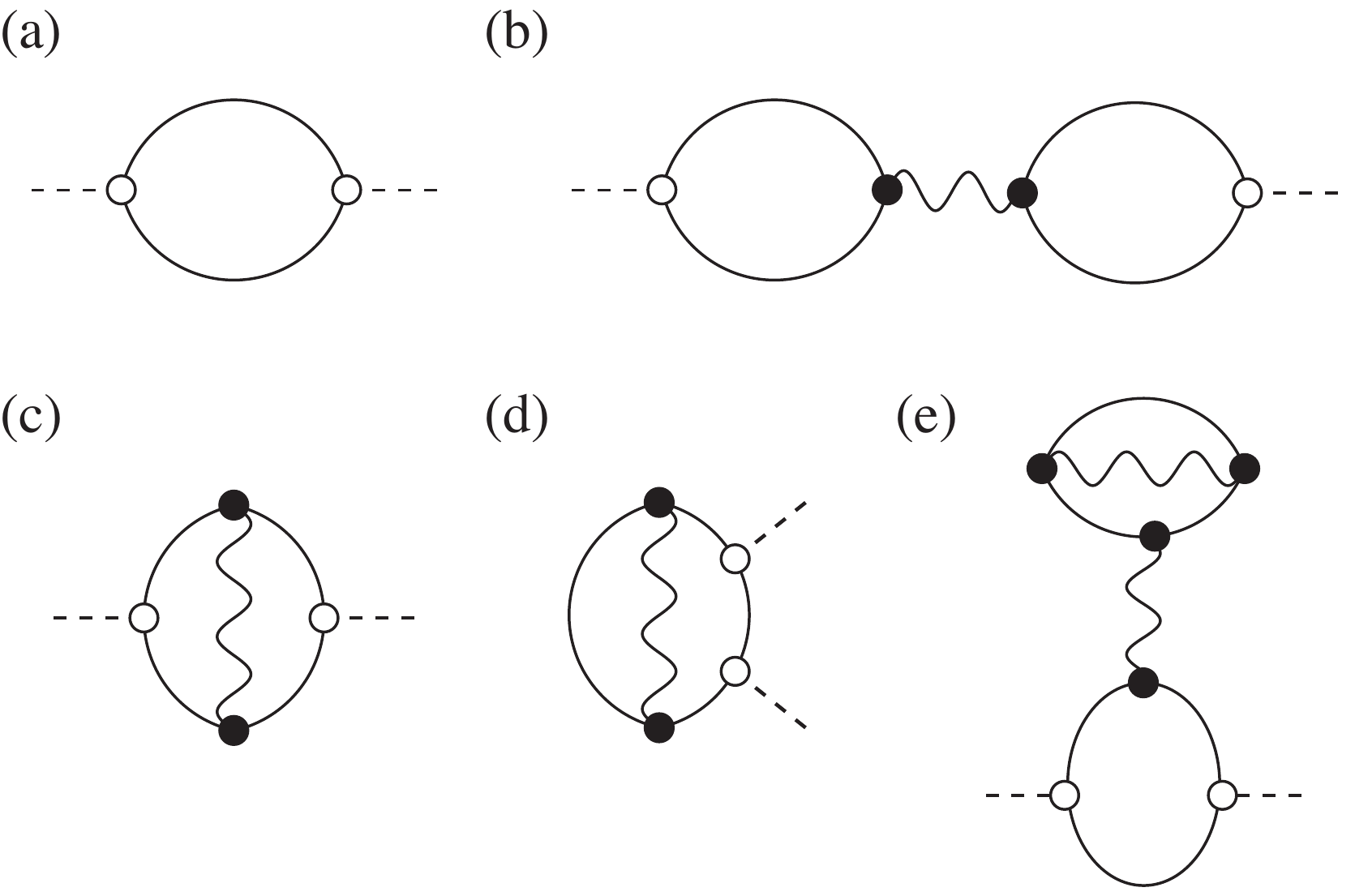}
 \caption{Diagrammatic representation of (a) saddle point contribution and  (b-e) the $1/N$ corrections. 
 In our calculation we only include the contribution (b) for reasons explained in the text. The diagram (c) corresponds to a vertex correction
 relative to (a), while the diagrams (d) and (e) include a Hartree-Fock correction of the single-spinon propagator. The dashed lines represent the external lines, the full lines represent  spinon propagators at the SP level and the wavy lines represent the RPA propagator.~\citep{Auerbach1994}}
 \label{fig2:feymann}
 \end{figure}

\subsubsection{Saddle Point Approximation }
\noindent In this approximation the dynamical spin susceptibility is computed by considering the saddle point effective action of Eq.~\eqref{Seff-sp} 
and the single spinon Green's function of Eq.~\eqref{green}. Each contribution is given by
\begin{equation} \label{chi1sp}
\chi_{_I \mu\mu}^{\rm sp}(\bm q,i\omega) = \frac{1}{2}\tr
 \left[ \mathcal{G}^{\rm sp}  u^{\mu}(\bm q,i\omega) \mathcal{G}^{\rm sp} u^{\mu}(-\bm q,-i\omega) \right],
\end{equation}\\
and
\begin{equation}
 \chi_{_{II}\mu\mu}^{\rm sp}(\bm q,i\omega)\! =\! \frac{1}{4}\tr\left[\mathcal{G}^{\rm sp} u^{\mu}(\bm q,i\omega) \right] 
 \tr\left[ \mathcal{G}^{\rm sp} u^{\mu}(-\bm q,-i\omega) \right].
\end{equation}
\noindent $\chi_{_{II} \mu\mu}^{\rm sp}$ vanishes because  $ \mathcal{G}^{\rm sp} u^{\mu} $ is traceless for each $\mu$ value. $\chi_{_I}^{\rm sp}$ is represented diagramatically in Fig.~1~(a) and it coincides with the $\chi_{}^{MF}$ that is obtained from  the SBMFT.~\cite{Mezio2011}

\subsubsection{Gaussian Fluctuations Approximation }
\noindent To compute  Gaussian corrections to $\chi_{ \; \mu\mu}^{\rm sp}(\bm q,i\omega)$,  the Green's function $\mathcal{G}$ that appears in Eqs.~(\ref{chi1}) and Eq. (\ref{chi2}) must be expanded around the saddle point
\begin{equation}\label{green-exp}
 \mathcal{G} \simeq \mathcal{G}^{\rm sp} - \sum_{\alpha} \bigg[\mathcal{G} \frac{\partial \mathcal{G}^{-1}}{\partial \phi_{\alpha}} \mathcal{G} \bigg]\bigg|_{\rm sp} \Delta \phi_{\alpha} . 
  \end{equation}
By replacing Eqs. \eqref{green-exp} and \eqref{Seff-fl} 
in Eq.~\eqref{chi1}, the Gaussian correction $\chi^{\rm fl}_I$ to the dynamical spin susceptibility results
\begin{widetext}
\begin{eqnarray}
\chi_{_{I} \mu\nu}^{\rm fl}(\bm q, i\omega) &=&  \sum_{\alpha_1 \alpha_2} \Bigg[ \frac{1}{2\mathcal{Z}^{(2)}}  \int D[\bar\phi,\phi] \ \Delta \overline \phi_{\alpha_1} \Delta \phi_{\alpha_2} \ e^{-\sum\limits_{\alpha \alpha'} \bar \phi_{\alpha} S_{\alpha \alpha'}^{(2)} \phi_{\alpha'}}      \Bigg] \times \nonumber \\
&\times& \Big\{Tr\big[ \mathcal{G}^{\rm sp} v_{\phi_{\alpha_1}} \ \mathcal{G}^{\rm sp} v_{\phi_{\alpha_2}} \mathcal{G}^{\rm sp} u^{\nu}(-\bm q, -i\omega) \mathcal{G}^{\rm sp} u^{\mu}(\bm q, i\omega) \big] + \nonumber \\ &+&  
Tr\big[ \mathcal{G}^{\rm sp} v_{\phi_{\alpha_1}} \ \mathcal{G}^{\rm sp} v_{\phi_{\alpha_2}} \mathcal{G}^{\rm sp} u^{\mu}(\bm q, i\omega) \mathcal{G}^{\rm sp} u^{\nu}(-\bm q, -i\omega) \big] +  \nonumber \\
&+& \tr\big[ \mathcal{G}^{\rm sp} v_{\phi_{\alpha_1}} \ \mathcal{G}^{\rm sp} u^{\mu}(\bm q, i\omega) \mathcal{G}^{\rm sp} v_{\phi_{\alpha_2}}  \mathcal{G}^{\rm sp} u^{\nu}(-\bm q, -i\omega) \big] \Big\}=  \nonumber \\
&=& \sum_{\alpha_{1}\alpha_{2}} D_{\alpha_{2} \alpha_{1}}(\bm q, i\omega) \big[ \Gamma_{\alpha_{1} \alpha_{2} \nu \mu}(\bm q, i\omega) 
+  \Gamma_{\alpha_{1} \alpha_{2} \mu \nu}(\bm q, i\omega) + \Gamma_{\alpha_{1} \mu \alpha_{2}\nu}(\bm q, i\omega)  \big],
\label{chiflIlast}
\end{eqnarray}

\noindent where 
\begin{equation} \label{RPA}
 D_{\alpha_{2} \alpha_{1}}(\bm q, i\omega)= \Bigg[ \frac{1}{\mathcal{Z}^{(2)}}  \int D[\bar\phi,\phi] \ \Delta \overline \phi_{\alpha_1} \Delta \phi_{\alpha_2} \ e^{-\sum\limits_{\alpha \alpha'} \bar \phi_{\alpha} S_{\alpha \alpha'}^{(2)} \phi_{\alpha'}}  \Bigg] = \big[ (S^{(2)})^{-1} \big]_{\alpha_{2} \alpha_{1}}
\end{equation}

\noindent is the RPA propagator and 
\begin{equation}
 \Gamma_{\alpha_{1} \alpha_{2} \mu \nu}(\bm q, i\omega) = \frac{1}{2} \tr\big[ \mathcal{G}^{\rm sp} v_{\phi_{\alpha_1}} \ \mathcal{G}^{\rm sp} v_{\phi_{\alpha_2}} \mathcal{G}^{\rm sp} u^{\mu}(\bm q, i\omega) \mathcal{G}^{\rm sp} u^{\nu}(-\bm q, -i\omega) \big].
\end{equation}

\noindent Replacing Eqs. \eqref{green-exp} and \eqref{Seff-fl} in Eq.~\eqref{chi2},  we obtain the Gaussian correction to the dynamical spin susceptibility 

\begin{eqnarray}
\chi_{_{II}\; \mu\nu}^{\rm fl}(\bm q, i\omega) &=&  \sum_{\alpha_1 \alpha_2} \frac{1}{2}\tr\big[ \mathcal{G}^{\rm sp} \ v_{\phi_{\alpha_1}} \ \mathcal{G}^{\rm sp} \ u^{\mu}(\bm q, i\omega) \big] \times 
\Bigg[ \frac{1}{\mathcal{Z}^{(2)}}  \int D[\bar\phi,D\phi] \ \Delta \overline \phi_{\alpha_1} \Delta \phi_{\alpha_2} \ e^{-\sum\limits_{\alpha \alpha'} \bar \phi_{\alpha} S_{\alpha \alpha'}^{(2)} \phi_{\alpha'}}      \Bigg] \times   \nonumber \\
&&\frac{1}{2} \tr\big[ \mathcal{G}^{\rm sp} \ v_{\phi_{\alpha_2}} \ \mathcal{G}^{\rm sp} \ u^{\nu}(-\bm q, -i\omega) \big]  = \sum_{\alpha_{1}\alpha_{2}} \Lambda_{\mu \alpha_{1}}(\bm q, i\omega) \ D_{\alpha_{2} \alpha_{1}}(\bm q, i\omega) \ \Lambda_{\nu \alpha_{2}}(-\bm q, -i\omega) \ ,
\label{chiflIIlast}
\end{eqnarray}
\end{widetext}
\noindent where 
\begin{equation}\label{lambda}
 \Lambda_{\mu \alpha}(\bm q, i\omega)\! =\!  \frac{\partial^2 S_{\rm eff}}{\partial j^{\mu} \partial \phi_{\alpha}}\! = \!\frac{1}{2} \tr\big[ \mathcal{G}^{\rm sp} v_{\phi_\alpha} \mathcal{G}^{\rm sp} u^{\mu}(\bm q, i\omega) \big] .
\end{equation}
The term $\chi_{_{II}}^{\rm fl}$ \eqref{chiflIIlast} is diagramatically represented in Fig.~\ref{fig2:feymann}~(b), while Fig.~\ref{fig2:feymann}~(c) and (d) are the diagrams corresponding to the terms $\chi_{_{I}}^{\rm fl}$ (\ref{chiflIlast}). 
In the context of a large-$N$ expansion, these Gaussian corrections to the dynamical 
susceptibility correspond to $1/N$ contributions.~\cite{Auerbach1988}
However, it can be shown~\cite{Shindou2013} that the full Gaussian corrections 
do not collect all the $1/N$ terms, because the diagram shown in Fig.~\ref{fig2:feymann}~(e), arising from  $S_{int}$  in Eq.~\eqref{Seffexp}, is also of order $1/N$. 
This diagram must then be added to $\chi_{_{I}}^{\rm fl}$ of Eq.~\eqref{chiflIlast} in order to produce the full $1/N$ correction to the dynamical spin susceptibility.
\\

In the following, we will only include the $1/N$ correction $\chi_{_{II}}^{\rm fl}$. There are two reasons for only including this contribution in a first approach to the problem. At the SP level, the local constraint of the SBs is relaxed to a global one, allowing for unphysical bosonic density fluctations (states that are outside the physical Hilbert space). In other words, the density (charge) susceptibility, $\chi^{\rm sp}_{\rm ch}$,  which  should be zero due to the local constraint \eqref{constr}, becomes finite at the SP level. Arovas and Auerbach demonstrated~\cite{Arovas1988} that the inclusion of the diagram shown in  Fig.~\ref{fig2:feymann}~(b)  cancels exactly the finite SP charge susceptibility, i.e., $\chi_{\rm ch}=\chi^{\rm sp}_{\rm ch}+\chi^{\rm fl}_{\rm ch}=0$. On the other hand, the inclusion of  diagrams (c) and (d) of Fig.~\ref{fig2:feymann} requires extra counter terms  to fulfil $\chi_{\rm ch}=0$.~\cite{Auerbach1994} The second reason for restricting to the diagram shown in  Fig.~\ref{fig2:feymann}~(b)  is that it is the only one that can introduce poles (collective modes) in the corrected dynamical susceptibility. Note that the poles of this diagram coincide with the poles of the RPA propagator because both $\chi_{_{II}}^{\rm fl}$ and $D_{\alpha_1, \alpha_2}$ are evaluated at the same 
${\bm q}$ and $\omega$.

Another important observation is that the diagram shown in  Fig.~\ref{fig2:feymann}~(b)  vanishes for a singlet
ground state ($h=0$). The simple reason is that $\Lambda_{\mu}$ of Eq.~\eqref{lambda} can be interpreted as a crossed susceptibility for two fields that belong to different representations of the symmetry group, SU(2), of ${\cal H}$ ( $j^{\mu}$ are components of a vector field,  while the auxiliary fields $\phi_{\alpha}$ are scalars).
Consequently, $\Lambda_{\mu}$ vanishes for a singlet ground state, leading to the cancellation of $\chi_{_{II}}^{\rm fl}$.
This result was found by Arovas and Auerbach thirty years ago~\cite{Arovas1988} and it is most likely the reason why  further attempts of computing $1/N$  corrections to $\chi$ are not found in the subsequent literature. A key observation of this manuscript is that  the diagram shown in  Fig.~\ref{fig2:feymann}~(b) {\it does not vanish} for the broken symmetry (magnetically ordered) ground state.

Some remarks are in order regarding the zero gauge modes and the computation of $\chi_{_{II}}^{\rm fl}$ through  Eq.~\eqref{chiflIIlast}. In principle, the computation of the two Gaussian integrals involved in the RPA propagator~\eqref{RPA} requires the use of the Faddeev-Popov prescription. However, this prescription can be circumvented due to the orthogonality between $\Lambda_{\mu}$ 
and the zero gauge modes. 
This orthogonality arises from the gauge invariance of the effective action, which implies $\frac{\partial S_{\rm eff}}{\partial \theta}=0$ (also $\frac{\partial S_{\rm eff}}{\partial \overline\theta}=0$) or 
\begin{equation}
\sum_{\alpha} \frac{\partial S_{\rm eff}}{\partial \phi_{\alpha}} \frac{\partial \phi_{\alpha}}{\partial \theta}=0.
\end{equation} 
\noindent This equation is valid for any value of the auxiliary fields. By taking the derivative  with respect to the external source $j^{\mu}$, we get 
\begin{equation}
\sum_{\alpha} \frac{\partial^2 S_{\rm eff}}{\partial j^{\mu}\partial \phi_{\alpha}} \frac{\partial \phi_{\alpha}}{\partial \theta}
+ \frac{\partial S_{\rm eff}}{\partial \phi_{\alpha}} \frac{\partial^2 \phi_{\alpha}}{\partial j^{\mu}\partial \theta}=0.
\end{equation} 

\noindent The orthogonality between $\Lambda_{\mu \alpha}=\frac{\partial^2 S_{\rm eff}}{\partial j^{\mu}\partial \phi_{\alpha}}|_{\rm sp}$ and the right zero gauge mode $\vec\phi^R_{0}$ is obtained after evaluating this equation  at the SP values of the auxiliary fields and noticing that  the second term vanishes because of the SP condition  (\ref{spcond}). The same holds for the left zero gauge mode if the effective action is derived with respect to $\overline\theta$.  If the fluctuations of the auxiliary fields are decomposed into components parallel and perpendicular to the   zero gauge mode directions, $\vec{\Delta \phi}= \vec{\phi_{\|}}+\vec{\phi_{\perp}}$, 
Eq.~\eqref{chiflIIlast} can be rewritten as follows 
\begin{widetext}
 \begin{equation}
\chi_{_{II} {\mu \nu}}^{\rm fl}(\bm q, i\omega) = \lim_{\epsilon \rightarrow 0}   
\frac{\int D[\bar\phi_{\|},\phi_{\|}] D[\bar\phi_{\perp},\phi_{\perp}] \Lambda_{\mu \alpha_1} \times [\overline \phi_{\|} +\overline \phi_{\perp}]_{\alpha_1} [\phi_{\|}+\phi_{\perp}]_{\alpha_2}  \times \Lambda_{\nu \alpha_2} 
e^{- \epsilon {\bar\phi_{\|}}^{\dagger} {\phi_{\|}}- {\bar \phi_{\perp}}^{\dagger}  S_{\perp}^{(2)} \phi_{\perp} }}
{\int D [\bar\phi_{\|},\phi_{\|}] D[\bar\phi_{\perp},\phi_{\perp}]  e^{- \epsilon {\bar\phi_{\|}}^{\dagger} 
{\phi_{\|}}- {\bar\phi_{\perp}}^{\dagger}  S_{\perp}^{(2)} \phi_{\perp} }},
\label{regu}
\end{equation}
\noindent where the zero modes have been Gaussian regularized by means of the finite positive constant $\epsilon$ in both integrals, which must be sent to zero at the end of the calculation.~\cite{Negele1998} 
The indices ${\bm q}$ and $i \omega$ have been eliminated to  make the notation more compact and  summation over repeated indices is assumed. The orthogonality between $\Lambda_{\mu}$ and $\phi_{\|}$ implies that Eq.~\eqref{regu} can be factorized as
 \begin{equation}
\chi_{_{II} \mu \nu}^{\rm fl}(\bm q, i\omega) =    
\frac{\int D[\bar\phi_{\perp},\phi_{\perp}] \Lambda_{\mu \alpha_1} \times \overline \phi_{\perp{\alpha_1}} \phi_{\perp \alpha_2}  \times \Lambda_{\nu \alpha_2} 
e^{- {\bar\phi_{\perp}}^{\dagger}  S_{\perp}^{(2)} \phi_{\perp} }}
{\int D[\bar\phi_{\perp},\phi_{\perp}]  e^{- {\bar\phi_{\perp}}^{\dagger}  S_{\perp}^{(2)} \phi_{\perp} }} \lim_{\epsilon\rightarrow0}
\frac{\int D[\bar\phi_{\|},\phi_{\|}] e^{- \epsilon {\bar\phi_{\|}}^{\dagger} {\phi_{\|}} }}
{\int D [\bar\phi_{\|},\phi_{\|}]  e^{- \epsilon {\bar\phi_{\|}}^{\dagger} 
{\phi_\|} }}.
\label{final}
\end{equation}
\end{widetext}
\noindent Therefore, $\chi^{\rm fl}_{_{II}}$ of Eq.~\eqref{chiflIIlast} can be computed with the matrix $S^{(2)}_{\perp}$ which is the projection of the fluctuation matrix $S^{(2)}$ onto the subspace
perpendicular to the zero gauge modes. The same result is obtained if, instead of the Gaussian regularization, we consider a {\it positive} zero eigenvalue $\lambda_0$ and at the end of the calculation we take the limit $\lambda_0 \to 0$, in the line of Appendix \ref{appendix_fp}. This procedure avoids the use of the Faddeev-Popov prescription. We can conclude that the zero gauge modes do not contribute to the dynamical spin susceptibility because they do not couple to the external magnetic field $j^{\mu}$ contained in $\Lambda_{\mu}$.~\cite{Shindou2013}

\subsection{Summary of the calculation of Gaussian fluctuations}

Here we outline a summary of the main steps followed for the computation of the Gaussian  corrections to the different quantities considered in this manuscript:\\

i) Starting from the partition function (\ref{generatriz}), the effective action is expanded around its saddle point (\ref{Seffexp}) up to quadratic order~(\ref{Seff-fl}).\\

ii) The saddle point approximation leads to a set of self-consistent equations~\eqref{selfeqs} for the  seven parameters $A_{\delta}$, $B_{\delta}$ $\lambda$ (${\delta}>0$ takes three possible values in the triangular lattice). These equations correspond to the SP condition~\eqref{spcond} after considering the ansatz ~\eqref{ansat}.\\

iii) The seven mean field parameters of the SP solution  are plugged in the fluctuation matrix $S^{(2)}$, which for each $(\bm k, i\omega)$ is a complex non-Hermitian matrix of dimension $13\times13$ [remember that $\vec\phi^\dagger \!\!\!=\!\!\!(\overline{W}^{A\;\omega}_{{\bm k} \;\delta}, {W}^{A\;-\omega}_{{-\bm k} \;\delta},\overline{W}^{B\;\omega}_{{\bm k} \;\delta},{W}^{B\;-\omega}_{{-\bm k} \;\delta}, \lambda^{-\omega}_{{-\bm k} })$]. The  Gaussian fluctuation approximation is stable if all the eigenvalues of the  fluctuation matrix have a positive real part (positive-stable condition).\\

iv) Confirming the presence of the zero gauge modes of the fluctuation matrix is an important sanity check for the correct computation of the fluctuation matrix $S^{(2)}$. This can be done by multiplying the fluctuation matrix (which is computed numerically) by the analytical expression of the zero gauge mode $\vec\phi^R$ \eqref{rightzero}.\\

v) The Faddeev-Popov prescription  is applied after confirming the existence of zero gauge modes to carry on the Gaussian integral over genuine fluctuations of the auxiliary fields. Alternatively, the Faddeev-Popov prescription can be avoided by using the truncated matrix $S^{(2)}_{\rm tr}$. \\

vi) The ground state energy, $E_{\rm gs}(h)=E^{\rm sp}_{\rm gs}(h)+E^{\rm fl}_{\rm gs}(h)$, is  obtained by computing  $E^{\rm sp}_{\rm gs}(h)$ and $E^{\rm fl}_{\rm gs}(h)$ of Eqs.~\eqref{egsp} and ~\eqref{EGS}, respectively.\\

vii) 
The local magnetization, $m(h)=m^{\rm sp}_{\rm gs}(h)+m^{\rm fl}(h)$, is obtained by taking the numerical derivative of $E_{\rm gs}(h)$ with respect to $h$.\\ 

viii) The dynamical structure factor is obtained via  $\mathcal{S}(\bm q, \omega) = -\frac{1}{\pi} {\rm Im} \chi(\bm q, \omega)$, 
after analytic continuation $i\omega\to \omega+i \eta^+$ of the dynamical susceptibility 
$\chi(\bm q, i\omega)=\chi_{_{I}\; \mu\mu}^{\rm sp}(\bm q,i\omega)+\chi_{_{II}\; \mu\mu}^{\rm fl}(\bm q,i\omega)$, 
where Eqs.~\eqref{chi1sp} and \eqref{chiflIIlast} are used, respectively. It is worth emphasizing that there is 
no need to perform the Faddeev-Popov trick in this case because the zero gauge modes do not couple to the external sources.

\section{Spontaneous SU(2) Symmetry Breaking}

In this section we investigate the consequences of the spontaneous $SU(2)$ symmetry breaking on the dynamical spin susceptibility. 
The correct description of this phenomenon requires to carry on the thermodynamic limit of Eq.~\eqref{chiflIIlast} in the right order:
\begin{equation}
\lim_{h \rightarrow 0} \lim_{N_s \rightarrow \infty}\! \mathcal{\chi}_{_{II}}^{\rm fl}(\bm k, i\omega; h)= \!\!
\lim_{N_s \rightarrow \infty}\! \mathcal{\chi}_{_{II}}^{\rm fl}(\bm k, i\omega; h\!\sim \!N_s^{-1}).
\label{limits}
\end{equation}
To trace back the origin of the symmetry breaking contribution, it is instructive to compare the saddle point Green's functions in the 
thermodynamic limit with ($h\ne0$) and without ($h=0$)  the symmetry breaking field. In both cases, the spinons (bosons) condense at 
$T=0$. Correspondingly, the Green's function in the thermodynamic limit includes contributions from condensed and non-condensed  bosons:
\begin{equation}
\mathcal{G}^{\rm sp}(\bm k, i\omega) = \mathcal{G}^{\rm sp}_{\rm n}(\bm k, i\omega) +  \mathcal{G}^{\rm sp}_{\rm c}(\bm k, i\omega).
\label{gnc}
\end{equation}
The energy of the non-condensed bosons remains finite in the thermodynamic limit. Thus, the two limits, $h \to 0$ and $N_s \to \infty$, commute 
for $\mathcal{G}^{\rm sp}_{\rm n}(\bm k, i\omega)$. In contrast, the energy of the condensed bosons is ${\cal O}(N_s^{-1})$, as required by the 
macroscopic occupation number, implying that $\mathcal{G}^{\rm sp}_{\rm c}(\bm k, i\omega)$ is linear in $N_s$. Because of this factor of $N_s$, 
the change of $\mathcal{G}^{\rm sp}_{\rm c}(\bm k, i\omega)$ induced by the ${\cal O}(1/N_s)$  symmetry breaking field  in  Eq.~\eqref{limits} 
remains finite in the thermodynamic limit. In other words, the symmetry breaking field only modifies $\mathcal{G}^{\rm sp}_{\rm c}(\bm k, i\omega)$ 
in the thermodynamic limit.

The Green's function for the $SU(2)$ symmetric case ($h=0$) is
\begin{equation}\label{greenrel}
 \mathcal{G}^{\rm sp}_{0}(\bm k, i\omega) \equiv \lim\limits_{N_s\rightarrow \infty} \ \lim\limits_{h\rightarrow 0}  \mathcal{G}^{\rm sp}(\bm k, i\omega; h).
\end{equation}
The contribution from the condensate is 
\begin{eqnarray}
\mathcal{G}^{\rm sp}_{0-\rm c}  (\bm k, i\omega) & =  \delta_{{\bm k},{\bm 0}} \left(\frac{g_{0}^{-}(\bm k=\bm 0)}{i \omega - \varepsilon_{\bm 0}^{}} +
\frac{ g_{0}^{+}(\bm k=\bm 0)}{i \omega + \varepsilon_{\bm 0}} \right)  \nonumber \\
&\;\; +  \delta_{{\bm k},{\bm Q}}  \left( \frac{ g_{0}^{-}(\bm k=\bm Q)}{i \omega - \varepsilon_{\bm Q}^{}} + \frac{ g_{0}^{+}(\bm k=\bm Q)}{i \omega + \varepsilon_{\bm Q}} \right) \nonumber \\
&\!+  \delta_{{\bm k},{\bar{\bm Q}}}  \left(\frac{ g_{0}^{-}(\bm k=\bar{\bm Q})}{i \omega - \varepsilon_{\bm \bar{\bm Q}}} 
+ \frac{ g_{0}^{+}(\bm k=\bar{\bm Q})}{i \omega + \varepsilon_{\bar{\bm Q}}} \right)\ , 
\end{eqnarray}
with, ${\bm Q}$, $\bar{\bm Q} \equiv -{\bm Q}$ and ${\bm 0}$ being the condensation wave vectors, and
\begin{widetext}
\begin{equation}
  g_{0}^{-}(\bm k=\bm 0) =  -\left(
  \begin{array}{cccc}
   v_{\bm 0}^{_2}   &  z_{\bm 0}^{}    & 0                & 0               \\
   z_{\bm 0}^{}     &  u_{\bm 0}^{_2}  & 0                & 0               \\
   0                &  0               &  v_{\bm 0}^{_2}  &  z_{\bm 0}^{}   \\
   0                &  0               &  z_{\bm 0}^{}    &  u_{\bm 0}^{_2} \\
  \end{array} \right), 
  \ g_{0}^{+}(\bm k=\bm 0) =   \left(
  \begin{array}{cccc}
   u_{\bm 0}^{_2}   &  z_{\bm 0}^{}    &  0               &  0              \\
   z_{\bm 0}^{}     &  v_{\bm 0}^{_2}  &  0               &  0              \\
   0                &  0               &  u_{\bm 0}^{_2}  &  z_{\bm 0}^{}   \\
   0                &  0               &  z_{\bm 0}^{}    &  v_{\bm 0}^{_2} \\
  \end{array} \right),
  \end{equation}

\begin{equation}
  g_{0}^{-}(\bm k=\bm Q) =  -\left(
  \begin{array}{cccc}
   0   &  0    & 0       & 0               \\
   0   &  0    & 0       & 0               \\
   0   &  0   &  v_{\bar{\bm Q} -}^2  &  z_{\bar{\bm Q} -}   \\
   0                &  0               &  z_{\bar{\bm Q} -}    &  u_{\bar{\bm Q} - }^{2} \\
  \end{array} \right), 
  \ g_{0}^{+}(\bm k=\bm Q) =   \left(
  \begin{array}{cccc}
   0  &  0   &  0               &  0              \\
   0  &  0  &  0               &  0              \\
   0                &  0               &  u_{\bar{\bm Q} -}^2   &  z_{\bar{\bm Q} - }  \\
   0                &  0               &  z_{\bar{\bm Q} -}     &  v_{\bar{\bm Q}-}^{2} \\
  \end{array} \right),
  \end{equation}
  
  \begin{equation}
  g_{0}^{-}(\bm k=\bar{\bm Q}) =  -\left(
  \begin{array}{cccc}
   v_{\bar{\bm Q} -}^{2}  &  z_{\bar{\bm Q}-}    & 0       & 0               \\
   z_{\bar{\bm Q} -}    &  u_{\bar{\bm Q} -}^{2}    & 0       & 0               \\
   0                &  0               &   0   &  0 \\
   0                &  0               &   0   &  0 \\
  \end{array} \right)  \ , 
  \ g_{0}^{+}(\bm k=\bar{\bm Q}) =   \left(
  \begin{array}{cccc}
   u_{\bar{\bm Q}-}^2   &  z_{\bar{\bm Q} -}    &  0               &  0              \\
   z_{\bar{\bm Q} -}     &  v_{\bar{\bm Q} -}^{2}  &  0               &  0              \\
   0                &  0               &  0  &  0  \\
   0                &  0               &  0  &  0 \\
  \end{array} \right) \ ,
  \end{equation}
\end{widetext}
with $ v_{\bm 0}^{2} = \lim\limits_{h\rightarrow 0} \ v_{\bm 0 \pm}^{2} $, $ u_{\bm 0}^{2} = \lim\limits_{h\rightarrow 0} \ u_{\bm 0 \pm}^{2} $, 
$ z_{\bm 0} = \lim\limits_{h\rightarrow 0} \ z_{\bm 0 \pm} $, $\varepsilon_{\bm 0} = \lim\limits_{h\rightarrow 0} \ \varepsilon_{\bm 0 \pm} \sim N_s^{-1} $, 
and $ \varepsilon_{\pm \bm Q} = \lim\limits_{h\rightarrow 0} \varepsilon_{\pm \bm Q -} \sim N_s^{-1} $.\\

Only the ${\bm k}={\bm 0}$ point of the lower band  condenses for finite external field $h$  because $\varepsilon_{\bm 0 -}\sim \frac{1}{N_s}$. 
It can be seen that the condensation does not occur at the ${\bm k} = {\bm 0}$ point of the higher band and at the $\pm {\bm Q}$ points because 
$\varepsilon_{\bm 0 +}, \varepsilon_{\pm \bm Q -} \!\sim\! \frac{1}{\sqrt{N_s}}$, which does not produce macroscopic occupation number 
in the thermodynamic limit. The condensate contribution to  the Green's function in the symmetry breaking ($h \neq 0$) case  is then modified 
relative to the $h=0$ case: 
\begin{equation}
\mathcal{G}^{\rm sp}_{\rm sb-c}({\bm k}, i\omega) = \delta_{{\bm k},{\bm 0}} \left( \frac{ g_{\rm sb}^{-}({\bm k}=
{\bm 0}) } {i \omega - \varepsilon_{\bm 0 -}} + \frac{  g_{\rm sb}^{+} ({\bm k}={\bm 0}) }{i \omega + \varepsilon_{\bm 0 -}} \right),  
\end{equation}
with
\begin{widetext}
\begin{equation}
g_{\rm sb}^{-}({\bm k}={\bm 0}) = - \frac{1}{2} \left(
\begin{array}{cccc}
v_{\bm 0 -}^{2}             &  z_{\bm 0 -}              & v_{\bm 0 -}^{2}           & z_{\bm 0 -}   \\
z_{\bm 0 -}               &  u_{\bm 0 -}^{2}            & z_{\bm 0 -}             &   u_{\bm 0 -}^{2}   \\
v_{\bm 0 -}^{2}             & z_{\bm 0 -}              &  v_{\bm 0 -}^{2}            &  z_{\bm 0 -}   \\
z_{\bm 0 -}               & u_{\bm 0 -}^{2}            &  z_{\bm 0 -}              &  u_{\bm 0 -}^{2} \\
\end{array} \right)   , 
\;\;\;\;\;g_{\rm sb}^{+} ({\bm k}={\bm 0})  =  \frac{1}{2} \left(
\begin{array}{cccc}
u_{\bm 0 -}^{2}             & z_{\bm 0 -}              &  u_{\bm 0 -}^{2}            &  z_{\bm 0 -}   \\
z_{\bm 0 -}               &   v_{\bm 0 -}^{2}            &  z_{\bm 0 -}             &  v_{\bm 0 -}^{2}   \\
u_{\bm 0 -}^{2}             &  z_{\bm 0 -}              & u_{\bm 0 -}^{2}            & z_{\bm 0 -}   \\
z_{\bm 0 -}               &  v_{\bm 0 -}^{2}            & z_{\bm 0 -}              & v_{\bm 0 -}^{2} \\
\end{array} \right) \ . 
\end{equation}
\end{widetext}
By taking the thermodynamic limit of $ \Lambda_{\mu \phi}(\bm q,i\omega)$ [see Eq.~\eqref{lambda}] in the presence of symmetry breaking field $h$, we obtain
\begin{eqnarray}
 \Lambda_{\mu \phi}(\bm q,i\omega) &=& \frac{1}{2} \tr\left[ \ \mathcal{G}^{\rm sp}_{\rm n} \ v_{\phi} \ \mathcal{G}^{\rm sp}_{\rm n} \ u^{\mu}(\bm q, i \omega) \right]  \nonumber \\
 & + & \frac{1}{2} \tr \left[\mathcal{G}_{\rm n}^{\rm sp} \ v_{\phi} \ \mathcal{G}_{\rm sb-c}^{\rm sp} \ u^{\mu}(\bm q, i \omega) \right]  \nonumber \\ 
 & + & \frac{1}{2} \tr \left[\mathcal{G}_{\rm sb-c}^{\rm sp} \ v_{\phi} \ \mathcal{G}_{\rm n}^{\rm sp} \ u^{\mu}(\bm q, i \omega) \right] \\ 
 & + & \frac{1}{2} \tr \left[\mathcal{G}_{\rm sb-c}^{\rm sp} \ v_{\phi} \ \mathcal{G}_{\rm sb-c}^{\rm sp} \ u^{\mu}(\bm q, i \omega) \right]\delta_{\bm{q,0}} \nonumber \ .
\end{eqnarray}
\noindent The first line vanishes because the non-condensed part of the Green's function preserves the $SU(2)$ symmetry.~\footnote{Once again this is true because $\Lambda_{\mu}$ is crossed susceptibility for two fields that belong to different representations of SU(2).} The remaining lines give non-zero contributions because $\mathcal{G}_{\rm sb-c}^{\rm sp}$ is not invariant under the  $SU(2)$ symmetry group.\\

For the polarization matrix $\Pi_{\phi_{\alpha_1} \phi_{\alpha_2}}$, we have
\begin{eqnarray}\label{polar}
 \Pi_{\phi_{\alpha_1} \phi_{\alpha_2}}\!\!\!\! &=& \!\! \frac{1}{4} \tr\left[\mathcal{G}^{\rm sp}_{\rm n} \; v_{\phi_{\alpha_1}} \! \mathcal{G}^{\rm sp}_{\rm n} \ v_{\phi_{\alpha_2}} \right] \nonumber \\ 
 &+& \!\! \frac{1}{4} \tr \left[\mathcal{G}^{\rm sp}_{\rm n} \ v_{\phi_{\alpha_1}}  \mathcal{G}_{\rm sb-c}^{\rm sp} \ v_{\phi_{\alpha_2}} \right]  \nonumber   \\
 &+&\!\! \frac{1}{4} \tr \left[\mathcal{G}_{\rm sb-c}^{\rm sp} \ v_{\phi_{\alpha_1}} \ \mathcal{G}^{\rm sp}_{\rm n} \ v_{\phi_{\alpha_2}} \right] \\
 &+& \!\! \frac{1}{4} \tr \left[\mathcal{G}_{\rm sb-c}^{\rm sp} \ v_{\phi_{\alpha_1}} \ \mathcal{G}_{\rm sb-c}^{\rm sp} \ v_{\phi_{\alpha_2}} \right] \delta_{\bm{q,0}} \nonumber \ .
\end{eqnarray}
In this case all the lines are non-zero because $\Pi_{\phi_{\alpha_1} \phi_{\alpha_2}}$ is a scalar under spin rotations.

\section{Results}

This Section includes the results of the formalism presented in Sec. II for the TLHM.
The  calculations are carried on finite size triangular lattices with $N_s=l^2+m^2+n^2$ sites ($l,m,n$ are  integer numbers), that have the same discrete symmetries of the infinite triangular lattice, imposing periodic boundary conditions. We have used cluster sizes up to $N_s=43200$ and $\eta^+=0.01$ for the analytic continuation. 

In the following, we present the results of the local magnetization and the dynamical structure factor, and we analyze the long-wavelength limit of the SB theory.

\subsection{Local magnetization}

\begin{figure}[ht]
\vspace*{0.cm}
 \includegraphics*[width=0.48\textwidth,angle=-0]{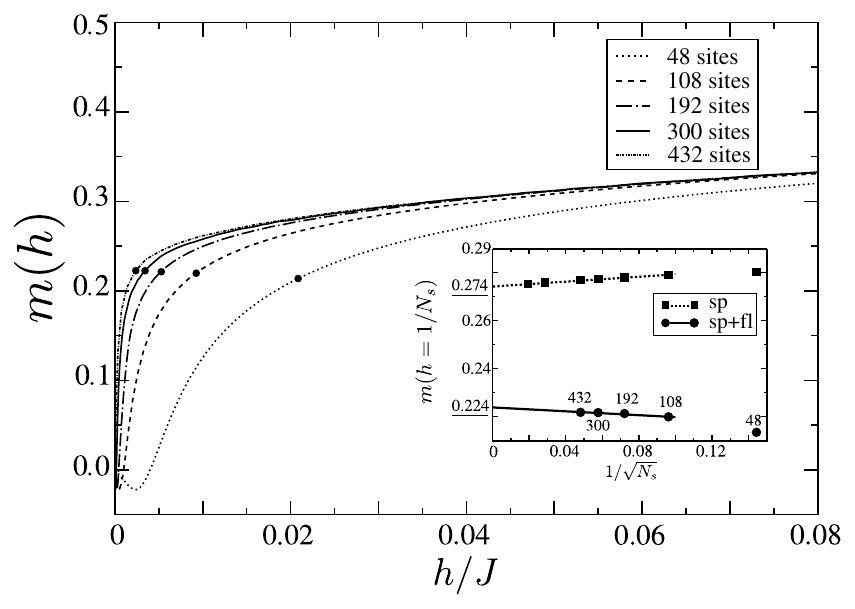}
\caption{$1/N$ correction of the local magnetization $m$ as a function of the symmetry breaking field $h$ for several lattice sizes. Full circles correspond to $h=1/N_s$. Inset: extrapolation to thermodynamic limit of the  magnetization for the SP approximation (full squares) and the SP plus Gaussian fluctuations correction (full circles), using $h=1/N_s$. }
\label{fig3:mag}
\end{figure}

Fig.~\ref{fig3:mag} shows the local magnetization $m(h)=m^{\rm sp}(h)+m^{\rm fl}(h)$ 
of the $120^{\circ}$ ordering as a function of the symmetry breaking field $h$ 
for several lattice sizes. These curves are obtained by taking the numerical
derivative  with respect to $h$ of the ground state energy computed at the 
Gaussian level, $E_{\rm gs}(h)=E^{\rm sp}_{\rm gs}(h)+E^{\rm fl}_{\rm gs}(h)$, [see Eqs. \eqref{egsp} and \eqref{EGS}]. The full circles indicate the magnetization value corresponding to $h=1/N_s$. In the inset, these points are extrapolated to the thermodynamic limit along with the SP results (full squares, not shown in the main figure). 
For the SP extrapolation we can access very large cluster sizes, while for the Gaussian fluctuation approximation we are able to go up to $N_s=432$ because of the inacurracies inherent to the numerical derivation. However, it is worth mentioning that the Gaussian correction to the ground state energy can be computed in a reliable way for very large system sizes.~\cite{Manuel1998}

Notably, the Gaussian corrections reduce the SP magnetization from $m^{\rm sp}=0.274$ to $m=0.224$. The latter agrees quite well with the value $m=0.205\pm 0.005$ obtained with the most sophisticated methods like QMC~\cite{Capriotti1999} and DMRG.~\cite{White2007} 
Then, at the Gaussian level, the SB theory seems to support the hypothesis of proximity of the $120^{\circ}$ N\'eel order to a QMP. For small system sizes (48 and 108 lattices) and $h$ much smaller than the finite size spinon gap, $1/N_s$, there is a small region where $m$ has a negative dip. This is a finite size effect that disappears upon increasing $N_s$. 

\subsection{Dynamical structure factor}

\begin{figure*}[!t]
 \includegraphics[width=18cm]{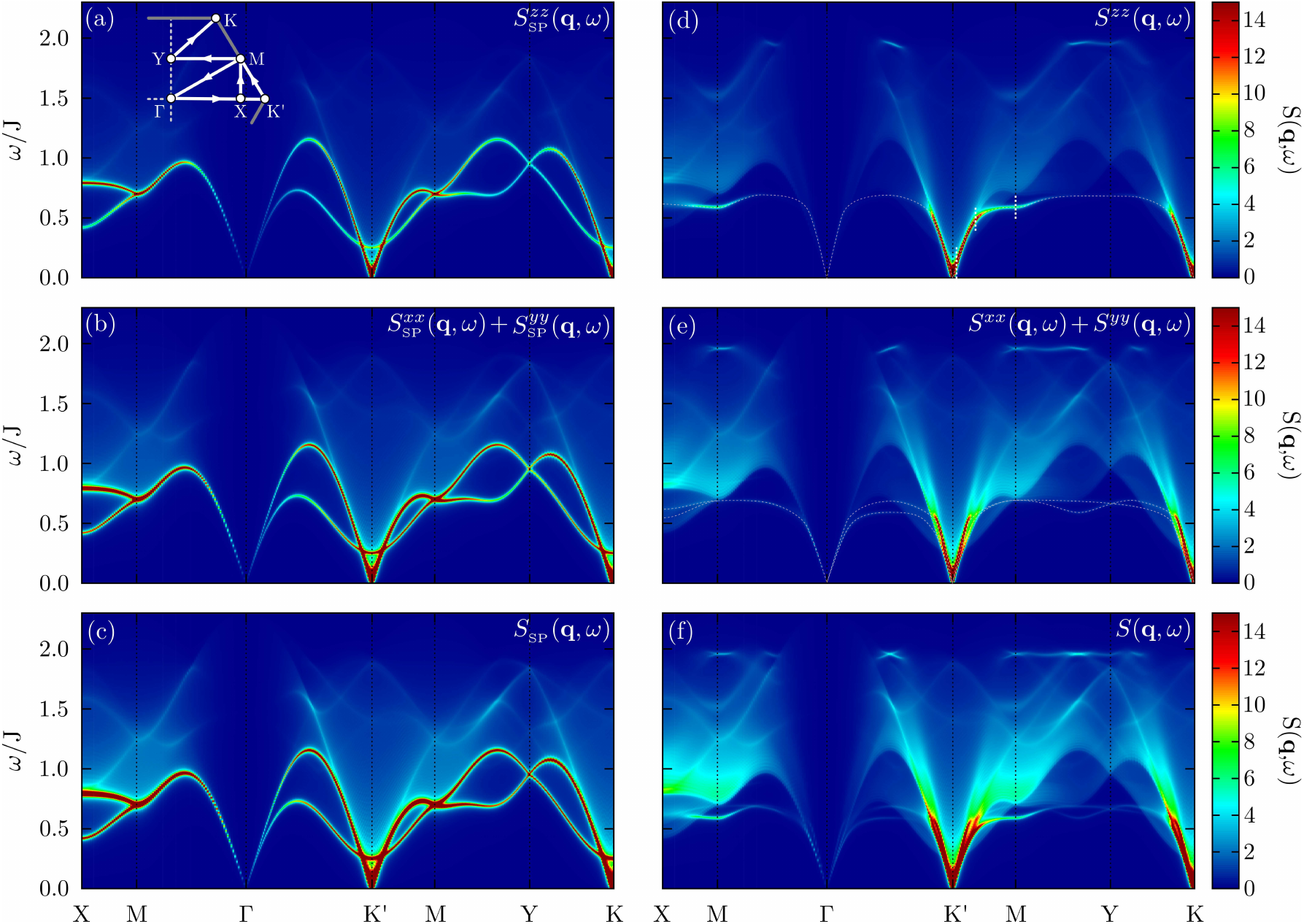} 
 \centering
 \caption{(Color online) First column: dynamical structure factor obtained at the saddle point level: (a) $S_{\rm sp}^{zz}(q,\omega)$, (b) $S_{\rm sp}^{xx}(q,\omega)=S^{yy}_{\rm sp}(q,\omega)$ and (c) $S_{\rm sp}(q,\omega)= S_{\rm sp}^{xx}(q,\omega) + S_{\rm sp}^{yy}(q,\omega)+S_{\rm sp}^{zz}(q,\omega)$.
   Second column: dynamical structure factor obtained after including the $1/N$ correction (Gaussian fluctuations): (d) $S^{zz}(q,\omega)$, (e) $S^{xx}(q,\omega)=S^{yy}(q,\omega)$ and (f) $S(q,\omega)= S^{xx}(q,\omega) + S^{yy}(q,\omega)+S^{zz}(q,\omega)$. White dashed lines indicate the magnons branches. The path within the hexagonal Brillouin zone is indicated in the inset of panel (a). The results correspond to the triangular lattice size of $N_s=120\times120\times3=43200$, a value of the magnetic field $h=1/N_s$, and analytic continuation $i\omega \rightarrow \omega + i\eta^+$ with $\eta^+=0.01$.} 
 \label{fig4:chi}
 \end{figure*}

Fig.~\ref{fig4:chi} shows a comparison between the $S({\bm q}, \omega)$ 
obtained at the SP level [see Eq.~\eqref{chi1sp}] and after including the 
$1/N$ Gaussian correction shown in Fig.~\ref{fig2:feymann}(b) 
[see Eq.~\eqref{chiflIIlast}]. Figs. \ref{fig4:chi}(a-c) clearly show that by properly accounting for the spontaneous symmetry breaking, we  obtain different responses for the out of plane, $S^{zz}({\bm q}, \omega)$, and in-plane, $S^{xx}({\bm q}, \omega)=S^{yy}({\bm q}, \omega)$, components of the magnetic structure factor. 
This natural response of a magnetically ordered ground state should be contrasted with the isotropic one that is obtained with the singlet ground state.~\cite{Mezio2011,Mezio2012}
As we have already mentioned, the SP solution exhibits a branch of spurious modes arising from density fluctuations that violate the bosonic number constraint \eqref{constr}. However, the main weakness of the SP solution is the lack of  magnon modes expected for a magnetically ordered state. 
The spectrum consists only of a two-spinon continuum (branch cut) because the excitations of  the single-spinon condensate 
are  {\it non-interacting} spinon modes.~\cite{Chubukov1995}

\begin{figure}[t]
\vspace*{0.cm}
 \includegraphics*[width=0.48\textwidth,angle=0]{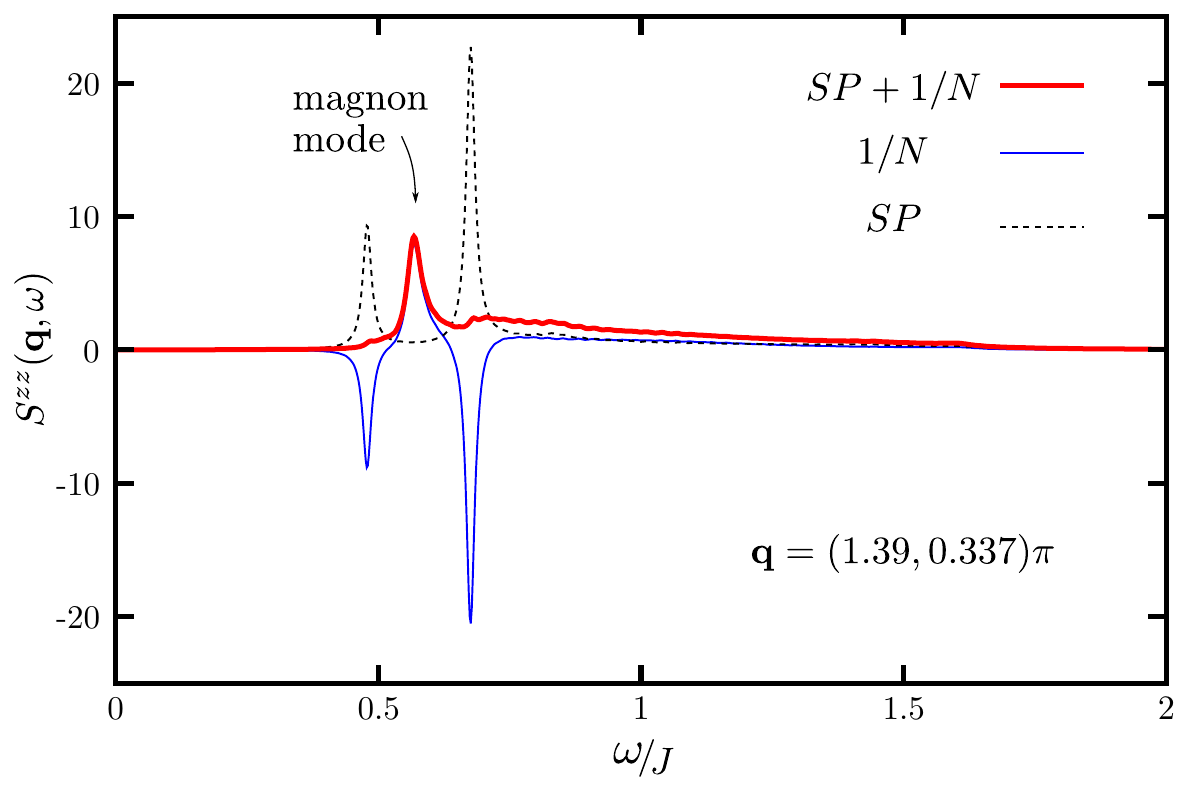}
\caption{(Color online) Intensity of $S^{zz}({\bm q},\omega)$ as a function of the energy $\omega/J$ for ${\bm k}=(1.139,0.337)\pi$ (between $K^{\prime}$ and $M$). The dashed (black) and thin (blue) lines correspond to saddle point and $1/N$ correction, separetly. The thick (red) line corresponds to SP plus 1/N correction (Gaussian fluctuations). }
\label{fig5:corte}
\end{figure}

The $1/N$ contribution modifies $S({\bm q}, \omega)$ in a dramatic way [see Figs.~\ref{fig4:chi}(d-f)]. The auxiliary gauge field 
fluctuations bind the spinons into magnons (collective modes) that appear as isolated poles below the two-spinon continuum (TSC).~\cite{Chubukov1995,Chubukov1996}
In addition, the cancellation of the density fluctuations~\cite{Auerbach1994} produced by the diagram depicted in Fig.~\ref{fig2:feymann}(b) removes  the spurious modes of the SP solution. This cancellation does not persist if we include the other $1/N$ corrections shown in Fig.~\ref{fig2:feymann}(c-e).  The effects of the $1/N$ correction can be better appreciated in  the frequency dependence of  $S^{zz}({\bm q},\omega)$ for a fixed value of momentum ${\bm k}=(1.139,0.337)\pi$ between ${\bm K}^{\prime}$ and ${\bm M}$ (see Fig.~\ref{fig5:corte}). The main contributions of the SP solution are cancelled exactly  by the $1/N$ correction. 
This cancellation is accompanied by the emergence of simple poles associated with the collective modes of the ordered state. A similar behavior (not shown in Fig.~\ref{fig5:corte}) is obtained for the  $S^{xx}({\bm q},\omega)$ and $S^{yy}({\bm q},\omega)$ components. The removal of the spurious modes and the emergence of magnon poles are the most important {\it qualitative} changes relative to the SP solution.

Closer inspection of the bottom and the top of the TSC  in Figs. \ref{fig4:chi}(d-f) reveals two additional differences. The SP solution exhibits a large spectral weight at the bottom of the TSC, which is transferred to the magnon peak after including the Gaussian correction. In addition, a weak but sharp isolated pole also appears right above the top edge of the TSC. This sharp feature is expected to become overdamped upon inclusion of four and higher spinon excitations resulting from higher orders in the $1/N$ expansion. \\

\begin{figure}[ht]
\vspace*{0.cm}
 \includegraphics*[width=0.48\textwidth,angle=0]{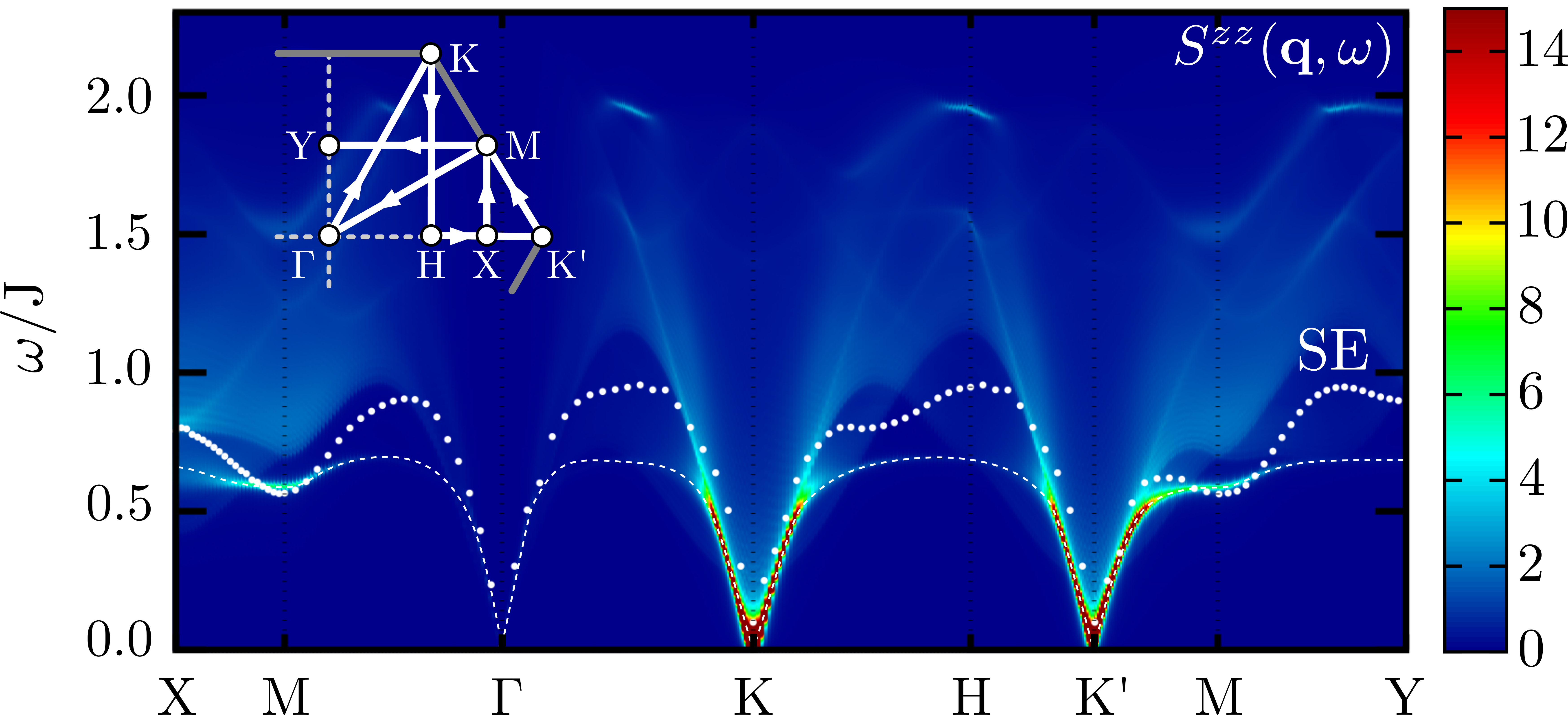}
\caption{(Color online) Comparison of $S^{zz}({\bm q},\omega)$ obtained after including the 1/N correction (Gaussian fluctuations) with the magnon dispersion relation predicted by series expansions~\cite{Zheng2006}(SE) (white circles). The dashed line represents the dispersion of the magnon poles. }
\label{fig6:se}
\end{figure}

Fig.~\ref{fig6:se} shows a  comparison of the resulting single-magnon dispersion  (white dashed line), coming from the out of plane $S^{zz}(\bm k, \omega),$ with the one obtained from series expansions~\cite{Zheng2006} (white circles). The strong downward renormalization with respect to LSWT  predicted by SE is reproduced by the SB theory at the Gaussian level, along with the appearance of rotonic excitations around the $M$ point. In particular, the {\it quantitative} agreement becomes very good  when the magnon peaks are sharper, i.e.,  
around the momenta ${\bm K}$, ${\bm K}^{\prime}$ and ${\bm M}$. 
Consistently with the fact that the ground state of the TLHM is proximate to a quantum melting point, this result supports our original hypothesis of describing the spin-$1$ magnon excitation as a two-spinon bound state.

On the other hand, it is also important to analyze the qualitative differences between  Figs.~\ref{fig4:chi}(d-f) and the $S({\bm q}, \omega)$ obtained from a large $S$ expansion.~\cite{Mourigal2013a} The first  obvious qualitative difference is  the structure of the high-energy continuum. In semi-classical approaches,  this continuum arises from two or more magnon modes. In large-$N$ expansions,  it is dominated by two-spinon modes. Consequently, it extends over a wider energy range that  is three times bigger than the single magnon bandwith for the case under consideration [see Figs.~\ref{fig4:chi}(d-f)].
\begin{figure}[ht]
\includegraphics[width=\columnwidth]{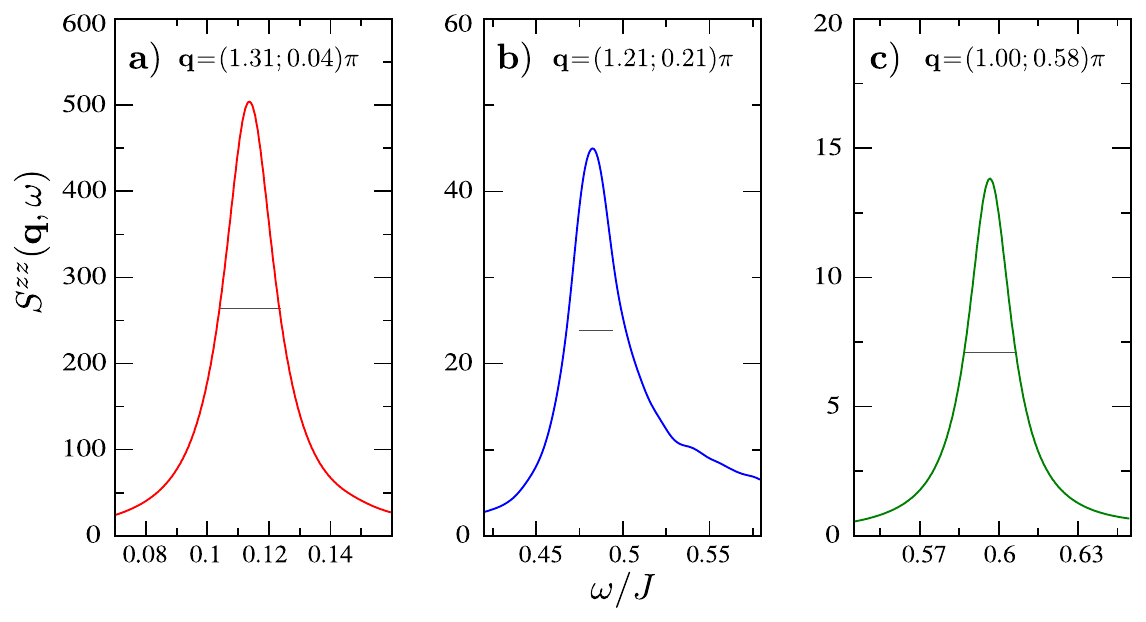}
 \caption{(Color online) Magnon quasi-particle peak for wave vectors (a) ${\bm q}=(1.31,0.04)\pi$ and (b) ${\bm q}=(1.21,0.21)\pi$ near the ordering wave-vector ${\bf K^{\prime}}$. (c) Single magnon peak at the ${\bf M}$ point. These wave vectors are indicated in Fig.~\ref{fig4:chi}(d) with vertical dashed white lines. The segments (thin black line) display the value $2\eta^+$ used for the analytic continuation. In all cases $\eta^+=0.01$. }
 \label{fig7:peaks}
 \end{figure}

Another qualitative difference with large-$S$ expansions is the origin of the magnon quasi-particle peak broadening. As we discuss in the next subsection, the magnon branch gets inside the two-spinon continuum in the long wavelength limit. The kinematic conditions then allow for single-magnon to two spinon decay that broadens the magnon peak.~\cite{notemd} Upon moving away from the long wavelength limit (region around the $\Gamma$ and the $K$ points) the single-magnon branch is shifted below the two-spinon continuum [see Fig.~\ref{fig4:chi}(d-f)] and  the intrinsic broadening  disappears. A simple phase space argument shows that the broadening of the magnon peak must also go to zero in the $q\to 0$ limit, around $\Gamma$ and $\pm K$ points. Consequently, the strongest effects of the single-magnon to two-spinon decay are expected to occur near the momentum space region where the single-magnon branch is about to emerge from the two spinon continuum. These effects can be observed in Figs.~\ref{fig7:peaks}(a-c) which display the magnon peak for three representative wave vectors.
The wave vector ${\bm q}=(1.31,0.04)\pi$ in Fig.~\ref{fig7:peaks}(a) is very close to the $K$ point. This explains the small broadening that is obtained for this particular wave-vector. As expected from the above-mentioned argument, the magnon peak acquires a much broader structure for ${\bm q}=(1.21,0.21)\pi$ [see Fig.~\ref{fig7:peaks}(b)]. As shown in Fig.~\ref{fig4:chi}(f),  the magnon peak is emerging from the two-spinon continuum at this particular wave vector. Finally, the broadening of the magnon peak disappears at the ${\bm M}$ point [see Fig.~\ref{fig7:peaks}(c)] because the kinematic conditions no longer allow for single-magnon to two spinon decay (the magnon mode is below the edge of the two-spinon continuum).\\

\noindent Furthermore, it is well known~\cite{Auerbach1988} that the SP solution violates the sum rule  $\int d\omega \sum_{\bm q} S({\bm q}, \omega)=N S(S+1)$  by a factor of 1.5 due to the violation of the local constraint of the SBs. We find that this factor reduces to 1.2 upon including the Gaussian correction shown in Fig.~\ref{fig2:feymann}b.\\

\noindent Finally, by taking the large-$S$ limit, for $N=2$, we have found that the $1/N$ correction recovers the dynamical structure factor predicted by 
LSW (to be published elsewhere). This result should be contrasted with the failure of the mean field SB theory to  recover the semiclassical dispersion for spiral 
states.~\cite{Coleman1994,Mattsson1995}

\subsection{Long Wavelength Limit}

Linear spin wave theory is expected to work well in the long wavelength limit.~\cite{Dombre89} 
Consequently, although the current approach is motivated by the experimental observation of  anomalies in the dynamical spin structure factor 
that appear at wavelengths comparable to the lattice parameter, it is still interesting to analyze the outcome of our approach in the 
long wavelength limit. In this limit the  spectrum consists of the three low-energy Goldstone modes  around $\bm 0$ and $\pm {\bm Q}:$ (i) 
$\rvert \bm k \rvert < \Lambda$; (ii) $\rvert \bm k - \bm Q \rvert < \Lambda$; (iii) $\rvert \bm k + \bm Q \rvert < \Lambda$, where $\Lambda$ 
is a momentum cut-off below which the dispersion relation is practically linear. These collective modes (magnons) are obtained as  poles of 
the RPA propagator
\begin{equation}
D(\bm q, i\omega) =\left[\Pi_0 - \Pi(\bm q, i\omega)\right]^{-1},
\end{equation}
where the polarization operator, $\Pi(\bm q, i\omega)$, is determined by the degrees of freedom with wavelength longer than $\Lambda^{-1}$. 
The  effective low-energy action  for the spinons and their coupling to the gauge field is obtained by performing a gradient expansion of the effective action 
(see Appendix \ref{app1}).

\subsubsection{Around the \texorpdfstring{$\Gamma$}{G} point}
To obtain results in the long wavelength limit, we must expand the polarization operator in powers of $q$ and $\omega$.  According to Eq.~\eqref{polar}, the polarization operator can be decomposed into three contributions. The first line of Eq.~\eqref{polar} corresponds to a  contribution from the non-condensed bosons only. We will refer to this contribution as
${\Pi}_{nn}$. The second and the third line of Eq.~\eqref{polar} correspond to a mixed contribution from the condensed and non-condensed bosons and will be denoted as ${\Pi}_{cn}$.
Finally, the last line of Eq.~\eqref{polar} corresponds to a contribution from the condensed bosons only, that will be denoted as ${\Pi}_{cc}$.

The leading order contribution to $\Pi(\bm q, i\omega)$ is  ${\cal O} (q^{-2})$
\begin{eqnarray}
\Pi_{cn}^{(-2)}(\bm q, i\omega) =\frac{\phi Z/2}{c^2 q^2 + \omega^2} \Theta,
\end{eqnarray}
where $\phi $ refers to the density of condensed bosons, $Z=\lambda + \gamma_{\bm Q /2}^B $, and 
$\Theta=\theta \rvert u_1 \rangle \langle v_1 \rvert $ after a  singular value decomposition (SVD) 
(see Appendix \ref{app1}). Given that $\Pi_{cn}^{(-2)}(\bm q, i\omega)$ diverges for 
$\rvert \bm q \rvert, \omega \rightarrow 0$, it is clear that the magnon mode belongs to the null 
space ${\cal P}$ of  $\Pi_{cn}^{(-2)}(\bm q, i\omega)$.  Therefore, to extract the magnon pole, we 
just need to consider the next order  contributions to the polarization matrix projected into the null  
subspace ${\cal P}$:
\begin{equation}
{\bar \Pi} = P_L \cdot \Pi \cdot P_R,
\end{equation}
with the left/right projectors
\begin{eqnarray}
P_L &=& \sum_{\sigma>1} \rvert u_{\sigma} \rangle \langle u_{\sigma} \rvert, \\
P_R &=& \sum_{\sigma>1} \rvert v_{\sigma} \rangle \langle v_{\sigma} \rvert,
\end{eqnarray}
where $\langle u_{\sigma}\rvert u_1\rangle =0 $ and $\langle v_{\sigma}\rvert v_1\rangle =0 $ for $\sigma>1$. 
It turns out that the projection of the ${\cal O} (q^{-1})$ contributions, that we denote as $\Pi^{(-1)}$, is 
equal to zero. However, $\Pi^{(-2)}$  still connects the subspace ${\cal P}$ with the orthogonal  subspace, 
leading to a ${\cal O}(1)$ contribution within the subspace ${\cal P}$ that is obtained via a second order process:
\begin{eqnarray}\label{subsub1}
{\bar \Pi}^{(0)}_1 (\bm q,i\omega) =&  - P_L \cdot \frac{\Pi_{cn}^{(-1)} \rvert v_1 \rangle  \langle u_1 \rvert  
\Pi_{cn}^{(-1)}}  { \langle u_1 \rvert \Pi_{cn}^{(-2)} \rvert v_1 \rangle }    \cdot P_R,
\end{eqnarray}
where
 \begin{eqnarray}
{\bar \Pi}_{cn}^{(-1)}(\bm q, i\omega) ={\phi q \over c^2 q^2 + \omega^2} \left( { i\omega \over 2q  } {\bar \Gamma}
+ {Z \over 4} \hat{\bm q} \cdot {\bar  {\bm  M}} \right).
\end{eqnarray}
Here we used the notation ${\bar A} \equiv  P_L \cdot  A \cdot P_R$ for any matrix $A$. $\bar \Gamma$ and $\bar M$ 
are defined in in the Appendix \ref{app1}. The non-condensed bosons give no contribution to this second order process.

In addition to ${\bar \Pi}^{(0)}_1 (\bm q,i\omega)$, we must consider the other  ${\cal O}(1)$ contributions to the polarization operator.  
The contribution ${\bar \Pi}_{nn}^{(0)}(\bm q, i\omega)$ from the non-condensed bosons is a regular integral  that depends on the cutoff $\Lambda$. 
Finally, the ${\cal O}(1)$ contribution arising from the combination of  condensed and non-condensed bosons is
\begin{eqnarray}\label{subsub3}
{\bar \Pi_{cn}}^{(0)}(\bm q, i\omega) \!\!= \!\!\frac{\phi q^2/2}{c^2 q^2 +\omega^2} \left( \bar {\Sigma} 
 \! + \! {i\omega \over q}  \hat{\bm q} \cdot  \bar{{\bm R}} + {Z\over 2} \hat{\bm q} \cdot \bar{{\bm D}} \cdot \hat{\bm q} \right),
\end{eqnarray}

\noindent where these matrices $\Sigma$, $\bm R$ and  $\bm D$  are defined in Appendix~\ref{app1}.

 We note that $\Pi_{cc}$ is not included in this analysis because it only gives a finite contribution {\it at the condensate wave vector}, while we are only interested in the behavior of $\Pi$ {\it around} these points ($q$ can be arbitrarily small but always finite)

In summary, the magnon mode in the long wavelength limit is obtained from the solution of the equation:

\begin{eqnarray}
\det \left[ \left( {\bar \Pi}_0 - {\bar \Pi}_{nn}^{(0)} - {\bar \Pi}_{cn}^{(0)} - {\bar \Pi}^{(0)}_1 \right)  \right] = 0.
\end{eqnarray}

\noindent  The left hand side of this equation is a function of $\omega/q$. Thus the root of this function gives $\omega = v_{\Gamma} q$, with magnon velocity $v_{\Gamma} = 1.087 J$. The fact that $v_{\Gamma}$ turns out to be a real number confirms that the magnon mode is a well-defined quasi-particle  in the long wavelength limit. The magnon velocity  at the $\Gamma$ point obtained with a $1/S$ expansion up to ${\cal O} (S^{-1})$ is  $1.15J$.~\cite{Chubukov1994,Chernyshev2009}

Near the magnon pole, the polarization operator takes the  form: 

\begin{eqnarray}
\lim_{\rvert \bm q \rvert, \omega \rightarrow 0} \Pi(\bm q, \omega > 0) = \alpha \left(1-{\omega \over vq }\right) \rvert \mu_m \rangle \langle \nu_m \rvert  + ..., 
\label{Pilw}
\end{eqnarray}

\noindent where $\rvert \mu_m\rangle $ and $\rvert \nu_m\rangle $ refer to the magnon mode in the null subspace ${\cal P}$ and ``...'' abbreviates the regular contributions from the orthogonal subspaces.
According to our derivation, the leading order correction to Eq.~\eqref{Pilw} is ${\cal O}(q)$. The pole equation  is modified by the matrix element of $ \Delta \Pi(\bm q,\omega) \equiv \Pi(\bm q,\omega) - \lim_{\rvert \bm q \rvert, \omega \rightarrow 0} \Pi(\bm q,\omega)$ in the subspace of the magnon mode described by the left (or right) state $\rvert \mu_m \rangle$ (or $\rvert \nu_m \rangle$), namely

\begin{eqnarray}
 \alpha \left(1-{i\omega \over v q }\right) = \langle \mu_m \rvert  \Delta \Pi(\bm q,\omega) \rvert \nu_m \rangle.
\end{eqnarray}

\noindent The contribution from the condensed bosons is analytic at $\rvert \bm q \rvert =0$ for a given ratio of $\omega/q$. Consequently, the leading order contribution to the matrix element $\langle u_m \rvert  \Delta \Pi_{cn}(\bm q,\omega) \rvert v_m \rangle$ is

\begin{eqnarray}
 \langle \mu_m \rvert  \Delta \Pi_{cn}(\bm q,\omega) \rvert \nu_m \rangle \propto q F\left(\frac{\omega}{cq}\right)\ .
\end{eqnarray}

\noindent This contribution gives rise to a non-linear correction $\sim q^2$ to the magnon dispersion, but no contribution to the magnon decay  because of the energy mismatch upon splitting the magnon into a non-condensed  (with momentum $\bm q$) and a condensed  (with $\bm 0$ momentum) spinon.

The contribution from the non-condensed bosons, $\Pi_{nn}^{\rm reg}(\bm q,\omega)$,  is not guaranteed to be analytic in $\bm q$ at $\bm q = \bm 0$ for given ratio of $\omega/q$. Therefore, we assume

\begin{eqnarray}
 \langle \mu_m \rvert  \Delta \Pi_{nn}(\bm q,\omega) \rvert \nu_m \rangle \propto q^{\gamma} H\left(\frac{\omega}{cq}\right). 
\end{eqnarray}

\noindent  Because the magnon velocity is higher than the spinon velocity, the kinematic conditions enable the decay of one magnon into two slow spinons.~\cite{notemd} This process gives rise to an imaginary part of $H({\omega \over cq})$. Consequently, the magnon pole moves away from the real axis leading to a finite decay rate $\Gamma_q \propto q^{1+\gamma}$.  A numerical  solution of the determinant of $\Pi_0 - \Pi(\bm q, i\omega)$ gives  $\gamma = 1$, implying that $\Gamma_q \propto q^2$.

\subsubsection{Around the \texorpdfstring{$\pm K$}{+-K} points}

The  magnon modes around $\pm K$ are connected by inversion symmetry. Thus, without loss of generality, we only consider the magnon dispersion around the $K$ point. The structure of the polarization operator $\Pi( \bm q + \bm Q, i\omega)$ is much simpler than the one obtained  for the $\Gamma$ point. The contribution from the non-condensed bosons is regular in the long wavelength limit, 

\begin{eqnarray}
\lim_{\bm q \rightarrow 0, \omega \rightarrow 0} \Pi_{nn}(\bm q + \bm Q,i\omega) = \Pi_{nn}(\bm Q,0),
\end{eqnarray}

\noindent while the contribution from the condensed bosons is

\begin{eqnarray}
\Pi_{cn}(\bm q + \bm Q,i\omega) ={\phi q^2/2 \over c^2 q^2 + \omega^2} \left( \Sigma^K  
+  Z \hat{\bm q} \cdot \bm Q^K \cdot \hat{\bm q}   \right).
\end{eqnarray}

By applying the procedure that we described for the $\Gamma$ point, we obtain a magnon  velocity  $v_K = 1.033 J $, which is very close to the value $v_K=0.9948J$  obtained by non-linear spin wave theory [up to ${\cal O}(S^{-1})$].~\cite{Chubukov1994,Chernyshev2009} The  non-linear correction to the magnon  dispersion is ${\cal O}(q^2)$, while the decay rate $\Gamma_q$ turns out to be proportional to $q^{5/2}$.

\section{Discussion}

In summary, we have demonstrated that the Gaussian  corrections of the Schwinger boson approach to the TLHAM
eliminates  serious limitations of the SP approximation and provides much better description  of the order parameter and the dynamical response of magnetically ordered states. This description becomes particularly appealing in the proximity of transitions into spin liquids. Here we have focused on the $1/N$ correction introduced by the diagram shown in Fig.~\ref{fig2:feymann}b. The main reason is that this is the only diagram that generates poles in $S({\bm q}, \omega)$ at the $1/N$ level. The rest of the $1/N$ diagrams shown in  Fig.~\ref{fig2:feymann}(c-e) renormalizes the interaction vertex, as well as the single-spinon propagator. It is important to note that these diagrams generate four-spinon contributions that will extend the high-energy spectral weight beyond  the two-spinon  continuum shown in Fig.~\ref{fig4:chi}. 

\begin{figure}[t]
 \includegraphics[width=7cm]{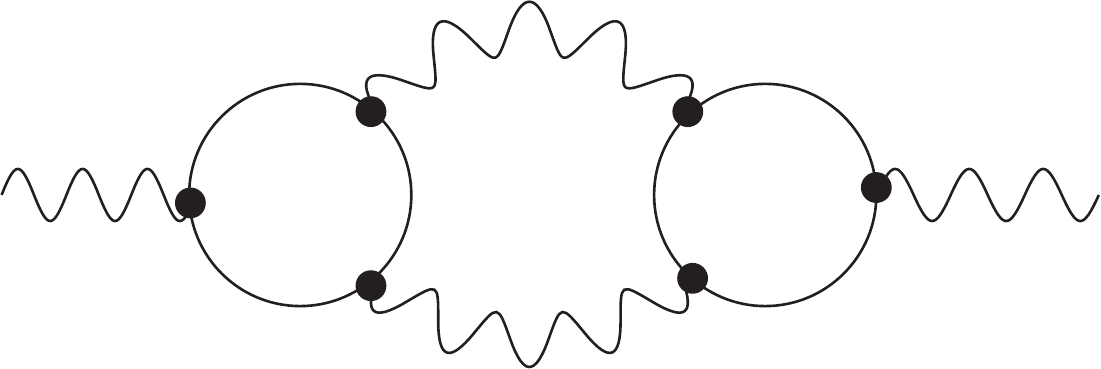}
 \caption{Diagram of order $1/N^2$ for the RPA propagator that accounts for the single-magnon to two magnon decay process.
Note the similarity between this diagram and the lowest order diagram in a $1/S$ expansion that accounts for the single-magnon to two-magnon decay (each wavy line must be interpreted as a magnon propagator, while each internal loop  must be interpreted as a cubic vertex).
 }
 \label{fig8:decaer}
 \end{figure}

The magnon excitations obtained from the $1/N$ correction considered in this manuscript consist in two-spinon bound states. Its dispersion agrees well with the magnon dipersion obtained from 
series expansions~\cite{Zheng2006} in the regions where the magnon spectral weight is high. Moreover, the magnon velocities obtained by taking the long wavelength limit around the ${\Gamma}$ and the ${\pm K}$ points agree very well with the result obtained from linear spin wave theory plus $1/S$ corrections.~\cite{Chubukov1994,Chernyshev2009} At the $1/N$ level, the magnon decay occurs via emission of two-spinons. Given that spinons are not low-energy modes of the Higgs phase (they are gapped out by the Higgs mechanism), this mechanism should be replaced by the more traditional single magnon to two magnon-decay in the long wavelength limit of the theory. However, to capture the single-magnon to two magnon decay within this  formalism, it is necessary to include $1/N^2$ corrections, such as the diagram shown in Fig.~\ref{fig8:decaer}. While the inclusion of two-magnon and four spinon processes is beyond the scope of this manuscript, we must keep in mind that these corrections are necessary to reproduce some  aspects of $S({\bm q}, \omega)$, such as the magnon broadening in the long wavelength limit or high energy contributions arising from the four-spinon continuum.

Recent inelastic neutron scattering measurements in Ba$_2$CoSb$_2$O$_9$~\cite{Ito2017} have revealed a three-stage energy structure in  $S({\bm q}, \omega)$ composed of single-magnon low-energy excitations and two high-energy dispersive excitation continua whose peaks are separated by an energy scale of order $J$.  Based on the results obtained in this manuscript, we speculate that the two stage high-energy structure arises from two-spinon and four-spinon contributions. Testing this conjecture requires not only the inclusion of additional diagrams, but also an  extension of the formalism presented in this manuscript to the case of a 3D lattice (vertically staked triangular layers) with  anisotropic  exchange interaction (Ba$_2$CoSb$_2$O$_9$ has a small easy-plane exchange anisotropy).

Because of the proximity of the $120^{\circ}$ N\'eel order ground state to a spin liquid state, we have used the $S=1/2$ triangular AFM Heisenberg model as an example of application. However, this formalism  can be applied to   other magnetically ordered states in the proximity of a spin liquid phase.  The current quest for materials that can realize quantum spin liquids  calls for approaches that capture the signatures of fractional excitations in  the dynamical  response. Given that most of these materials exhibit some form of magnetic ordering at low enough temperatures,  our approach is addressing the increasing demand
of modeling and understanding the dynamical response of quantum magnets in the proximity of a quantum melting point.

\begin{acknowledgments}
We wish to thank C. J. Gazza and D. A. Tennant for useful conversations. This work was partially supported by CONICET under grant Nro 364 (PIP2015).
S-S. Z. and CDB were partially supported from the LANL Directed Research and Development program.  
\end{acknowledgments}

\appendix

\section{Complex Gaussian integrals} 
\label{appendix_gaussian}
In order to compute the Gaussian correction to the partition function, we need to derive the necessary condition for the convergence and the value of the complex Gaussian integral 
\begin{equation}
Z = \int D[\vec\phi^\dagger,\vec\phi]e^{-\vec\phi^\dagger \cdot \bm A \cdot \vec\phi}, 
\label{zori}
\end{equation}
where $\vec\phi = \left(\phi_1,\cdots,\phi_n\right) \in {\mathbb C}^n,$ $\vec\phi^\dagger$ is the Hermitian conjugate of $\vec\phi$, 
the measure is given by 
$D[\vec\phi^\dagger,\vec\phi] = \prod_{\mu=1}^n \frac{d\bar{\phi}_\mu d\phi_\mu}{2\pi i},$ 
and $\bm A$ is a $n\times n$ complex matrix diagonalizable, not necessarily Hermitian. 

As $\bm A$ is diagonalizable, there exists a non-singular matrix $\bm R$ such that $\bm A$ can be transformed into a diagonal matrix 
$\bm \Lambda (\Lambda_{\mu \nu}=\lambda_\mu \delta_{\mu \nu})$ through the similarity transformation 
\begin{equation}
 \bm A = \bm R \cdot \bm \Lambda \cdot \bm R^{-1}. 
 \label{diagonalization}
\end{equation}
The diagonal entries of $\bm \Lambda$ are the (complex) eigenvalues of $\bm A$, while the vector columns of 
 $\bm R$ are the corresponding right-eigenvectors, $\bm A \cdot \vec \phi^{R}_{\mu}=\lambda_{\mu} \vec \phi^{R}_{\mu}.$
On the other hand, the rows of $\bm L \equiv \bm R^{-1}$ are the left-eigenvectors of $\bm A$, that is $\vec\phi^L_\mu \cdot \bm A = \lambda_\mu \vec\phi^L_\mu$.
Notice that if $\bm A$ is non-Hermitian, $\bm R$ is not unitary and, as a consequence, the right-eigenvectors are not orthogonal among themselves. 
The same is valid for the set of left-eigenvectors. However, as $\bm L \cdot \bm R = \bm I$, right-eigenvectors are orthonormalized with respect to the 
left-eigenvectors. 

Using Eq.~\eqref{diagonalization} and performing the linear transformation $\vec\phi  = \bm R \cdot \vec z,$ , whose Jacobian is given by $\det (\bm R^\dagger \cdot \bm R),$ $Z$ becomes
 \begin{equation}
  Z = \det{\bm G} \int D[\vec z^\dagger,\vec z] e^{-\vec z^\dagger\cdot \bm G \cdot \bm \Lambda \cdot \vec z}, 
  \label{zintgl}
     \end{equation}
 where the measure is $D[\vec z^\dagger, \vec z] =\prod_{\mu=1}^n \frac{d\bar{z}_\mu dz_\mu}{2\pi i}$ and we have defined 
 the Gram matrix $\bm G \equiv \bm R^\dagger \cdot \bm R,$ whose elements are given by the inner product of the right-eigenvectors, 
 $G_{\mu \nu} = \vec\phi^{R\dagger}_{\mu}\cdot \vec\phi^R_\nu$. $\bm G$ is Hermitian, and positive-definite as 
 the bilinear form 
$$ \vec z^\dagger \cdot \bm G \cdot \vec z = \left|\left|\sum_{\mu}z_\mu \vec\phi^{R}_\mu\right|\right|^2 > 0\;\; {\rm for}\;\;{\bm z \neq \bm 0}.  $$

Now, we will prove that, given a $n\times n$ diagonal matrix $\bm \Lambda$ whose diagonal entries $\lambda_\mu$ all have a positive real part, 
and a $n\times n$ Hermitian positive-definite matrix $\bm G$:
\begin{equation}
 I_n(\bm G, \bm \Lambda)  \equiv \int D[\vec z^\dagger,\vec z] e^{-\vec z^\dagger \cdot \bm G \cdot \bm \Lambda \cdot \vec z} = \frac{1}{\det (\bm G  \cdot \bm \Lambda)}. 
 \label{gausiana_gl}
\end{equation}
We prove this result by induction. 
For $n=1$, 
$$ I_{1}(\bm G, \bm \Lambda) = \int \frac{d\bar{z}_1 dz_1}{2\pi i} e^{-G_{11}\lambda_1 |z_1|^2}$$
where $\bm G$ is the positive real number $G_{11}.$ The analytical continuation of the well-known integral 
$\int \frac{d\bar{z}dz}{2\pi i}e^{-a|z|^2} = \frac{1}{a},$ easily computed for real $a >0,$ shows that 
$I_1$ converges if $\re \lambda_1 > 0$, and its value is given by 
$$ 
I_1(\bm G, \bm \Lambda) = \frac{1}{G_{11}\lambda_1} \equiv \frac{1}{\det (\bm G \cdot \bm \Lambda)}.
$$
So, for $n=1$ the integral (\ref{gausiana_gl}) is valid. 
Now we calculate the integral for a generic $n$, assuming its validity for $n-1$. 

\begin{eqnarray}
 I_n(\bm G, \bm \Lambda) &  = & \int \prod_{\mu=1}^{n-1}\frac{d\bar{z}_\mu dz_\mu}{2\pi i}    
 \exp\left[-\sum_{\mu\nu=1}^{n-1}\bar{z}_\mu G_{\mu \nu}\lambda_\nu z_\nu\right] \times 
\nonumber\\
& & \times \int \frac{d\bar{z}_n dz_n}{2\pi i} \exp\left[-S^{(n)}(\vec z^\dagger,\vec z)\right],\label{intzn}
\end{eqnarray}
where 
\begin{eqnarray}
       S^{(n)}(\vec z^\dagger, \vec z) & = & G_{nn}\lambda_n |z_n|^2 + \\
       & & +\bar{z}_n \sum_{\nu=1}^{n-1}G_{n\nu}\lambda_\nu z_\nu + 
       z_n \sum_{\mu =1}^{n-1} G_{\mu n}\lambda_n \bar{z}_\mu.\nonumber
      \end{eqnarray}
As $\bm G$ is positive-definite, all its diagonal elements are positive real numbers. Hence, the integral over 
$\bar{z}_n, z_n$ in Eq.~\eqref{intzn} is convergent if $\re \lambda_n > 0.$ Using  the well known integral
$\int \frac{d\bar{z} dz}{2\pi i} e^{-a|z|^2 + J \bar{z} + J' z} = \frac{1}{a}\exp[\frac{JJ'}{a}],$
valid for $\re a >0$ and complex $J, J'$, the integral is given by 
\begin{equation}
 \frac{1}{G_{nn}\lambda_n} \exp\left[\sum_{\mu \nu =1}^{n-1}\bar{z}_\mu \frac{G_{\mu n}G_{n \nu}}{G_{nn}}\lambda_\nu z_\nu \right]. 
\end{equation}
After performing the integration over $\bar{z}_n, z_n$, $I_n$ (\ref{intzn}) is given by a $(n-1)$-dimensional complex Gaussian integral  
\begin{equation}
 I_{n}(\bm G,\bm \Lambda)  = \frac{1}{G_{nn}\lambda_n} I_{n-1}((\bm G/G_{nn}),\bm {\tilde \Lambda})
 \label{intzn1}
\end{equation}
where the $(n-1)\times (n-1)$ matrix $(\bm G/G_{nn})$ is the so-called 
Schur complement of $\bm G$ with respect to $G_{nn}$,~\cite{Horn2012} defined as  
\begin{equation}
(\bm G/G_{nn})_{\mu \nu} = G_{\mu \nu} -  \frac{G_{\mu n}G_{n \nu}}{G_{nn}},
\label{defschur}
\end{equation}
while $\bm \tilde \Lambda$ is the diagonal $(n-1)\times(n-1)$ matrix that results from taking out the $n$-th row 
and the $n$-th column in $\bm \Lambda.$
$(\bm G/G_{nn})$ is a Hermitian and positive-definite matrix.
To prove this last statement, we take into account that $\bm G$ is positive-definite, that 
is 
\begin{eqnarray*}
 \vec z^\dagger \cdot \bm G \cdot \vec z & = & \sum_{\mu,\nu=1}^{n-1}\bar{z}_\mu G_{\mu \nu} z_\nu + 
 \bar{z}_n \sum_{\nu =1}^{n-1}G_{n\nu} z_\nu + \\ 
 & + & \sum_{\mu=1}^{n-1} \bar{z}_\mu G_{\mu n} z_n + G_{nn}|z_n|^2 > 0,
 \end{eqnarray*}
for all non-zero $\vec z \in {\mathbb C}^{n}$. In particular, if we take $z_n = -\frac{1}{G_{nn}}\sum_{\nu=1}^{n-1}G_{n \nu}z_{\nu},$ we get 
$\vec z^{\dagger} \cdot (\bm G/G_{nn}) \cdot \vec z > 0,$ for all non-zero $\vec z \in {\mathbb C}^{n-1}.$
As $(\bm G/G_{nn})$ is Hermitian positive-definite, and the real part of all the diagonal entries of $\bm {\tilde \Lambda}$ are positive, 
the $(n-1)$-dimensional complex integral in (\ref{intzn1}), by hypothesis, equals  $1/\det((\bm G/G_{nn}) \cdot \bm {\tilde \Lambda})$, and 
it results in 
\begin{equation}
 I_{n}(\bm G,\bm \Lambda) = \frac{1}{G_{nn}\det (\bm G/G_{nn})} \times \frac{1}{\det\bm \Lambda}.
 \label{in}
\end{equation}
The Schur's identity~\cite{Horn2012} tell us that, if $G_{nn} \neq 0$,  
\begin{equation}
 \det\bm G = G_{nn} \det (\bm G/G_{nn}). 
\label{schuridentity}
\end{equation}
Replacing the identity in Eq.~\eqref{in}, we end the proof of the Gaussian integral
\begin{equation}
\int D[\vec z^\dagger,\vec z] e^{-\vec z^\dagger \cdot \bm G \cdot \bm \Lambda \cdot \vec z} = \frac{1}{\det (\bm G  \cdot \bm \Lambda)}. 
\end{equation}

The application of above equation to the integral $Z$ as expressed in Eq.~\eqref{zintgl} allows us, on one hand, 
to establish the convergence condition of the original Gaussian integral \eqref{zori}. 
That is, all the eigenvalues of the $\bm A$ matrix should have a positive real part. A matrix with this property is called a positive-stable
matrix.
On the other hand, we  get  
\begin{equation}
Z = \int D[\vec\phi^\dagger,\vec\phi]e^{-\vec\phi^\dagger \cdot \bm A \cdot \vec\phi} = \frac{1}{\det\bm A}, 
\end{equation}
since $\det \bm \Lambda = \det \bm A$. This result is the generalization of the usual complex Gaussian integral, 
where it is requested that $\bm A$ or its real part be positive-definite matrices. 
It can be shown that, while any matrix with a positive-definite real part is a positive-stable matrix, 
the converse is not true.

An alternative way to arrive at the positive-stable condition for the convergence of the Gaussian integral is
to  use the analytical continuation of the Gaussian integral with a Hermitian matrix.~\cite{Kamenev2009}

\section{Faddeev-Popov treatment of zero modes}
\label{appendix_fp}
In this Appendix, we show how to derive the partition function in the presence of zero gauge modes, 
by means of the  Faddeev-Popov prescription. 

Let $Z$ be the Gaussian integral 
\begin{equation}
 Z = \int D[\vec\phi^\dagger, \vec\phi] e^{-\vec\phi^\dagger\cdot \bm A\cdot \vec\phi} 
 \label{zint}
\end{equation}
where $\bm A$ is a diagonalizable matrix that has a zero eigenvalue, with its corresponding right
$\vec{\phi}^R_0$ and left $\vec{\phi}^L_0$ eigenvectors. In our case, 
this zero mode is the consequence of the invariance of $Z$ 
under a $U(1)$ gauge transformation characterized by the (complex) phase $\theta$, whose expression is 
\begin{equation}
 \vec\phi \to \vec\phi(\theta) = \vec\phi + \theta \vec\phi^R_0,\;\;
 \vec\phi^\dagger \to \vec\phi^\dagger(\bar{\theta}) = \vec\phi^\dagger 
 + \bar{\theta} \vec\phi_0^L.
 \label{gaugeinf}
\end{equation}
In fact, the exponent of the Gaussian integral is gauge invariant, $\vec\phi^\dagger(\bar{\theta}) \cdot \bm A \cdot \vec\phi(\theta) = 
\vec\phi^\dagger\cdot \bm A \cdot \vec\phi$, and the transformation \eqref{gaugeinf} has a unit Jacobian.

For a positive-stable $\bm A$, the Gaussian integral (\ref{zint}) is given by $\det^{-1}({\bm A})$ (see Appendix \ref{appendix_gaussian}). As a consequence, 
the presence of a zero eigenvalue implies its divergence,  
and this signals the absence of a restoring force along the zero mode direction, due precisely to the gauge symmetry. 
However, using the Faddeev-Popov trick we can 
exactly extract from $Z$ the contribution of the gauge group volume $\int \frac{d\bar{\theta} d\theta}{2\pi i}$ --that counts the 
redundant gauge degree of freedom that give rise to such divergence--, remaining a physically sound result from the integration of 
the genuine Gaussian fluctuations.

To proceed with the Faddeev-Popov trick, first we define a Dirac delta function on the complex plane 
by means of the integral representation
\begin{equation}
 \delta(\bar{z}, z) = \int \frac{d\bar{\alpha} d\alpha}{2\pi i} e^{i(\bar{\alpha} z + \bar{z} \alpha)}, 
\end{equation}
such that $\delta(\bar{z}, z)$ satisfies, for a given function $F$, the usual relation
\begin{equation}
 \int \frac{d\bar{z} dz}{2\pi i} \delta(\bar{z},z)F(\bar{z},z) = F(0,0).
 \label{delta}
\end{equation}
Indeed, 
\begin{equation}
 \delta(\bar{z},z) = \int \frac{d\alpha_R d\alpha_I}{\pi}e^{2 i (\alpha_R x + \alpha_I y)} =  \pi \delta(x)\delta(y),
 \label{deltaz}
\end{equation}
where $\alpha=\alpha_R + i \alpha_I$ and $z=x+iy$. Replacing (\ref{deltaz}) in Eq.~\eqref{delta}, we get the usual 
property of the Dirac delta function. 

To get rid of the (divergent) gauge fluctuations, we impose {\it natural} gauge fixing conditions that constraint the fluctuations to be orthogonal 
to the zero mode,~\cite{Trumper1997}
\begin{equation} 
g(\bar\theta) = \vec\phi^{\dagger}(\bar{\theta}) \cdot 
\hat\phi^R_0 = 0,\;\; h(\theta) = \hat\phi^L_0\cdot \vec\phi(\theta) = 0,
\label{gaugecond}
\end{equation}
where we have defined the zero mode ``versors'' as
$ \hat\phi^{L}_0 \equiv \vec\phi^L_0/\sqrt{\vec\phi^L_0 \cdot \vec\phi^R_0}$ and 
$\hat\phi^{R}_0 \equiv \vec\phi^R_0/\sqrt{\vec\phi^L_0 \cdot \vec\phi^R_0}. $
Taking into account Eq.~\eqref{gaugeinf}, the gauge fixing conditions are given by 
\begin{eqnarray}
g(\bar{\theta}) & =  & \vec\phi^\dagger\cdot \hat\phi^R_0 + \bar{\theta}\sqrt{\vec\phi^L_0 \cdot \vec\phi^R_0},
\label{condtheta}\\
 h(\theta) & =  & \hat\phi^L_0\cdot \vec\phi + \theta \sqrt{\vec\phi^L_0 \cdot \vec\phi^R_0}, 
 \nonumber
\end{eqnarray}
and they are imposed in $Z$ through the Faddeev-Popov trick, that consists in the construction of the unit as 
\begin{equation}
1 = \Delta_{\rm FP}(\vec\phi^\dagger,\vec\phi) \times \int \frac{d\bar{\theta} d\theta}{2\pi i} \delta(g(\bar{\theta}),h(\theta)), 
 \label{conditionNH}
\end{equation}
where $\Delta_{\rm FP}$ is the so-called Faddeev-Popov determinant. To compute it, we first replace the Dirac delta distribution by its integral representation and express the gauge conditions by means of Eq.~\eqref{condtheta}:
\begin{widetext}
\begin{eqnarray}
 \int \frac{d\bar{\theta} d\theta}{2\pi i} \delta(g(\bar{\theta}),h(\theta)) & = &
\int \frac{d\bar{\alpha} d\alpha}{2\pi i} e^{i\left(\bar{\alpha}\vec\phi^\dagger \cdot \hat{\phi}^R_0 + \alpha \hat{\phi}^L_0 \cdot 
 \vec\phi\right)}  \int\frac{d\bar{\theta}d\theta}{2\pi i} e^{i\sqrt{\vec\phi^L_0 \cdot \vec\phi^R_0}\left(\bar{\theta}\alpha + \theta
 \bar\alpha\right)} = \nonumber \\ 
 & = & \int \frac{d\bar{\alpha} d\alpha}{2\pi i} e^{i\left(\bar{\alpha}\vec\phi^\dagger \cdot \hat{\phi}^R_0 + \alpha \hat{\phi}^L_0 \cdot 
 \vec\phi\right)} \delta\left(\sqrt{\vec\phi^L_0 \cdot \vec\phi^R_0}\bar\alpha,\sqrt{\vec\phi^L_0 \cdot \vec\phi^R_0}\alpha\right) 
 = \frac{1}{\vec\phi^L_0 \cdot \vec\phi^R_0}.\nonumber\end{eqnarray}
\end{widetext}
Here we have used the property $\delta(a\bar{z},bz) = \frac{1}{ab}\delta(\bar{z},z)$ valid for
complex $a$ and $b$ with $\re (ab) > 0$.
Hence, we arrive at the expression for the Faddeev-Popov determinant
\begin{equation}
 \Delta_{\rm FP} = \vec\phi^L_0 \cdot \vec\phi^R_0.
\end{equation}
After computing $\Delta_{\rm FP}$, we insert the unit \eqref{conditionNH} in the partition function (\ref{zint}) and replace again 
the Dirac delta that fixes  the gauge choice with its integral expression.
The partition function becomes
\begin{widetext}
\begin{equation}
 Z = \Delta_{\rm FP} \int \frac{d\bar{\theta} d\theta}{2\pi i}\int \frac{d\bar{\alpha} d\alpha}{2\pi i}
 \int D\left[\vec\phi^\dagger,\vec\phi\right]
              e^{i\left(\bar{\alpha} \hat\phi_0^{L}\cdot \vec \phi(\theta) 
              + \alpha\vec\phi^{\dagger}(\bar{\theta})\cdot \hat\phi^R_0\right)}            
\; e^{-\vec\phi^\dagger \cdot {\bm A}\cdot \vec\phi}.
\end{equation}

%
To extract the gauge-group volume from $Z$, we perform a gauge transformation $\vec\phi^\dagger \to \vec\phi^{\dagger}(-\bar{\theta}),\;\vec\phi \to \vec\phi(-\theta)$ 
for given phases $\bar{\theta}, \theta$.
Given that the action and the measure are gauge-invariant quantities, we get rid of the $\theta$-dependence of the integrand
and the (divergent) gauge-group volume can be factored as an (irrelevant) multiplicative constant out of the integral:
\begin{equation}
Z = \left[\int \frac{d\bar\theta d\theta}{2\pi i}\right] 
  \times \Delta_{\rm FP} 
           \int \frac{d\bar\alpha d\alpha}{2\pi i}
           \int D\left[\vec\phi^\dagger,\vec\phi\right]
               e^{i\left(\bar{\alpha} \hat\phi_0^{L}\cdot \vec \phi 
              + \alpha\vec\phi^{\dagger}\cdot \hat\phi^R_0\right)}            
 \;e^{-\vec\phi^\dagger \cdot {\bm A}\cdot \vec\phi}.
  \label{zvol}
 \end{equation}
 \end{widetext}
In what follows, we remove this gauge-group volume factor. 

Next, we  decompose the field vectors $\vec\phi^\dagger, \vec\phi$ in the basis of the right eigenvectors, separating explicitly the component parallel to the zero mode, 
\begin{equation}
 \vec\phi = \sum_{\mu \neq 0} z_{\mu} \hat{\phi}^R_\mu + z_0 \hat{\phi}^R_0,\;\;\;
 \vec\phi^\dagger = \sum_{\mu \neq 0} \bar{z}_{\mu} \hat{\phi}^{R\dagger}_\mu + \bar{z}_0 \hat{\phi}^{R\dagger}_0.
\end{equation}
After applying this decomposition and taking into account that the right-eigenvectors are orthogonalized with respect to the left eigenvectors, but not necessarily among each other, the exponent $i\left(\bar{\alpha} \hat\phi_0^{L} \cdot \vec\phi + \alpha \vec\phi^\dagger \cdot \hat\phi_0^R \right)$ in the integral \eqref{zvol} becomes 
\begin{equation}i \left(\bar{\alpha} z_0 + \alpha \bar{z}_0 G_{00} + \alpha(\vec z^\dagger_\perp \cdot \bm G)_0\right)
\end{equation}
where $\bm G$ is the (Hermitian positive-definite) Gram matrix of the right eigenvectors, 
$G_{\mu \nu} = \hat{\bm \phi}^{R\dagger}_{\mu}\cdot \hat{\bm \phi}^R_{\nu}$, and 
$\vec z_\perp$ is the vector of complex coordinates $\left\{z_\mu\right\}$ excluding $z_0$. 

On the other hand, the exponent $\vec\phi^\dagger \cdot \bm A \cdot \vec\phi$ becomes 
\begin{eqnarray}
  \vec z^\dagger \cdot \bm G \cdot \bm \Lambda \cdot \vec z & =  & 
 \lambda_0 G_{00}|z_0|^2 + \lambda_0 z_0(\vec z^\dagger_\perp \cdot \bm G)_0  + \nonumber\\
 & + & \bar{z}_0 (\bm G \cdot \bm \Lambda \cdot \vec z_\perp)_0 + \vec z^\dagger_\perp \cdot \bm G \cdot \bm \Lambda \cdot \vec z_\perp, \nonumber
\end{eqnarray}
where $\bm \Lambda$ is a diagonal matrix whose diagonal entries are the eigenvalues of $\bm A$. 

Then, we make the change of variables 
$\left\{\vec\phi^\dagger,\vec\phi\right\} \to \left\{\vec z^\dagger_\perp,\vec z_\perp, \bar{z}_0, z_0\right\}$
in $Z,$ whose Jacobian is $\det \bm G$, and we first integrate over $\bar z_0, z_0$. To properly treat the zero eigenvalue $\lambda_0$, as it has no ``restoring force",  
we assign a positive value to $\lambda_0$ and take the limit $\lambda_0 \to 0^+$ at the end of the calculation. 
After the change of variables, we get
\begin{widetext}
\begin{equation}
Z = \Delta_{\rm FP} \det {\bm G} \int D[\vec z^\dagger_\perp,\vec z_\perp]
    e^{-\vec z^\dagger_\perp \cdot \bm G \cdot \bm \Lambda \cdot\vec z_\perp} 
     \int \frac{d\bar\alpha d\alpha}{2\pi i}e^{i \alpha (\vec z^\dagger_\perp \cdot \bm G)_0}\times I_{0}
\end{equation}
where
          \begin{eqnarray}
          I_0 & \equiv & \int \frac{d\bar{z}_0 dz_0}{2\pi i}
             \exp\left[-\lambda_0 G_{00}|z_0|^2 
 +z_0\left(i\bar\alpha - \lambda_0 (\vec z^\dagger_\perp \cdot \bm G)_0\right) + 
 \bar{z}_0 \left(i\alpha G_{00} - (\bm G\cdot \bm \Lambda \cdot \vec z_\perp)_0\right)\right] = \nonumber\\
& = & \frac{1}{\lambda_0 G_{00}}\exp\left[-\frac{1}{\lambda_0}|\alpha|^2-i\alpha
(\vec z^\dagger_\perp \cdot \bm G)_0-i\bar\alpha \frac{(\bm G\cdot \bm \Lambda \cdot \vec z_\perp)_0}{\lambda_0 G_{00}} +
\frac{(\vec z^\dagger_\perp \cdot \bm G)_0(\bm G\cdot \bm \Lambda \cdot \vec z_\perp)_0}{G_{00}} 
\right].
 \end{eqnarray}
 \end{widetext}

We then collect all the terms that depend on $\bar\alpha, \alpha$ and perform the integral
\begin{equation}
 \int \frac{d\bar{\alpha}d\alpha}{2\pi i} \exp\left[-\frac{1}{\lambda_0}|\alpha|^2 - i\bar\alpha 
 \frac{(\bm G\cdot \bm \Lambda \cdot \vec z_\perp)_0}{\lambda_0 G_{00}}
 \right] = \lambda_0. 
\end{equation}
At this stage, $Z$ is expressed as 
\begin{equation}
 Z  =  \Delta_{\rm FP}  
 \frac{\det {\bm G}}{G_{00}}  
         \int D[\vec z^\dagger_\perp,\vec z_\perp]e^{-\vec z^\dagger_\perp \cdot (\bm G/G_{00}) \cdot \bm {\tilde \Lambda} 
           \cdot\vec z_\perp}, \label{z3}
            \end{equation}
where the Hermitian positive-definite matrix $(\bm G/G_{00})$ is the Schur complement [see Eq.~\eqref{defschur}] of  $\bm G$ with respect to $G_{00}$,  and $\bm {\tilde \Lambda}$ is the matrix that results from extracting the zero mode column and row  in $\bm \Lambda$. Notice that the regularization parameter $\lambda_0$ goes away as we integrate over the zero mode coordinates and $\alpha$, rendering  the $\lambda_0 \to 0^+$ limit trivial.
The last step is to perform the integral over $\vec z^\dagger_\perp, \vec z_\perp$ using the Gaussian integral~\eqref{gausiana_gl}:
\begin{equation}
 \int D[\vec z^\dagger_\perp,\vec z_\perp] e^{-\vec z^\dagger_\perp \cdot (\bm G/G_{00})  \cdot \bm 
 {\tilde \Lambda} \cdot \vec z} = \frac{1}{\det (\bm G/G_{00}) \det \bm {\tilde \Lambda}}.
 \end{equation}
As $\det{\bm G} = G_{00}\det (\bm G/G_{00})$ [see Eq.~\eqref{schuridentity}]  and $\det \bm{\tilde \Lambda} = \det \bm {A}_\perp$, we finally arrive at the formula 
\begin{equation}
   Z = \frac{\Delta_{\rm FP}}{\det \bm {A}_\perp} = \frac{\vec\phi^L_0 \cdot \vec\phi^R_0}{\det{\bm A}_{\perp}}.
       \end{equation}

Hence, the Gaussian correction to our Schwinger boson partition function, 
\begin{equation}
 {\mathcal Z}^{\rm fl} = \prod_{{\bf k},i\omega_n >0} \frac{\Delta_{\rm FP}(\bk,i\omega_n)}{\det {\bm S}^{(2)}_{\perp}(\bk,i\omega_n)},  
\end{equation}
  where the Faddeev-Popov determinant is given by Eq.~\eqref{detFP}. 
Notice that the partition function for all the Hamiltonian parameters set to zero (exchange interactions, external sources, breaking-symmetry magnetic field) is
\begin{equation}
 {\mathcal Z}_0 = \prod_{{\bf k},i\omega_n >0} \frac{\Delta_{\rm FP,0}(\bk,i\omega_n)}{\det {\bm S}^{(2)}_{\perp,0}(\bk,i\omega_n)} =
 \prod_{{\bf k},i\omega_n > 0} \omega_n^2, 
\end{equation}
since the saddle-point parameters $A_\delta = B_\delta = 0$, and the perpendicular matrix is the identity. 
Given that we want to evaluate the free energy relative to this zero of energy, we divide the partition function by
${\mathcal Z}_0$:
\begin{equation}
 \frac{{\mathcal Z}^{\rm fl}}{{\mathcal Z}_0} = \prod_{\bk i\omega_n > 0} \frac{\Delta_{\rm FP}(\bk,i\omega_n)}{w_n^2 \det {\bm S}^{(2)}_{\perp}(\bk,i\omega_n)}.
\end{equation}

\section{Relationship between the determinants of the perpendicular and truncated matrices}
\label{appendix_determinants}

Let $\bm A$ be a diagonalizable $n \times n$ complex matrix that is taken to the diagonal form $\bm \Lambda$ through the similarity transformation \eqref{diagonalization} 
$\bm A = \bm R \cdot \bm \Lambda \cdot \bm R^{-1},$
where the diagonal entries of $\bm \Lambda$ ($\Lambda_{\mu \nu} = \lambda_\mu \delta_{\mu \nu}$) are the eigenvalues, while 
the columns of $\bm R$ are the right-eigenvectors and the rows of $\bm L = \bm R^{-1}$ are the left-eigenvectors of $\bm A$.
We assume that the $n$-th eigenvalue of ${\bm A}$ is zero, being $\hat\phi^R_0$ and $\hat\phi^L_0$ the right- and left- zero modes, respectively. 
As $\bm L \cdot \bm R =\bm I$, the zero modes satisfy $\hat \phi^L_0 \cdot \hat \phi^R_0 =1$.

The {\it truncated} matrix $\bm A_{\rm tr}$ is defined as the $(n-1)\times (n-1)$ matrix that results from taking out the $n$-th column and the 
$n$-th row of $\bm A$, while the {\it perpendicular} matrix ${\bm A}_{\perp}$ is the 
 $(n-1)\times (n-1)$ diagonal matrix whose elements are the same as the non-zero eigenvalues of ${\bm A}$, 
 that is, $A_{\perp \alpha\beta} = \lambda_\alpha \delta_{\alpha \beta}$ for $\alpha, \beta =1,\cdots,n-1$.\\
 
We will prove the following relationship between determinants:
\begin{equation}
 \det{\bm A}_{\rm tr} = \phi^L_{0n}\phi^R_{0n}\; \det{\bm A}_{\perp}, 
\end{equation}
where $\phi^R_{0n}, \phi^L_{0n}$ are the $n$-th components of the ``normalized'' zero modes, $\hat\phi^L_0, \hat\phi^R_0$.\\

We start by separating explicitly   the $n$-th column and the $n$-th row of $\bm A$
\begin{equation}
 {\bm A} = \left( \begin{array}{cc}
 {\bm A}_{\rm tr} & \vec A_{\perp n} \\ 
                    \vec A_{n\perp} & A_{nn}
   \end{array}\right),
      \end{equation}
where $\vec A_{\perp n} \!=\!\left(\begin{array}{c} 
A_{1n} \\ \cdots \\ A_{n-1\; n}\end{array}\right)$ and $\vec A_{ n\perp } \!=\!\left(A_{n1}, \cdots, A_{n,n-1} \right).$
Analogously, we write the vectors as $\vec \phi = \left(\begin{array} {c} \vec \phi_{\perp} \\ \phi_{n}\end{array}\right).$
Using these definitions, the right zero mode equation ${\bm A}\cdot \hat{\phi}^R_0= 0$ can be written as 
\begin{eqnarray*}
  {\bm A}_{\rm tr} \cdot \hat\phi^R_{0\perp} + {\vec A}_{\perp n}\phi^{R}_{0n} = \vec 0
        \label{sgn}
    \\
  \vec{A}_{n\perp} \cdot \hat\phi^R_{0\perp} + A_{nn}\phi^R_{0n} =  0, & &
  \label{snn} 
          \end{eqnarray*}
          while for $\hat\phi^L_0 \cdot \bm A = 0$ we have
          \begin{eqnarray*}
   \hat\phi^L_{0\perp} \cdot {\bm A}_{\rm tr}+ \phi^{L}_{0n}{\vec A}_{n \perp} = \vec 0\\
     \hat\phi^L_{0\perp}\cdot \vec{A}_{\perp n } + \phi^L_{0n} A_{nn} =  0. & &
          \end{eqnarray*}
These equations allow us to write all the elements of $\bm A$ in terms of the elements of the truncated matrix:
 \begin{eqnarray}
 \vec A_{\perp n} & = & -\frac{1}{\phi^R_{0n}} {\bm A}_{\rm tr} \cdot \hat\phi^R_{0\perp},\nonumber\\
 \vec A_{n \perp} & = & -\frac{1}{\phi^L_{0n}} \hat\phi^L_{0\perp}\cdot \bm A_{\rm tr},\label{SNG}\\
 A_{nn} &  = & \frac{1}{\phi^L_{0n}\phi^{R}_{0n}}\hat\phi^L_{0\perp}\cdot \bm A_{\rm tr}\cdot \hat\phi^R_{0\perp}.
 \nonumber
  \end{eqnarray}
Then, we right- and left-multiply $\bm A$ by the $n \times (n-1)$ matrix containing the first $n-1$ right-eigenvectors and the 
$(n-1)\times n$ matrix containing the first $n-1$ left-eigenvectors, respectively, 
in order to compute the $(n-1)\times(n-1)$ perpendicular matrix $\bm A_{\perp}$:
\begin{equation*}
 \left(\begin{array}{c} \hat\phi^L_1\\ \cdots\\ \hat\phi^L_{n-1} \end{array}\right) 
 \cdot {\bm A} \cdot 
 \left(\hat\phi^R_1, \cdots, \hat\phi^R_{n-1}\right)  =
        \left(\begin{array}{ccc}
              \lambda_1 & 0 & \cdots  \\ 
               \cdots & \cdots & \cdots   \\
               \cdots & 0  & \lambda_{n-1}
             \end{array}\right) = \bm A_\perp.
\end{equation*}
By replacing Eqs.~(\ref{SNG}) in the above equation and after a little algebra, we get 
 \begin{equation}
  (\bm L/L_{nn}) \cdot \bm A_{\rm tr} \cdot (\bm R/R_{nn}) = \bm A_\perp, 
  \label{schur}
 \end{equation}
 where $(\bm L/L_{nn})$ and $(\bm R/R_{nn})$ are Schur complements as defined in Eq.~\eqref{defschur}.
 The Schur's identity tells us that 
 $ \det \bm L = L_{nn}\det (\bm L/L_{nn})$ and $\det\bm R = R_{nn}\det (\bm R/R_{nn}).
 $
 As $\det \bm L \times \det \bm R =1,$ and $L_{nn}=\phi^L_{0n},$ $R_{nn}=\phi^R_{0n}$,  we obtain  the desired relationship 
 between the determinants of the truncated and perpendicular matrices from Eq.~\eqref{schur}
 \begin{equation}
  \det \bm A_{\rm tr} = \phi^L_{0n}\phi^R_{0n}\det \bm A_{\perp}.
 \end{equation}

For the  SB case of interest, when we use the non-Hermitian fluctuation matrix ${\bf S}^{(2)}$, with its (non-normalized) right- 
and left-zero modes [Eqs.\eqref{rightzero} and \eqref{leftzero}], we obtain
 $\phi^R_{0n} =   \frac{i\omega_n}{\sqrt{\Delta_{\rm FP}}}$ and $\phi^L_{0n} = -\frac{i\omega_n}{\sqrt{\Delta_{\rm FP}}},$
 with the Faddeev-Popov determinant given by Eq.~\eqref{detFP}. This results in the relation 
 \begin{equation}
  \det {{\bm S}^{(2)}_{\rm tr}}(\bk,i\omega_n) = \frac{\omega_n^2}{\Delta_{\rm FP}(\bk,i\omega_n)} \det {\bm S}^{(2)}_{\perp}(\bk,i\omega_n).
 \end{equation}
A similar relation holds if we use the Hermitian fluctuation matrix $\tilde {\bm S}^{(2)}$. In this case, 
the left-zero mode is just the Hermitian conjugate of the right-zero mode, 
$\vec\phi^L_0 = \vec\phi^{R\dagger}_0$, so the Faddeev-Popov determinant results $\Delta^{\rm H}_{\rm FP}(\bk,i\omega)= \vec\phi^{R\dagger}_0(\bk,i\omega_n) \cdot \vec\phi^R_0(\bk,i\omega_n)$. In both cases, non-Hermitian and Hermitian fluctuation matrices, we have numerically checked that the 
above relation between determinants is satisfied. 

\section{Long wavelength limit of the Schwinger boson theory}
\label{app1}

We derive the Schwinger boson theory in the  long wavelength limit by expanding the  spinon dispersion  around its gapless points 
($\Gamma$ and $\pm K$ points 
of the Brillouin zone) to provide asymptotic forms of the single-spinon Green's function and the polarization operator.
 
\subsection{Linearized spinon dispersion and Green's function}
The spinon dispersion  is approximated by 
\begin{eqnarray}
 \varepsilon_{\bm k \sigma} = ck ,\;\; \sigma =\pm,
 \label{dispersionGamma}
 \end{eqnarray}
around the $\Gamma$ point,  where the spin velocity is
\begin{eqnarray}
 c = \sqrt{{\frac{2}{3}} \left[ \left(\gamma_{{\bm Q}/2}^A\right)^2 - \gamma_{{\bm Q}/2}^{B} \left(\lambda+\gamma_{{\bm Q}/2}^{B}\right) \right] }.
 \end{eqnarray}
The two gapless branches have the same velocity.

Around the $\pm K$ point, there is only one gapless branch with the same velocity $c$:
\begin{eqnarray}\label{dispersionK}
 \varepsilon_{\bm k \pm \bm Q,-} = ck.
 \end{eqnarray}

After taking the thermodynamic limit according to Eq.~\eqref{gnc}, the spinon Green's function  can be separated
into the condensed and non-condensed contributions 
\begin{eqnarray}
\mathcal{G}^{\rm sp}(\bm k, i\omega) & \!\! = \!\! & \mathcal{G}^{\rm sp}_{\rm n}(\bm k, i\omega) \! + \! (2\pi)^2 \delta(\bm k) \mathcal{G}^{\rm sp}_{\rm c}(\bm 0, i\omega).
\end{eqnarray}
The first term describes the non-condensed bosons with $\bm k \neq \bm 0$, while the second term describes the condensed bosons at $\bm k = \bm 0$:
\begin{eqnarray}
\mathcal{G}^{\rm sp}_{\rm c}(\bm 0, i\omega) = {g_c} \phi \left( {\frac{1}{\varepsilon_{\bm 0} -i\omega} } + {\frac{1}{\varepsilon_{\bm 0} +i\omega} }  \right).
\end{eqnarray}

\noindent Here $\phi$ is the density of  condensed spinons at  the saddle point level, $\varepsilon_{\bm 0} = \lim\limits_{N_s\rightarrow \infty} 
{\lambda+\gamma_{ { \bm Q} /2}^B \over 4\phi N_s}$, and

\begin{equation}
 g_{\rm c} = {1\over 2} \left(
  \begin{array}{cccc}
  1             &   1            &   -1            &  -1  \\
  1             &   1            &   -1            &  -1  \\
  -1            &   -1           &    1            &  1  \\
  -1            &   -1           &    1            &  1  \\
  \end{array} \right) \ .
\end{equation}

The low energy sector of the non-condensed boson contains three types with momenta $\left| \bm k \right| < \Lambda$, $\left|\bm k - \bm Q \right|< \Lambda$ or 
$\left|\bm k + \bm Q \right| < \Lambda$. Correspondingly, we derive the  asymptotic form of the Green's function in the long wavelength limit around each of the three wave vectors.

\subsubsection{Around \texorpdfstring{$\Gamma$}{G} point: \texorpdfstring{$\left| {\bf k} \right| < \Lambda$}{|k| < Lambda}}

The leading order contribution to the Green's function has the form

\begin{eqnarray}
\mathcal{G}^{\rm sp}_{\rm n}(\bm k, i\omega) =  \sum_{\alpha=1,2} { Z \over c^2 k^2 +\omega^2 } I_{\alpha},
\end{eqnarray}

\noindent where $Z=\lambda+\gamma_{ \bar{ \bm Q} /2}^B$ and

\begin{equation}
I_{1}=\left(\begin{array}{cccc}
1 & 1 & 0 & 0\\
1 & 1 & 0 & 0\\
0 & 0 & 0 & 0\\
0 & 0 & 0 & 0
\end{array}\right),\;I_{2}=\left(\begin{array}{cccc}
0 & 0 & 0 & 0\\
0 & 0 & 0 & 0\\
0 & 0 & 1 & 1\\
0 & 0 & 1 & 1
\end{array}\right).
\end{equation}

\subsubsection{Around \texorpdfstring{$\pm K$}{+-K} point: \texorpdfstring{$\left|\bf k \pm \bf Q \right| < \Lambda$}{|k +- Q| < Lambda}}

Given that there is only one gapless branch for these two wave vectors, we have

\begin{eqnarray}
\mathcal{G}^{\rm sp}_{\rm n}(\bm k + \bm Q, i\omega) &=&  { Z \over c^2 k^2 +\omega^2 } I_{1}^K, \\
\mathcal{G}^{\rm sp}_{\rm n}(\bm k - \bm Q, i\omega) &=&  { Z \over c^2 k^2 +\omega^2 } I_{2}^K, 
\end{eqnarray}
with

\begin{equation}
I_{1}^{K}=\left(\begin{array}{cccc}
  1 & -1 & 0 & 0 \\
 -1 &  1 & 0 & 0 \\
  0 &  0 & 0 & 0 \\
  0 &  0 & 0 & 0
\end{array}\right), \;I_{2}^K=\left(\begin{array}{cccc}
0 & 0 & 0 & 0\\
0 & 0 & 0 & 0\\
0 & 0 & 1 & -1\\
0 & 0 & -1 & 1
\end{array}\right).
\end{equation}

\subsection{Polarization operator}

The RPA  propagator  is determined by the polarization operator $\Pi(\bm q, i\omega)$, whose 
computation  in the long wavelength limit, $\bm q \rightarrow \bm 0$, follows from Eq.~\eqref{polar}.

\subsubsection{Around \texorpdfstring{$\Gamma$}{G} point: \texorpdfstring{$\left|\bf k \right| < \Lambda$}{|k| < Lambda}}

We first consider the leading order contribution  ${\cal O} (q^{-2})$ arising  from the condensed bosons:

\begin{eqnarray}
\Pi_{cn}^{(-2)}(\bm q, i\omega) ={\phi Z/2 \over c^2 q^2 + \omega^2} \Theta,
\end{eqnarray}

\noindent where 

\begin{eqnarray}
\Theta_{\alpha \beta} = \text{Tr}\left[ I_p v_{\alpha}(\bm 0, \bm 0) I_p v_{\beta} (\bm 0, \bm 0)  \right],
\end{eqnarray}

\noindent  and $I_p = I_1 +I_2$. The matrix $\Theta$ contains only one non-zero matrix element: $\Theta=\theta \rvert u_1 \rangle \langle v_1 \rvert $. The subspace ${\cal Q}_{\perp}$ with projector ${\cal P}_{\perp}=\rvert u_1 \rangle \langle v_1 \rvert $ contains no pole, implying that  the magnon pole must appear in the orthogonal subspace ${\cal Q}$  with projector ${\cal P} = \sum_{\nu >1}  \rvert u_{\nu} \rangle \langle v_{\nu} \rvert$, where $\langle u_{\nu} \rvert u_1\rangle = 0$ and $\langle v_{\nu} \rvert v_1\rangle = 0$ for $\nu >1$.

We next consider ${\cal O}(q^{-1})$ terms. The non-condensed bosons give a contribution

 \begin{eqnarray}
\Pi_{nn}^{(-1)}(\bm q, i\omega) ={ Z^2 \over 2 c^3 q } \Phi_0({i\omega \over cq})  \Theta,
\end{eqnarray}

\noindent where 

\begin{equation}
\Phi_0(x) = \int_{0}^{\infty}\frac{d^{2}\bm{k}}{(2\pi)^{2}}\frac{\rvert \bm{k}  \rvert+\rvert\bm{k}+\hat{\bm{q}}\rvert }
  {\rvert\bm{k} \rvert \rvert\bm{k}+ \hat{\bm{q}}\rvert \left[\left(\rvert\bm{k} \rvert+\rvert\bm{k} + \hat{\bm{q}}\rvert\right)^{2}-x^{2}\right]},
\end{equation}

\noindent is a dimensionless function and $\hat{\bm{q}}$ is the unit vector along the $\bm{q}$ direction. Because this term has the same matrix structure as the leading order contribution $\Pi_{cn}^{(-2)}(\bm q, i\omega)$, it can be neglected in the long wavelength limit. The combination of  condensed bosons with  non-condensed bosons
with momentum $\bm q$, gives an additional ${\cal O}(q^{-1})$ contribution 

\begin{eqnarray}
\Pi_{cn}^{(-1)}(\bm q, i\omega) ={\phi q \over c^2 q^2 + \omega^2} \left( { i\omega \over 2q  } \Gamma
+ {Z \over 4} \hat{\bm q} \cdot \bm{ M} \right),
\end{eqnarray}

\noindent where

\begin{eqnarray}
\Gamma & = & \text{Tr}\left[ A_p v_{\alpha}(\bm 0, \bm 0) I_p v_{\beta} (\bm 0, \bm 0)  \right]\nonumber \\ 
	&& - \text{Tr}\left[ I_p v_{\alpha}(\bm 0, \bm 0) A_p v_{\beta} (\bm 0, \bm 0)  \right],
\end{eqnarray}

\begin{equation}
A_{p}={1\over 2}
\left(\begin{array}{cccc}
 -1 & 0 & 0 & 0 \\
 0 &  1 & 0 & 0 \\
  0 &  0 & -1 & 0 \\
  0 &  0 & 0 & 1
\end{array}\right),
\end{equation}

\noindent and

\begin{eqnarray}
{\bm M}_{\alpha\beta} &=& -\bm \delta_{\alpha} \left\{ \text{Tr}\left[ I_p v^{(10)}_{\alpha}(\bm 0, \bm 0) I_p v_{\beta} (\bm 0, \bm 0)  \right] \right. -  \nonumber \\
&& \left. \text{Tr}\left[ I_p v_{\alpha}(\bm 0, \bm 0) I_p v^{(01)}_{\beta} (\bm 0, \bm 0)  \right] \right\} \nonumber \\
&& + \bm \delta_{\beta} \left\{ \text{Tr}\left[ I_p v_{\alpha}(\bm 0, \bm 0) I_p v^{(10)}_{\beta} (\bm 0, \bm 0)  \right] \right. -  \nonumber \\
&& \left. \text{Tr}\left[ I_p v^{(01)}_{\alpha}(\bm 0, \bm 0) I_p v_{\beta} (\bm 0, \bm 0)  \right] \right\}.
\end{eqnarray}

\noindent Here $v^{10}_{\alpha}(\bm k, \bm q) = \partial v_{\alpha}(\bm k, \bm q) /\partial ({\bm k \cdot \bm \delta _{\alpha}}) $ and $v^{01}_{\alpha}(\bm k, \bm q) = \partial v_{\alpha}(\bm k, \bm q) /\partial ({\bm k \cdot \bm \delta _{\beta}}) $ refer to the first derivative of the internal vertex.

As explained in the main text,  ${\cal P}^{\dagger} \Pi_{nn}^{(-1)}(\bm q, i\omega) {\cal P}^{\;} = {\cal P}^{\dagger}\Pi_{cn}^{(-1)}(\bm q, i\omega) {\cal P}^{\;}=0$. Thus  the magnon pole arises from ${\cal O}(q^{0})$ contributions to the polarization matrix. The first ${\cal O}(q^{0})$ contribution arises from the second order process in  $\Pi_{cn}^{(-1)}(\bm q, i\omega)$ mentioned in the main text. Here, we provide the explicit form of the remaining ${\cal O}(q^{0})$ contributions. The non-condensed bosons give a contribution

 \begin{eqnarray}
\Pi_{nn}^{(0)}(\bm q, i\omega) &=&  {Z \over 2 c^2} \Phi_1 ({i\omega \over cq}) \Gamma +  {Z^2 \over 4 c^3}  \Phi_0({i\omega \over cq}) \hat{\bm q}\cdot  {\bm M} \nonumber \\
&&+ \Pi_{nn}^{\rm reg}(\bm 0, 0),
\end{eqnarray}

\noindent where 

\begin{equation}
\Phi_1(x) = \int_{0}^{\infty}\frac{d^{2}\bm{k}}{(2\pi)^{2}} \frac{ x }
  {\rvert\bm{k}+ \hat{\bm{q}} \rvert  \left[\left(\rvert\bm{k} \rvert+\rvert\bm{k} + \hat{\bm{q}}\rvert\right)^{2}-x^{2}\right]},
\end{equation}

\noindent is a dimensionless function. The first two terms are projected to zero under the action of  ${\cal P}$, implying that they do not affect the position of the magnon pole in the long wavelength limit. The last term, $\Pi_{nn}^{\rm reg}(\bm 0, 0)$, is a regular integral, which depends on the cutoff $\Lambda$.

The last ${\cal O}(q^{0})$ contribution arises from a combination of condensed bosons with  non-condensed bosons with momentum $\bm q$. After applying the projector ${\cal P}$, we obtain

\begin{eqnarray}
\Pi_{cn}^{(0)}(\bm q, i\omega) \! = \! {\phi q^2/2 \over c^2 q^2 +\omega^2} [ \Sigma  + \!  {i\omega \over q}  \hat{\bm q} \! \cdot \!  {\bm R} \!+\! {Z\over 2} \hat{\bm q} \! \cdot \! {\bm D} \cdot \hat{\bm q} ].
\end{eqnarray}

\noindent where

\begin{eqnarray}
\Sigma_{\alpha\beta} & = & \text{Tr}\left[ g_c v_{\alpha}(\bm 0, \bm 0) B_p v_{\beta} (\bm 0, \bm 0) \right]  \nonumber \\
&=&  \text{Tr}\left[ B_p v_{\alpha}(\bm 0, \bm 0) g_c v_{\beta} (\bm 0, \bm 0) \right],
\end{eqnarray}

\noindent with 

\begin{equation}
B_{p}={1\over 3}
\left(\begin{array}{cccc}
 -\gamma_{\bm Q/2}^B &  \gamma_{\bm Q/2}^A & 0 & 0 \\
 \gamma_{\bm Q/2}^A &  - \gamma_{\bm Q/2}^B & 0 & 0 \\
  0 &  0 & -\gamma_{\bm Q/2}^B & \gamma_{\bm Q/2}^A \\
  0 &  0 & \gamma_{\bm Q/2}^A & -\gamma_{\bm Q/2}^B
\end{array}\right),
\end{equation}

\begin{eqnarray}
{\bm R}_{\alpha\beta} &=& -\bm \delta_{\alpha} \left\{ \text{Tr}\left[ A_p v^{(10)}_{\alpha}(\bm 0, \bm 0) g_c v_{\beta} (\bm 0, \bm 0)  \right] \right. + \nonumber \\
&& \left. \text{Tr}\left[ g_c v^{(01)}_{\alpha}(\bm 0, \bm 0) A_p v_{\beta} (\bm 0, \bm 0)  \right] \right\} \nonumber \\
&& - \bm \delta_{\beta} \left\{ \text{Tr}\left[ A_p v_{\alpha}(\bm 0, \bm 0) g_c v^{(01)}_{\beta} (\bm 0, \bm 0)  \right] \right. +  \nonumber \\
&& \left. \text{Tr}\left[ g_c v_{\alpha}(\bm 0, \bm 0) A_p v^{(10)}_{\beta} (\bm 0, \bm 0)  \right] \right\} ,
\end{eqnarray}
and
\begin{eqnarray}
{\bm D}_{\alpha\beta} &=& \bm \delta_{\alpha} \bm \delta_{\beta} \left\{ \text{Tr}\left[ I_p v^{(10)}_{\alpha}(\bm 0, \bm 0) g_c v^{(01)}_{\beta} (\bm 0, \bm 0)  \right] \right. + \nonumber \\
&& \left. \text{Tr}\left[ g_c v^{(01)}_{\alpha}(\bm 0, \bm 0) I_p v^{(10)}_{\beta} (\bm 0, \bm 0)  \right] \right\}.\\
\nonumber
\end{eqnarray}

\subsubsection{Around \texorpdfstring{$\pm K$}{+-K} point: \texorpdfstring{$\left|\bf q \pm \bf Q \right| < \Lambda$}{|q+-Q|< Lambda}}
The $\pm K$ points are related by inversion symmetry. Around these points,  the singular ${\cal O}(q^{-1})$ and ${\cal O}(1)$ 
contributions to the polarization operator  from the non-condensed bosons and the singular  ${\cal O}(q^{-2})$ and ${\cal O}(q^{-1})$ 
contributions from the condensed bosons combined with non-condensed bosons with momentum $\bm q$ are all equal to zero in the 
long wavelength limit. 
The $\sim {\cal O}(1)$ contribution  from the non-condensed bosons  is a regular integral $\Pi_{nn}^{\rm reg}(\bm Q, 0) $, while the contribution 
from the condensed bosons combined with non-condensed bosons with momentum $\bm q$ is
\begin{equation}
\Pi_{cn}(\bm q + \bm Q,i\omega) \!=\!{\phi q^2/2 \over c^2 q^2 + \omega^2} \left( \Sigma^K  
+  Z \hat{\bm q} \cdot \bm D^K \cdot \hat{\bm q}\right),
\label{polK}
\end{equation}
where
\begin{eqnarray}
\Sigma^K_{\alpha\beta} & = & \text{Tr}\left[ g_c v_{\alpha}(\bm 0, \bm Q) B_1^K v_{\beta} (\bm Q, \bm 0) \right]  \nonumber \\
&=&  \text{Tr}\left[ B_2^K v_{\alpha}(-\bm Q, \bm 0) g_c v_{\beta} (\bm 0, -\bm Q) \right],
\end{eqnarray}
with 
\begin{equation}
B_{1}^K={1\over 3}
\left(\begin{array}{cccc}
 -\gamma_{\bm Q/2}^B &  \gamma_{\bm Q/2}^A & 0 & 0 \\
 \gamma_{\bm Q/2}^A &  - \gamma_{\bm Q/2}^B & 0 & 0 \\
  0 &  0 & 0 & 0 \\
  0 &  0 & 0 & 0
\end{array}\right),
\end{equation}
\begin{equation}
B_{2}^K={1\over 3}
\left(\begin{array}{cccc}
 0 & 0 & 0 & 0 \\
 0 &  0 & 0 & 0 \\
  0 &  0 & -\gamma_{\bm Q/2}^B & \gamma_{\bm Q/2}^A \\
  0 &  0 & \gamma_{\bm Q/2}^A & -\gamma_{\bm Q/2}^B
\end{array}\right),
\end{equation}
and
\begin{eqnarray}
{\bm D}^K_{\alpha\beta} &=& \bm \delta_{\alpha} \bm \delta_{\beta} \text{Tr}\left[ I_2^K v^{(10)}_{\alpha}(-\bm Q, \bm 0) g_c v^{(01)}_{\beta} (\bm 0, -\bm Q)  \right] \nonumber \\
&=& \bm \delta_{\alpha} \bm \delta_{\beta}  \text{Tr}\left[ g_c v^{(01)}_{\alpha}(\bm 0, \bm Q) g_c v^{(10)}_{\beta} (\bm Q, \bm 0)  \right].
\end{eqnarray} 
%
%


The polarization operator around the $-K$ point is obtained by applying the spatial inversion transformation to \eqref{polK}.

\bibliographystyle{apsrev4-1}
\bibliography{ref}

\begin{thebibliography}{87}%
\makeatletter
\providecommand \@ifxundefined [1]{%
 \@ifx{#1\undefined}
}%
\providecommand \@ifnum [1]{%
 \ifnum #1\expandafter \@firstoftwo
 \else \expandafter \@secondoftwo
 \fi
}%
\providecommand \@ifx [1]{%
 \ifx #1\expandafter \@firstoftwo
 \else \expandafter \@secondoftwo
 \fi
}%
\providecommand \natexlab [1]{#1}%
\providecommand \enquote  [1]{``#1''}%
\providecommand \bibnamefont  [1]{#1}%
\providecommand \bibfnamefont [1]{#1}%
\providecommand \citenamefont [1]{#1}%
\providecommand \href@noop [0]{\@secondoftwo}%
\providecommand \href [0]{\begingroup \@sanitize@url \@href}%
\providecommand \@href[1]{\@@startlink{#1}\@@href}%
\providecommand \@@href[1]{\endgroup#1\@@endlink}%
\providecommand \@sanitize@url [0]{\catcode `\\12\catcode `\$12\catcode
  `\&12\catcode `\#12\catcode `\^12\catcode `\_12\catcode `\%12\relax}%
\providecommand \@@startlink[1]{}%
\providecommand \@@endlink[0]{}%
\providecommand \url  [0]{\begingroup\@sanitize@url \@url }%
\providecommand \@url [1]{\endgroup\@href {#1}{\urlprefix }}%
\providecommand \urlprefix  [0]{URL }%
\providecommand \Eprint [0]{\href }%
\providecommand \doibase [0]{http://dx.doi.org/}%
\providecommand \selectlanguage [0]{\@gobble}%
\providecommand \bibinfo  [0]{\@secondoftwo}%
\providecommand \bibfield  [0]{\@secondoftwo}%
\providecommand \translation [1]{[#1]}%
\providecommand \BibitemOpen [0]{}%
\providecommand \bibitemStop [0]{}%
\providecommand \bibitemNoStop [0]{.\EOS\space}%
\providecommand \EOS [0]{\spacefactor3000\relax}%
\providecommand \BibitemShut  [1]{\csname bibitem#1\endcsname}%
\let\auto@bib@innerbib\@empty
\bibitem [{\citenamefont {Tokura}\ \emph {et~al.}(2017)\citenamefont {Tokura},
  \citenamefont {Kawasaki},\ and\ \citenamefont {Nagaosa}}]{Tokura2017}%
  \BibitemOpen
  \bibfield  {author} {\bibinfo {author} {\bibfnamefont {Y.}~\bibnamefont
  {Tokura}}, \bibinfo {author} {\bibfnamefont {M.}~\bibnamefont {Kawasaki}}, \
  and\ \bibinfo {author} {\bibfnamefont {N.}~\bibnamefont {Nagaosa}},\ }\href
  {\doibase 10.1038/nphys4274} {\bibfield  {journal} {\bibinfo  {journal}
  {Nature Physics}\ }\textbf {\bibinfo {volume} {13}},\ \bibinfo {pages} {1056}
  (\bibinfo {year} {2017})}\BibitemShut {NoStop}%
\bibitem [{\citenamefont {Han}\ \emph {et~al.}(2012)\citenamefont {Han},
  \citenamefont {Helton}, \citenamefont {Chu}, \citenamefont {Nocera},
  \citenamefont {Rodriguez-Rivera}, \citenamefont {Broholm},\ and\
  \citenamefont {Lee}}]{Han2012}%
  \BibitemOpen
  \bibfield  {author} {\bibinfo {author} {\bibfnamefont {T.-H.}\ \bibnamefont
  {Han}}, \bibinfo {author} {\bibfnamefont {J.~S.}\ \bibnamefont {Helton}},
  \bibinfo {author} {\bibfnamefont {S.}~\bibnamefont {Chu}}, \bibinfo {author}
  {\bibfnamefont {D.~G.}\ \bibnamefont {Nocera}}, \bibinfo {author}
  {\bibfnamefont {J.~A.}\ \bibnamefont {Rodriguez-Rivera}}, \bibinfo {author}
  {\bibfnamefont {C.}~\bibnamefont {Broholm}}, \ and\ \bibinfo {author}
  {\bibfnamefont {Y.~S.}\ \bibnamefont {Lee}},\ }\href
  {http://dx.doi.org/10.1038/nature11659} {\bibfield  {journal} {\bibinfo
  {journal} {Nature}\ }\textbf {\bibinfo {volume} {492}},\ \bibinfo {pages}
  {406} (\bibinfo {year} {2012})}\BibitemShut {NoStop}%
\bibitem [{\citenamefont {Zhou}\ \emph {et~al.}(2012)\citenamefont {Zhou},
  \citenamefont {Xu}, \citenamefont {Hallas}, \citenamefont {Silverstein},
  \citenamefont {Wiebe}, \citenamefont {Umegaki}, \citenamefont {Yan},
  \citenamefont {Murphy}, \citenamefont {Park}, \citenamefont {Qiu},
  \citenamefont {Copley}, \citenamefont {Gardner},\ and\ \citenamefont
  {Takano}}]{Zhou2012}%
  \BibitemOpen
  \bibfield  {author} {\bibinfo {author} {\bibfnamefont {H.~D.}\ \bibnamefont
  {Zhou}}, \bibinfo {author} {\bibfnamefont {C.}~\bibnamefont {Xu}}, \bibinfo
  {author} {\bibfnamefont {A.~M.}\ \bibnamefont {Hallas}}, \bibinfo {author}
  {\bibfnamefont {H.~J.}\ \bibnamefont {Silverstein}}, \bibinfo {author}
  {\bibfnamefont {C.~R.}\ \bibnamefont {Wiebe}}, \bibinfo {author}
  {\bibfnamefont {I.}~\bibnamefont {Umegaki}}, \bibinfo {author} {\bibfnamefont
  {J.~Q.}\ \bibnamefont {Yan}}, \bibinfo {author} {\bibfnamefont {T.~P.}\
  \bibnamefont {Murphy}}, \bibinfo {author} {\bibfnamefont {J.-H.}\
  \bibnamefont {Park}}, \bibinfo {author} {\bibfnamefont {Y.}~\bibnamefont
  {Qiu}}, \bibinfo {author} {\bibfnamefont {J.~R.~D.}\ \bibnamefont {Copley}},
  \bibinfo {author} {\bibfnamefont {J.~S.}\ \bibnamefont {Gardner}}, \ and\
  \bibinfo {author} {\bibfnamefont {Y.}~\bibnamefont {Takano}},\ }\href
  {\doibase 10.1103/PhysRevLett.109.267206} {\bibfield  {journal} {\bibinfo
  {journal} {Phys. Rev. Lett.}\ }\textbf {\bibinfo {volume} {109}},\ \bibinfo
  {pages} {267206} (\bibinfo {year} {2012})}\BibitemShut {NoStop}%
\bibitem [{\citenamefont {Banerjee}\ \emph {et~al.}(2016)\citenamefont
  {Banerjee}, \citenamefont {Bridges}, \citenamefont {Yan}, \citenamefont
  {Aczel}, \citenamefont {Li}, \citenamefont {Stone}, \citenamefont {Granroth},
  \citenamefont {Lumsden}, \citenamefont {Yiu}, \citenamefont {Knolle},
  \citenamefont {Bhattacharjee}, \citenamefont {Kovrizhin}, \citenamefont
  {Moessner}, \citenamefont {Tennant}, \citenamefont {Mandrus},\ and\
  \citenamefont {Nagler}}]{Banerjee2016a}%
  \BibitemOpen
  \bibfield  {author} {\bibinfo {author} {\bibfnamefont {A.}~\bibnamefont
  {Banerjee}}, \bibinfo {author} {\bibfnamefont {C.~A.}\ \bibnamefont
  {Bridges}}, \bibinfo {author} {\bibfnamefont {J.~Q.}\ \bibnamefont {Yan}},
  \bibinfo {author} {\bibfnamefont {A.~A.}\ \bibnamefont {Aczel}}, \bibinfo
  {author} {\bibfnamefont {L.}~\bibnamefont {Li}}, \bibinfo {author}
  {\bibfnamefont {M.~B.}\ \bibnamefont {Stone}}, \bibinfo {author}
  {\bibfnamefont {G.~E.}\ \bibnamefont {Granroth}}, \bibinfo {author}
  {\bibfnamefont {M.~D.}\ \bibnamefont {Lumsden}}, \bibinfo {author}
  {\bibfnamefont {Y.}~\bibnamefont {Yiu}}, \bibinfo {author} {\bibfnamefont
  {J.}~\bibnamefont {Knolle}}, \bibinfo {author} {\bibfnamefont
  {S.}~\bibnamefont {Bhattacharjee}}, \bibinfo {author} {\bibfnamefont {D.~L.}\
  \bibnamefont {Kovrizhin}}, \bibinfo {author} {\bibfnamefont {R.}~\bibnamefont
  {Moessner}}, \bibinfo {author} {\bibfnamefont {D.~A.}\ \bibnamefont
  {Tennant}}, \bibinfo {author} {\bibfnamefont {D.~G.}\ \bibnamefont
  {Mandrus}}, \ and\ \bibinfo {author} {\bibfnamefont {S.~E.}\ \bibnamefont
  {Nagler}},\ }\href {http://dx.doi.org/10.1038/nmat4604} {\bibfield  {journal}
  {\bibinfo  {journal} {Nat. Mater.}\ }\textbf {\bibinfo {volume} {15}},\
  \bibinfo {pages} {733} (\bibinfo {year} {2016})}\BibitemShut {NoStop}%
\bibitem [{\citenamefont {Ma}\ \emph {et~al.}(2016)\citenamefont {Ma},
  \citenamefont {Kamiya}, \citenamefont {Hong}, \citenamefont {Cao},
  \citenamefont {Ehlers}, \citenamefont {Tian}, \citenamefont {Batista},
  \citenamefont {Dun}, \citenamefont {Zhou},\ and\ \citenamefont
  {Matsuda}}]{Ma2016}%
  \BibitemOpen
  \bibfield  {author} {\bibinfo {author} {\bibfnamefont {J.}~\bibnamefont
  {Ma}}, \bibinfo {author} {\bibfnamefont {Y.}~\bibnamefont {Kamiya}}, \bibinfo
  {author} {\bibfnamefont {T.}~\bibnamefont {Hong}}, \bibinfo {author}
  {\bibfnamefont {H.~B.}\ \bibnamefont {Cao}}, \bibinfo {author} {\bibfnamefont
  {G.}~\bibnamefont {Ehlers}}, \bibinfo {author} {\bibfnamefont
  {W.}~\bibnamefont {Tian}}, \bibinfo {author} {\bibfnamefont {C.~D.}\
  \bibnamefont {Batista}}, \bibinfo {author} {\bibfnamefont {Z.~L.}\
  \bibnamefont {Dun}}, \bibinfo {author} {\bibfnamefont {H.~D.}\ \bibnamefont
  {Zhou}}, \ and\ \bibinfo {author} {\bibfnamefont {M.}~\bibnamefont
  {Matsuda}},\ }\href {\doibase 10.1103/PhysRevLett.116.087201} {\bibfield
  {journal} {\bibinfo  {journal} {Phys. Rev. Lett.}\ }\textbf {\bibinfo
  {volume} {116}},\ \bibinfo {pages} {087201} (\bibinfo {year}
  {2016})}\BibitemShut {NoStop}%
\bibitem [{\citenamefont {Paddison}\ \emph {et~al.}(2017)\citenamefont
  {Paddison}, \citenamefont {Daum}, \citenamefont {Dun}, \citenamefont
  {Ehlers}, \citenamefont {Liu}, \citenamefont {Stone}, \citenamefont {Zhou},\
  and\ \citenamefont {Mourigal}}]{Paddison2017}%
  \BibitemOpen
  \bibfield  {author} {\bibinfo {author} {\bibfnamefont {J.~A.~M.}\
  \bibnamefont {Paddison}}, \bibinfo {author} {\bibfnamefont {M.}~\bibnamefont
  {Daum}}, \bibinfo {author} {\bibfnamefont {Z.}~\bibnamefont {Dun}}, \bibinfo
  {author} {\bibfnamefont {G.}~\bibnamefont {Ehlers}}, \bibinfo {author}
  {\bibfnamefont {Y.}~\bibnamefont {Liu}}, \bibinfo {author} {\bibfnamefont
  {M.~B.}\ \bibnamefont {Stone}}, \bibinfo {author} {\bibfnamefont
  {H.}~\bibnamefont {Zhou}}, \ and\ \bibinfo {author} {\bibfnamefont
  {M.}~\bibnamefont {Mourigal}},\ }\href {\doibase 10.1038/nphys3971}
  {\bibfield  {journal} {\bibinfo  {journal} {Nature Physics}\ }\textbf
  {\bibinfo {volume} {13}},\ \bibinfo {pages} {117} (\bibinfo {year}
  {2017})}\BibitemShut {NoStop}%
\bibitem [{\citenamefont {Ito}\ \emph {et~al.}(2017)\citenamefont {Ito},
  \citenamefont {Kurita}, \citenamefont {Tanaka}, \citenamefont
  {Ohira-Kawamura}, \citenamefont {Nakajima}, \citenamefont {Itoh},
  \citenamefont {Kuwahara},\ and\ \citenamefont {Kakurai}}]{Ito2017}%
  \BibitemOpen
  \bibfield  {author} {\bibinfo {author} {\bibfnamefont {S.}~\bibnamefont
  {Ito}}, \bibinfo {author} {\bibfnamefont {N.}~\bibnamefont {Kurita}},
  \bibinfo {author} {\bibfnamefont {H.}~\bibnamefont {Tanaka}}, \bibinfo
  {author} {\bibfnamefont {S.}~\bibnamefont {Ohira-Kawamura}}, \bibinfo
  {author} {\bibfnamefont {K.}~\bibnamefont {Nakajima}}, \bibinfo {author}
  {\bibfnamefont {S.}~\bibnamefont {Itoh}}, \bibinfo {author} {\bibfnamefont
  {K.}~\bibnamefont {Kuwahara}}, \ and\ \bibinfo {author} {\bibfnamefont
  {K.}~\bibnamefont {Kakurai}},\ }\href {\doibase 10.1038/s41467-017-00316-x}
  {\bibfield  {journal} {\bibinfo  {journal} {Nat. Comm.}\ }\textbf {\bibinfo
  {volume} {8}},\ \bibinfo {pages} {235} (\bibinfo {year} {2017})}\BibitemShut
  {NoStop}%
\bibitem [{\citenamefont {Landau}(1937)}]{Landau1937}%
  \BibitemOpen
  \bibfield  {author} {\bibinfo {author} {\bibfnamefont {L.~D.}\ \bibnamefont
  {Landau}},\ }\href@noop {} {\bibfield  {journal} {\bibinfo  {journal} {Zh.
  Eksp. Theor. Fiz.}\ }\textbf {\bibinfo {volume} {7}},\ \bibinfo {pages} {19}
  (\bibinfo {year} {1937})}\BibitemShut {NoStop}%
\bibitem [{\citenamefont {Anderson}(1997)}]{Anderson1997}%
  \BibitemOpen
  \bibfield  {author} {\bibinfo {author} {\bibfnamefont {P.~W.}\ \bibnamefont
  {Anderson}},\ }\href@noop {} {\emph {\bibinfo {title} {Concepts in Solids}}}\
  (\bibinfo  {publisher} {World Scientific},\ \bibinfo {address} {Singapore},\
  \bibinfo {year} {1997})\BibitemShut {NoStop}%
\bibitem [{\citenamefont {Anderson}(1973)}]{Anderson1973}%
  \BibitemOpen
  \bibfield  {author} {\bibinfo {author} {\bibfnamefont {P.}~\bibnamefont
  {Anderson}},\ }\href {\doibase
  http://dx.doi.org/10.1016/0025-5408(73)90167-0} {\bibfield  {journal}
  {\bibinfo  {journal} {Mater. Res. Bull.}\ }\textbf {\bibinfo {volume} {8}},\
  \bibinfo {pages} {153 } (\bibinfo {year} {1973})}\BibitemShut {NoStop}%
\bibitem [{\citenamefont {ed.~by Lacroix~C.}(2010)}]{Lacroix2010}%
  \BibitemOpen
  \bibfield  {author} {\bibinfo {author} {\bibnamefont {ed.~by Lacroix~C.}},\
  }\href@noop {} {\emph {\bibinfo {title} {Introduction to Frustrated
  Magnetism}}}\ (\bibinfo  {publisher} {Springer},\ \bibinfo {address}
  {Heidelberg},\ \bibinfo {year} {2010})\BibitemShut {NoStop}%
\bibitem [{\citenamefont {Wen}(2002)}]{Wen2002}%
  \BibitemOpen
  \bibfield  {author} {\bibinfo {author} {\bibfnamefont {X.-G.}\ \bibnamefont
  {Wen}},\ }\href {\doibase 10.1103/PhysRevB.65.165113} {\bibfield  {journal}
  {\bibinfo  {journal} {Phys. Rev. B}\ }\textbf {\bibinfo {volume} {65}},\
  \bibinfo {pages} {165113} (\bibinfo {year} {2002})}\BibitemShut {NoStop}%
\bibitem [{\citenamefont {Sachdev}(2008)}]{Sachdev2008}%
  \BibitemOpen
  \bibfield  {author} {\bibinfo {author} {\bibfnamefont {S.}~\bibnamefont
  {Sachdev}},\ }\href {\doibase 10.1038/nphys894} {\bibfield  {journal}
  {\bibinfo  {journal} {Nature Physics}\ }\textbf {\bibinfo {volume} {4}},\
  \bibinfo {pages} {173} (\bibinfo {year} {2008})}\BibitemShut {NoStop}%
\bibitem [{\citenamefont {Normand}(2009)}]{Normand2009}%
  \BibitemOpen
  \bibfield  {author} {\bibinfo {author} {\bibfnamefont {B.}~\bibnamefont
  {Normand}},\ }\href {\doibase 10.1080/00107510902850361} {\bibfield
  {journal} {\bibinfo  {journal} {Contemporary Physics}\ }\textbf {\bibinfo
  {volume} {50}},\ \bibinfo {pages} {533} (\bibinfo {year} {2009})}\BibitemShut
  {NoStop}%
\bibitem [{\citenamefont {Balents}(2010)}]{Balents2010}%
  \BibitemOpen
  \bibfield  {author} {\bibinfo {author} {\bibfnamefont {L.}~\bibnamefont
  {Balents}},\ }\href {http://dx.doi.org/10.1038/nature08917} {\bibfield
  {journal} {\bibinfo  {journal} {Nature}\ }\textbf {\bibinfo {volume} {464}},\
  \bibinfo {pages} {199} (\bibinfo {year} {2010})}\BibitemShut {NoStop}%
\bibitem [{\citenamefont {Savary}\ and\ \citenamefont
  {Balents}(2017)}]{Savary2017}%
  \BibitemOpen
  \bibfield  {author} {\bibinfo {author} {\bibfnamefont {L.}~\bibnamefont
  {Savary}}\ and\ \bibinfo {author} {\bibfnamefont {L.}~\bibnamefont
  {Balents}},\ }\href {http://stacks.iop.org/0034-4885/80/i=1/a=016502}
  {\bibfield  {journal} {\bibinfo  {journal} {Rep. Prog. Phys.}\ }\textbf
  {\bibinfo {volume} {80}},\ \bibinfo {pages} {016502} (\bibinfo {year}
  {2017})}\BibitemShut {NoStop}%
\bibitem [{\citenamefont {Zhou}\ \emph {et~al.}(2017)\citenamefont {Zhou},
  \citenamefont {Kanoda},\ and\ \citenamefont {Ng}}]{Zhou2017}%
  \BibitemOpen
  \bibfield  {author} {\bibinfo {author} {\bibfnamefont {Y.}~\bibnamefont
  {Zhou}}, \bibinfo {author} {\bibfnamefont {K.}~\bibnamefont {Kanoda}}, \ and\
  \bibinfo {author} {\bibfnamefont {T.-K.}\ \bibnamefont {Ng}},\ }\href
  {\doibase 10.1103/RevModPhys.89.025003} {\bibfield  {journal} {\bibinfo
  {journal} {Rev. Mod. Phys.}\ }\textbf {\bibinfo {volume} {89}},\ \bibinfo
  {pages} {025003} (\bibinfo {year} {2017})}\BibitemShut {NoStop}%
\bibitem [{\citenamefont {Zheng}\ \emph {et~al.}(2006)\citenamefont {Zheng},
  \citenamefont {Fj\ae{}restad}, \citenamefont {Singh}, \citenamefont
  {McKenzie},\ and\ \citenamefont {Coldea}}]{Zheng2006}%
  \BibitemOpen
  \bibfield  {author} {\bibinfo {author} {\bibfnamefont {W.}~\bibnamefont
  {Zheng}}, \bibinfo {author} {\bibfnamefont {J.~O.}\ \bibnamefont
  {Fj\ae{}restad}}, \bibinfo {author} {\bibfnamefont {R.~R.~P.}\ \bibnamefont
  {Singh}}, \bibinfo {author} {\bibfnamefont {R.~H.}\ \bibnamefont {McKenzie}},
  \ and\ \bibinfo {author} {\bibfnamefont {R.}~\bibnamefont {Coldea}},\ }\href
  {\doibase 10.1103/PhysRevB.74.224420} {\bibfield  {journal} {\bibinfo
  {journal} {Phys. Rev. B}\ }\textbf {\bibinfo {volume} {74}},\ \bibinfo
  {pages} {224420} (\bibinfo {year} {2006})}\BibitemShut {NoStop}%
\bibitem [{\citenamefont {Huse}\ and\ \citenamefont {Elser}(1988)}]{Huse1988}%
  \BibitemOpen
  \bibfield  {author} {\bibinfo {author} {\bibfnamefont {D.~A.}\ \bibnamefont
  {Huse}}\ and\ \bibinfo {author} {\bibfnamefont {V.}~\bibnamefont {Elser}},\
  }\href {\doibase 10.1103/PhysRevLett.60.2531} {\bibfield  {journal} {\bibinfo
   {journal} {Phys. Rev. Lett.}\ }\textbf {\bibinfo {volume} {60}},\ \bibinfo
  {pages} {2531} (\bibinfo {year} {1988})}\BibitemShut {NoStop}%
\bibitem [{\citenamefont {Bernu}\ \emph {et~al.}(1992)\citenamefont {Bernu},
  \citenamefont {Lhuillier},\ and\ \citenamefont {Pierre}}]{Bernu1992}%
  \BibitemOpen
  \bibfield  {author} {\bibinfo {author} {\bibfnamefont {B.}~\bibnamefont
  {Bernu}}, \bibinfo {author} {\bibfnamefont {C.}~\bibnamefont {Lhuillier}}, \
  and\ \bibinfo {author} {\bibfnamefont {L.}~\bibnamefont {Pierre}},\ }\href
  {\doibase 10.1103/PhysRevLett.69.2590} {\bibfield  {journal} {\bibinfo
  {journal} {Phys. Rev. Lett.}\ }\textbf {\bibinfo {volume} {69}},\ \bibinfo
  {pages} {2590} (\bibinfo {year} {1992})}\BibitemShut {NoStop}%
\bibitem [{\citenamefont {Singh}\ and\ \citenamefont {Huse}(1992)}]{Singh1992}%
  \BibitemOpen
  \bibfield  {author} {\bibinfo {author} {\bibfnamefont {R.~R.~P.}\
  \bibnamefont {Singh}}\ and\ \bibinfo {author} {\bibfnamefont {D.~A.}\
  \bibnamefont {Huse}},\ }\href {\doibase 10.1103/PhysRevLett.68.1766}
  {\bibfield  {journal} {\bibinfo  {journal} {Phys. Rev. Lett.}\ }\textbf
  {\bibinfo {volume} {68}},\ \bibinfo {pages} {1766} (\bibinfo {year}
  {1992})}\BibitemShut {NoStop}%
\bibitem [{\citenamefont {Capriotti}\ \emph {et~al.}(1999)\citenamefont
  {Capriotti}, \citenamefont {Trumper},\ and\ \citenamefont
  {Sorella}}]{Capriotti1999}%
  \BibitemOpen
  \bibfield  {author} {\bibinfo {author} {\bibfnamefont {L.}~\bibnamefont
  {Capriotti}}, \bibinfo {author} {\bibfnamefont {A.~E.}\ \bibnamefont
  {Trumper}}, \ and\ \bibinfo {author} {\bibfnamefont {S.}~\bibnamefont
  {Sorella}},\ }\href {\doibase 10.1103/PhysRevLett.82.3899} {\bibfield
  {journal} {\bibinfo  {journal} {Phys. Rev. Lett.}\ }\textbf {\bibinfo
  {volume} {82}},\ \bibinfo {pages} {3899} (\bibinfo {year}
  {1999})}\BibitemShut {NoStop}%
\bibitem [{\citenamefont {White}\ and\ \citenamefont
  {Chernyshev}(2007)}]{White2007}%
  \BibitemOpen
  \bibfield  {author} {\bibinfo {author} {\bibfnamefont {S.~R.}\ \bibnamefont
  {White}}\ and\ \bibinfo {author} {\bibfnamefont {A.~L.}\ \bibnamefont
  {Chernyshev}},\ }\href {\doibase 10.1103/PhysRevLett.99.127004} {\bibfield
  {journal} {\bibinfo  {journal} {Phys. Rev. Lett.}\ }\textbf {\bibinfo
  {volume} {99}},\ \bibinfo {pages} {127004} (\bibinfo {year}
  {2007})}\BibitemShut {NoStop}%
\bibitem [{\citenamefont {Starykh}\ \emph {et~al.}(2006)\citenamefont
  {Starykh}, \citenamefont {Chubukov},\ and\ \citenamefont
  {Abanov}}]{Starykh06}%
  \BibitemOpen
  \bibfield  {author} {\bibinfo {author} {\bibfnamefont {O.~A.}\ \bibnamefont
  {Starykh}}, \bibinfo {author} {\bibfnamefont {A.~V.}\ \bibnamefont
  {Chubukov}}, \ and\ \bibinfo {author} {\bibfnamefont {A.~G.}\ \bibnamefont
  {Abanov}},\ }\href {\doibase 10.1103/PhysRevB.74.180403} {\bibfield
  {journal} {\bibinfo  {journal} {Phys. Rev. B}\ }\textbf {\bibinfo {volume}
  {74}},\ \bibinfo {pages} {180403} (\bibinfo {year} {2006})}\BibitemShut
  {NoStop}%
\bibitem [{\citenamefont {Chernyshev}\ and\ \citenamefont
  {Zhitomirsky}(2006)}]{Chernyshev2006}%
  \BibitemOpen
  \bibfield  {author} {\bibinfo {author} {\bibfnamefont {A.~L.}\ \bibnamefont
  {Chernyshev}}\ and\ \bibinfo {author} {\bibfnamefont {M.~E.}\ \bibnamefont
  {Zhitomirsky}},\ }\href {\doibase 10.1103/PhysRevLett.97.207202} {\bibfield
  {journal} {\bibinfo  {journal} {Phys. Rev. Lett.}\ }\textbf {\bibinfo
  {volume} {97}},\ \bibinfo {pages} {207202} (\bibinfo {year}
  {2006})}\BibitemShut {NoStop}%
\bibitem [{\citenamefont {Chernyshev}\ and\ \citenamefont
  {Zhitomirsky}(2009)}]{Chernyshev2009}%
  \BibitemOpen
  \bibfield  {author} {\bibinfo {author} {\bibfnamefont {A.~L.}\ \bibnamefont
  {Chernyshev}}\ and\ \bibinfo {author} {\bibfnamefont {M.~E.}\ \bibnamefont
  {Zhitomirsky}},\ }\href {\doibase 10.1103/PhysRevB.79.144416} {\bibfield
  {journal} {\bibinfo  {journal} {Phys. Rev. B}\ }\textbf {\bibinfo {volume}
  {79}},\ \bibinfo {pages} {144416} (\bibinfo {year} {2009})}\BibitemShut
  {NoStop}%
\bibitem [{\citenamefont {Zhitomirsky}\ and\ \citenamefont
  {Chernyshev}(2013)}]{Zhitomirsky2013}%
  \BibitemOpen
  \bibfield  {author} {\bibinfo {author} {\bibfnamefont {M.~E.}\ \bibnamefont
  {Zhitomirsky}}\ and\ \bibinfo {author} {\bibfnamefont {A.~L.}\ \bibnamefont
  {Chernyshev}},\ }\href {\doibase 10.1103/RevModPhys.85.219} {\bibfield
  {journal} {\bibinfo  {journal} {Rev. Mod. Phys.}\ }\textbf {\bibinfo {volume}
  {85}},\ \bibinfo {pages} {219} (\bibinfo {year} {2013})}\BibitemShut
  {NoStop}%
\bibitem [{\citenamefont {Read}\ and\ \citenamefont
  {Sachdev}(1991)}]{Read1991}%
  \BibitemOpen
  \bibfield  {author} {\bibinfo {author} {\bibfnamefont {N.}~\bibnamefont
  {Read}}\ and\ \bibinfo {author} {\bibfnamefont {S.}~\bibnamefont {Sachdev}},\
  }\href {\doibase 10.1103/PhysRevLett.66.1773} {\bibfield  {journal} {\bibinfo
   {journal} {Phys. Rev. Lett.}\ }\textbf {\bibinfo {volume} {66}},\ \bibinfo
  {pages} {1773} (\bibinfo {year} {1991})}\BibitemShut {NoStop}%
\bibitem [{\citenamefont {Sachdev}\ and\ \citenamefont
  {Read}(1991)}]{Sachdev1991}%
  \BibitemOpen
  \bibfield  {author} {\bibinfo {author} {\bibfnamefont {S.}~\bibnamefont
  {Sachdev}}\ and\ \bibinfo {author} {\bibfnamefont {N.}~\bibnamefont {Read}},\
  }\href {\doibase 10.1142/S0217979291000158} {\bibfield  {journal} {\bibinfo
  {journal} {International Journal of Modern Physics B}\ }\textbf {\bibinfo
  {volume} {05}},\ \bibinfo {pages} {219} (\bibinfo {year} {1991})}\BibitemShut
  {NoStop}%
\bibitem [{\citenamefont {Chubukov}\ \emph {et~al.}(1994)\citenamefont
  {Chubukov}, \citenamefont {Sachdev},\ and\ \citenamefont
  {Senthil}}]{Chubukov1994}%
  \BibitemOpen
  \bibfield  {author} {\bibinfo {author} {\bibfnamefont {A.~V.}\ \bibnamefont
  {Chubukov}}, \bibinfo {author} {\bibfnamefont {S.}~\bibnamefont {Sachdev}}, \
  and\ \bibinfo {author} {\bibfnamefont {T.}~\bibnamefont {Senthil}},\ }\href
  {\doibase 10.1016/0550-3213(94)90023-X} {\bibfield  {journal} {\bibinfo
  {journal} {Nuclear Physics B}\ }\textbf {\bibinfo {volume} {426}},\ \bibinfo
  {pages} {601} (\bibinfo {year} {1994})}\BibitemShut {NoStop}%
\bibitem [{\citenamefont {Manuel}\ and\ \citenamefont
  {Ceccatto}(1999)}]{Manuel1999}%
  \BibitemOpen
  \bibfield  {author} {\bibinfo {author} {\bibfnamefont {L.~O.}\ \bibnamefont
  {Manuel}}\ and\ \bibinfo {author} {\bibfnamefont {H.~A.}\ \bibnamefont
  {Ceccatto}},\ }\href {\doibase 10.1103/PhysRevB.60.9489} {\bibfield
  {journal} {\bibinfo  {journal} {Phys. Rev. B}\ }\textbf {\bibinfo {volume}
  {60}},\ \bibinfo {pages} {9489} (\bibinfo {year} {1999})}\BibitemShut
  {NoStop}%
\bibitem [{\citenamefont {Mishmash}\ \emph {et~al.}(2013)\citenamefont
  {Mishmash}, \citenamefont {Garrison}, \citenamefont {Bieri},\ and\
  \citenamefont {Xu}}]{Mishmash13}%
  \BibitemOpen
  \bibfield  {author} {\bibinfo {author} {\bibfnamefont {R.~V.}\ \bibnamefont
  {Mishmash}}, \bibinfo {author} {\bibfnamefont {J.~R.}\ \bibnamefont
  {Garrison}}, \bibinfo {author} {\bibfnamefont {S.}~\bibnamefont {Bieri}}, \
  and\ \bibinfo {author} {\bibfnamefont {C.}~\bibnamefont {Xu}},\ }\href
  {\doibase 10.1103/PhysRevLett.111.157203} {\bibfield  {journal} {\bibinfo
  {journal} {Phys. Rev. Lett.}\ }\textbf {\bibinfo {volume} {111}},\ \bibinfo
  {pages} {157203} (\bibinfo {year} {2013})}\BibitemShut {NoStop}%
\bibitem [{\citenamefont {Kaneko}\ \emph {et~al.}(2014)\citenamefont {Kaneko},
  \citenamefont {Morita},\ and\ \citenamefont {Imada}}]{Kaneko14}%
  \BibitemOpen
  \bibfield  {author} {\bibinfo {author} {\bibfnamefont {R.}~\bibnamefont
  {Kaneko}}, \bibinfo {author} {\bibfnamefont {S.}~\bibnamefont {Morita}}, \
  and\ \bibinfo {author} {\bibfnamefont {M.}~\bibnamefont {Imada}},\ }\href
  {\doibase 10.7566/JPSJ.83.093707} {\bibfield  {journal} {\bibinfo  {journal}
  {Journal of the Physical Society of Japan}\ }\textbf {\bibinfo {volume}
  {83}},\ \bibinfo {pages} {093707} (\bibinfo {year} {2014})},\ \Eprint
  {http://arxiv.org/abs/https://doi.org/10.7566/JPSJ.83.093707}
  {https://doi.org/10.7566/JPSJ.83.093707} \BibitemShut {NoStop}%
\bibitem [{\citenamefont {Li}\ \emph {et~al.}(2015)\citenamefont {Li},
  \citenamefont {Bishop},\ and\ \citenamefont {Campbell}}]{Li15}%
  \BibitemOpen
  \bibfield  {author} {\bibinfo {author} {\bibfnamefont {P.~H.~Y.}\
  \bibnamefont {Li}}, \bibinfo {author} {\bibfnamefont {R.~F.}\ \bibnamefont
  {Bishop}}, \ and\ \bibinfo {author} {\bibfnamefont {C.~E.}\ \bibnamefont
  {Campbell}},\ }\href {\doibase 10.1103/PhysRevB.91.014426} {\bibfield
  {journal} {\bibinfo  {journal} {Phys. Rev. B}\ }\textbf {\bibinfo {volume}
  {91}},\ \bibinfo {pages} {014426} (\bibinfo {year} {2015})}\BibitemShut
  {NoStop}%
\bibitem [{\citenamefont {Zhu}\ and\ \citenamefont {White}(2015)}]{Zhu15}%
  \BibitemOpen
  \bibfield  {author} {\bibinfo {author} {\bibfnamefont {Z.}~\bibnamefont
  {Zhu}}\ and\ \bibinfo {author} {\bibfnamefont {S.~R.}\ \bibnamefont
  {White}},\ }\href {\doibase 10.1103/PhysRevB.92.041105} {\bibfield  {journal}
  {\bibinfo  {journal} {Phys. Rev. B}\ }\textbf {\bibinfo {volume} {92}},\
  \bibinfo {pages} {041105} (\bibinfo {year} {2015})}\BibitemShut {NoStop}%
\bibitem [{\citenamefont {Hu}\ \emph {et~al.}(2015)\citenamefont {Hu},
  \citenamefont {Gong}, \citenamefont {Zhu},\ and\ \citenamefont
  {Sheng}}]{Hu15}%
  \BibitemOpen
  \bibfield  {author} {\bibinfo {author} {\bibfnamefont {W.-J.}\ \bibnamefont
  {Hu}}, \bibinfo {author} {\bibfnamefont {S.-S.}\ \bibnamefont {Gong}},
  \bibinfo {author} {\bibfnamefont {W.}~\bibnamefont {Zhu}}, \ and\ \bibinfo
  {author} {\bibfnamefont {D.~N.}\ \bibnamefont {Sheng}},\ }\href {\doibase
  10.1103/PhysRevB.92.140403} {\bibfield  {journal} {\bibinfo  {journal} {Phys.
  Rev. B}\ }\textbf {\bibinfo {volume} {92}},\ \bibinfo {pages} {140403}
  (\bibinfo {year} {2015})}\BibitemShut {NoStop}%
\bibitem [{\citenamefont {Iqbal}\ \emph {et~al.}(2016)\citenamefont {Iqbal},
  \citenamefont {Hu}, \citenamefont {Thomale}, \citenamefont {Poilblanc},\ and\
  \citenamefont {Becca}}]{Iqbal16}%
  \BibitemOpen
  \bibfield  {author} {\bibinfo {author} {\bibfnamefont {Y.}~\bibnamefont
  {Iqbal}}, \bibinfo {author} {\bibfnamefont {W.-J.}\ \bibnamefont {Hu}},
  \bibinfo {author} {\bibfnamefont {R.}~\bibnamefont {Thomale}}, \bibinfo
  {author} {\bibfnamefont {D.}~\bibnamefont {Poilblanc}}, \ and\ \bibinfo
  {author} {\bibfnamefont {F.}~\bibnamefont {Becca}},\ }\href {\doibase
  10.1103/PhysRevB.93.144411} {\bibfield  {journal} {\bibinfo  {journal} {Phys.
  Rev. B}\ }\textbf {\bibinfo {volume} {93}},\ \bibinfo {pages} {144411}
  (\bibinfo {year} {2016})}\BibitemShut {NoStop}%
\bibitem [{\citenamefont {Saadatmand}\ and\ \citenamefont
  {McCulloch}(2016)}]{Saadatmand16}%
  \BibitemOpen
  \bibfield  {author} {\bibinfo {author} {\bibfnamefont {S.~N.}\ \bibnamefont
  {Saadatmand}}\ and\ \bibinfo {author} {\bibfnamefont {I.~P.}\ \bibnamefont
  {McCulloch}},\ }\href {\doibase 10.1103/PhysRevB.94.121111} {\bibfield
  {journal} {\bibinfo  {journal} {Phys. Rev. B}\ }\textbf {\bibinfo {volume}
  {94}},\ \bibinfo {pages} {121111} (\bibinfo {year} {2016})}\BibitemShut
  {NoStop}%
\bibitem [{\citenamefont {Bauer}\ and\ \citenamefont
  {Fj\ae{}restad}(2017)}]{Bauer17}%
  \BibitemOpen
  \bibfield  {author} {\bibinfo {author} {\bibfnamefont {D.-V.}\ \bibnamefont
  {Bauer}}\ and\ \bibinfo {author} {\bibfnamefont {J.~O.}\ \bibnamefont
  {Fj\ae{}restad}},\ }\href {\doibase 10.1103/PhysRevB.96.165141} {\bibfield
  {journal} {\bibinfo  {journal} {Phys. Rev. B}\ }\textbf {\bibinfo {volume}
  {96}},\ \bibinfo {pages} {165141} (\bibinfo {year} {2017})}\BibitemShut
  {NoStop}%
\bibitem [{\citenamefont {Gong}\ \emph {et~al.}(2017)\citenamefont {Gong},
  \citenamefont {Zhu}, \citenamefont {Zhu}, \citenamefont {Sheng},\ and\
  \citenamefont {Yang}}]{Gong17}%
  \BibitemOpen
  \bibfield  {author} {\bibinfo {author} {\bibfnamefont {S.-S.}\ \bibnamefont
  {Gong}}, \bibinfo {author} {\bibfnamefont {W.}~\bibnamefont {Zhu}}, \bibinfo
  {author} {\bibfnamefont {J.-X.}\ \bibnamefont {Zhu}}, \bibinfo {author}
  {\bibfnamefont {D.~N.}\ \bibnamefont {Sheng}}, \ and\ \bibinfo {author}
  {\bibfnamefont {K.}~\bibnamefont {Yang}},\ }\href {\doibase
  10.1103/PhysRevB.96.075116} {\bibfield  {journal} {\bibinfo  {journal} {Phys.
  Rev. B}\ }\textbf {\bibinfo {volume} {96}},\ \bibinfo {pages} {075116}
  (\bibinfo {year} {2017})}\BibitemShut {NoStop}%
\bibitem [{\citenamefont {Zhu}\ \emph {et~al.}(2018)\citenamefont {Zhu},
  \citenamefont {Maksimov}, \citenamefont {White},\ and\ \citenamefont
  {Chernyshev}}]{Zhu18}%
  \BibitemOpen
  \bibfield  {author} {\bibinfo {author} {\bibfnamefont {Z.}~\bibnamefont
  {Zhu}}, \bibinfo {author} {\bibfnamefont {P.~A.}\ \bibnamefont {Maksimov}},
  \bibinfo {author} {\bibfnamefont {S.~R.}\ \bibnamefont {White}}, \ and\
  \bibinfo {author} {\bibfnamefont {A.~L.}\ \bibnamefont {Chernyshev}},\ }\href
  {\doibase 10.1103/PhysRevLett.120.207203} {\bibfield  {journal} {\bibinfo
  {journal} {Phys. Rev. Lett.}\ }\textbf {\bibinfo {volume} {120}},\ \bibinfo
  {pages} {207203} (\bibinfo {year} {2018})}\BibitemShut {NoStop}%
\bibitem [{\citenamefont {Coldea}\ \emph {et~al.}(2003)\citenamefont {Coldea},
  \citenamefont {Tennant},\ and\ \citenamefont {Tylczynski}}]{Coldea2003}%
  \BibitemOpen
  \bibfield  {author} {\bibinfo {author} {\bibfnamefont {R.}~\bibnamefont
  {Coldea}}, \bibinfo {author} {\bibfnamefont {D.~A.}\ \bibnamefont {Tennant}},
  \ and\ \bibinfo {author} {\bibfnamefont {Z.}~\bibnamefont {Tylczynski}},\
  }\href {\doibase 10.1103/PhysRevB.68.134424} {\bibfield  {journal} {\bibinfo
  {journal} {Phys. Rev. B}\ }\textbf {\bibinfo {volume} {68}},\ \bibinfo
  {pages} {134424} (\bibinfo {year} {2003})}\BibitemShut {NoStop}%
\bibitem [{\citenamefont {Gell-Mann}(1995)}]{Gell-Mann95}%
  \BibitemOpen
  \bibfield  {author} {\bibinfo {author} {\bibfnamefont {M.}~\bibnamefont
  {Gell-Mann}},\ }\href@noop {} {\emph {\bibinfo {title} {The Quark and the
  Jaguar}}}\ (\bibinfo  {publisher} {Springer-Verlag},\ \bibinfo {address} {New
  York},\ \bibinfo {year} {1995})\BibitemShut {NoStop}%
\bibitem [{\citenamefont {Greiner}\ and\ \citenamefont
  {Schafer}(1994)}]{Greiner94}%
  \BibitemOpen
  \bibfield  {author} {\bibinfo {author} {\bibfnamefont {W.}~\bibnamefont
  {Greiner}}\ and\ \bibinfo {author} {\bibfnamefont {A.}~\bibnamefont
  {Schafer}},\ }\href@noop {} {\emph {\bibinfo {title} {Quantum
  Chromodynamics}}}\ (\bibinfo  {publisher} {Springer-Verlag},\ \bibinfo
  {address} {Berlin Heidelberg},\ \bibinfo {year} {1994})\BibitemShut {NoStop}%
\bibitem [{\citenamefont {Shelton}\ \emph {et~al.}(1996)\citenamefont
  {Shelton}, \citenamefont {Nersesyan},\ and\ \citenamefont
  {Tsvelik}}]{Shelton1996}%
  \BibitemOpen
  \bibfield  {author} {\bibinfo {author} {\bibfnamefont {D.~G.}\ \bibnamefont
  {Shelton}}, \bibinfo {author} {\bibfnamefont {A.~A.}\ \bibnamefont
  {Nersesyan}}, \ and\ \bibinfo {author} {\bibfnamefont {A.~M.}\ \bibnamefont
  {Tsvelik}},\ }\href {\doibase 10.1103/PhysRevB.53.8521} {\bibfield  {journal}
  {\bibinfo  {journal} {Phys. Rev. B}\ }\textbf {\bibinfo {volume} {53}},\
  \bibinfo {pages} {8521} (\bibinfo {year} {1996})}\BibitemShut {NoStop}%
\bibitem [{\citenamefont {Lake}\ \emph {et~al.}(1978)\citenamefont {Lake},
  \citenamefont {Tsvelik}, \citenamefont {Notbohm}, \citenamefont
  {Alan~Tennant}, \citenamefont {Perring}, \citenamefont {Reehuis},
  \citenamefont {Sekar}, \citenamefont {Krabbes},\ and\ \citenamefont
  {Büchner}}]{Lake2009}%
  \BibitemOpen
  \bibfield  {author} {\bibinfo {author} {\bibfnamefont {B.}~\bibnamefont
  {Lake}}, \bibinfo {author} {\bibfnamefont {A.~M.}\ \bibnamefont {Tsvelik}},
  \bibinfo {author} {\bibfnamefont {S.}~\bibnamefont {Notbohm}}, \bibinfo
  {author} {\bibfnamefont {D.}~\bibnamefont {Alan~Tennant}}, \bibinfo {author}
  {\bibfnamefont {T.~G.}\ \bibnamefont {Perring}}, \bibinfo {author}
  {\bibfnamefont {M.}~\bibnamefont {Reehuis}}, \bibinfo {author} {\bibfnamefont
  {C.}~\bibnamefont {Sekar}}, \bibinfo {author} {\bibfnamefont
  {G.}~\bibnamefont {Krabbes}}, \ and\ \bibinfo {author} {\bibfnamefont
  {B.}~\bibnamefont {Büchner}},\ }\href {\doibase
  10.1016/0378-4371(78)90160-7} {\bibfield  {journal} {\bibinfo  {journal}
  {Nature Physics}\ }\textbf {\bibinfo {volume} {6}},\ \bibinfo {pages} {327 }
  (\bibinfo {year} {1978})}\BibitemShut {NoStop}%
\bibitem [{\citenamefont {Starykh}\ and\ \citenamefont
  {Reiter}(1994)}]{Starykh1994}%
  \BibitemOpen
  \bibfield  {author} {\bibinfo {author} {\bibfnamefont {O.~A.}\ \bibnamefont
  {Starykh}}\ and\ \bibinfo {author} {\bibfnamefont {G.}~\bibnamefont
  {Reiter}},\ }\href {\doibase 10.1103/PhysRevB.49.4368} {\bibfield  {journal}
  {\bibinfo  {journal} {Phys. Rev. B}\ }\textbf {\bibinfo {volume} {49}},\
  \bibinfo {pages} {4368} (\bibinfo {year} {1994})}\BibitemShut {NoStop}%
\bibitem [{\citenamefont {Powell}\ and\ \citenamefont
  {McKenzie}(2011)}]{Powell2011}%
  \BibitemOpen
  \bibfield  {author} {\bibinfo {author} {\bibfnamefont {B.~J.}\ \bibnamefont
  {Powell}}\ and\ \bibinfo {author} {\bibfnamefont {R.~H.}\ \bibnamefont
  {McKenzie}},\ }\href {http://stacks.iop.org/0034-4885/74/i=5/a=056501}
  {\bibfield  {journal} {\bibinfo  {journal} {Reports on Progress in Physics}\
  }\textbf {\bibinfo {volume} {74}},\ \bibinfo {pages} {056501} (\bibinfo
  {year} {2011})}\BibitemShut {NoStop}%
\bibitem [{\citenamefont {Chubukov}\ and\ \citenamefont
  {Starykh}(1995)}]{Chubukov1995}%
  \BibitemOpen
  \bibfield  {author} {\bibinfo {author} {\bibfnamefont {A.~V.}\ \bibnamefont
  {Chubukov}}\ and\ \bibinfo {author} {\bibfnamefont {O.~A.}\ \bibnamefont
  {Starykh}},\ }\href {\doibase 10.1103/PhysRevB.52.440} {\bibfield  {journal}
  {\bibinfo  {journal} {Phys. Rev. B}\ }\textbf {\bibinfo {volume} {52}},\
  \bibinfo {pages} {440} (\bibinfo {year} {1995})}\BibitemShut {NoStop}%
\bibitem [{\citenamefont {Lee}\ \emph {et~al.}(2006)\citenamefont {Lee},
  \citenamefont {Nagaosa},\ and\ \citenamefont {Wen}}]{Lee2006}%
  \BibitemOpen
  \bibfield  {author} {\bibinfo {author} {\bibfnamefont {P.~A.}\ \bibnamefont
  {Lee}}, \bibinfo {author} {\bibfnamefont {N.}~\bibnamefont {Nagaosa}}, \ and\
  \bibinfo {author} {\bibfnamefont {X.-G.}\ \bibnamefont {Wen}},\ }\href
  {\doibase 10.1103/RevModPhys.78.17} {\bibfield  {journal} {\bibinfo
  {journal} {Rev. Mod. Phys.}\ }\textbf {\bibinfo {volume} {78}},\ \bibinfo
  {pages} {17} (\bibinfo {year} {2006})}\BibitemShut {NoStop}%
\bibitem [{\citenamefont {Vojta}(2018)}]{Vojta2018}%
  \BibitemOpen
  \bibfield  {author} {\bibinfo {author} {\bibfnamefont {M.}~\bibnamefont
  {Vojta}},\ }\href {http://stacks.iop.org/0034-4885/81/i=6/a=064501}
  {\bibfield  {journal} {\bibinfo  {journal} {Reports on Progress in Physics}\
  }\textbf {\bibinfo {volume} {81}},\ \bibinfo {pages} {064501} (\bibinfo
  {year} {2018})}\BibitemShut {NoStop}%
\bibitem [{\citenamefont {Kamiya}\ \emph {et~al.}(2018)\citenamefont {Kamiya},
  \citenamefont {Ge}, \citenamefont {Hong}, \citenamefont {Qiu}, \citenamefont
  {Quintero-Castro}, \citenamefont {Lu}, \citenamefont {Cao}, \citenamefont
  {Matsuda}, \citenamefont {S.~Choi}, \citenamefont {Batista}, \citenamefont
  {Mourigal}, \citenamefont {D.~Zhou},\ and\ \citenamefont {Ma}}]{Kamiya2018}%
  \BibitemOpen
  \bibfield  {author} {\bibinfo {author} {\bibfnamefont {Y.}~\bibnamefont
  {Kamiya}}, \bibinfo {author} {\bibfnamefont {L.}~\bibnamefont {Ge}}, \bibinfo
  {author} {\bibfnamefont {T.}~\bibnamefont {Hong}}, \bibinfo {author}
  {\bibfnamefont {Y.}~\bibnamefont {Qiu}}, \bibinfo {author} {\bibfnamefont
  {D.}~\bibnamefont {Quintero-Castro}}, \bibinfo {author} {\bibfnamefont
  {Z.}~\bibnamefont {Lu}}, \bibinfo {author} {\bibfnamefont {H.}~\bibnamefont
  {Cao}}, \bibinfo {author} {\bibfnamefont {M.}~\bibnamefont {Matsuda}},
  \bibinfo {author} {\bibfnamefont {E.}~\bibnamefont {S.~Choi}}, \bibinfo
  {author} {\bibfnamefont {C.}~\bibnamefont {Batista}}, \bibinfo {author}
  {\bibfnamefont {M.}~\bibnamefont {Mourigal}}, \bibinfo {author}
  {\bibfnamefont {H.}~\bibnamefont {D.~Zhou}}, \ and\ \bibinfo {author}
  {\bibfnamefont {J.}~\bibnamefont {Ma}},\ }\href@noop {} {\bibfield  {journal}
  {\bibinfo  {journal} {Nature Communications}\ }\textbf {\bibinfo {volume}
  {9}} (\bibinfo {year} {2018})}\BibitemShut {NoStop}%
\bibitem [{\citenamefont {Mourigal}\ \emph {et~al.}(2013)\citenamefont
  {Mourigal}, \citenamefont {Fuhrman}, \citenamefont {Chernyshev},\ and\
  \citenamefont {Zhitomirsky}}]{Mourigal2013a}%
  \BibitemOpen
  \bibfield  {author} {\bibinfo {author} {\bibfnamefont {M.}~\bibnamefont
  {Mourigal}}, \bibinfo {author} {\bibfnamefont {W.~T.}\ \bibnamefont
  {Fuhrman}}, \bibinfo {author} {\bibfnamefont {A.~L.}\ \bibnamefont
  {Chernyshev}}, \ and\ \bibinfo {author} {\bibfnamefont {M.~E.}\ \bibnamefont
  {Zhitomirsky}},\ }\href {\doibase 10.1103/PhysRevB.88.094407} {\bibfield
  {journal} {\bibinfo  {journal} {Phys. Rev. B}\ }\textbf {\bibinfo {volume}
  {88}},\ \bibinfo {pages} {094407} (\bibinfo {year} {2013})}\BibitemShut
  {NoStop}%
\bibitem [{\citenamefont {Coldea}\ \emph {et~al.}(2001)\citenamefont {Coldea},
  \citenamefont {Tennant}, \citenamefont {Tsvelik},\ and\ \citenamefont
  {Tylczynski}}]{Coldea2001}%
  \BibitemOpen
  \bibfield  {author} {\bibinfo {author} {\bibfnamefont {R.}~\bibnamefont
  {Coldea}}, \bibinfo {author} {\bibfnamefont {D.~A.}\ \bibnamefont {Tennant}},
  \bibinfo {author} {\bibfnamefont {A.~M.}\ \bibnamefont {Tsvelik}}, \ and\
  \bibinfo {author} {\bibfnamefont {Z.}~\bibnamefont {Tylczynski}},\ }\href
  {\doibase 10.1103/PhysRevLett.86.1335} {\bibfield  {journal} {\bibinfo
  {journal} {Phys. Rev. Lett.}\ }\textbf {\bibinfo {volume} {86}},\ \bibinfo
  {pages} {1335} (\bibinfo {year} {2001})}\BibitemShut {NoStop}%
\bibitem [{\citenamefont {Isakov}\ \emph {et~al.}(2005)\citenamefont {Isakov},
  \citenamefont {Senthil},\ and\ \citenamefont {Kim}}]{Isakov05}%
  \BibitemOpen
  \bibfield  {author} {\bibinfo {author} {\bibfnamefont {S.~V.}\ \bibnamefont
  {Isakov}}, \bibinfo {author} {\bibfnamefont {T.}~\bibnamefont {Senthil}}, \
  and\ \bibinfo {author} {\bibfnamefont {Y.~B.}\ \bibnamefont {Kim}},\ }\href
  {\doibase 10.1103/PhysRevB.72.174417} {\bibfield  {journal} {\bibinfo
  {journal} {Phys. Rev. B}\ }\textbf {\bibinfo {volume} {72}},\ \bibinfo
  {pages} {174417} (\bibinfo {year} {2005})}\BibitemShut {NoStop}%
\bibitem [{\citenamefont {Arovas}\ and\ \citenamefont
  {Auerbach}(1988)}]{Arovas1988}%
  \BibitemOpen
  \bibfield  {author} {\bibinfo {author} {\bibfnamefont {D.~P.}\ \bibnamefont
  {Arovas}}\ and\ \bibinfo {author} {\bibfnamefont {A.}~\bibnamefont
  {Auerbach}},\ }\href {\doibase 10.1103/PhysRevB.38.316} {\bibfield  {journal}
  {\bibinfo  {journal} {Phys. Rev. B}\ }\textbf {\bibinfo {volume} {38}},\
  \bibinfo {pages} {316} (\bibinfo {year} {1988})}\BibitemShut {NoStop}%
\bibitem [{\citenamefont {Timm}\ \emph {et~al.}(1998)\citenamefont {Timm},
  \citenamefont {Girvin}, \citenamefont {Henelius},\ and\ \citenamefont
  {Sandvik}}]{Timm1998}%
  \BibitemOpen
  \bibfield  {author} {\bibinfo {author} {\bibfnamefont {C.}~\bibnamefont
  {Timm}}, \bibinfo {author} {\bibfnamefont {S.~M.}\ \bibnamefont {Girvin}},
  \bibinfo {author} {\bibfnamefont {P.}~\bibnamefont {Henelius}}, \ and\
  \bibinfo {author} {\bibfnamefont {A.~W.}\ \bibnamefont {Sandvik}},\ }\href
  {\doibase 10.1103/PhysRevB.58.1464} {\bibfield  {journal} {\bibinfo
  {journal} {Phys. Rev. B}\ }\textbf {\bibinfo {volume} {58}},\ \bibinfo
  {pages} {1464} (\bibinfo {year} {1998})}\BibitemShut {NoStop}%
\bibitem [{\citenamefont {Flint}\ and\ \citenamefont
  {Coleman}(2009)}]{Flint2009}%
  \BibitemOpen
  \bibfield  {author} {\bibinfo {author} {\bibfnamefont {R.}~\bibnamefont
  {Flint}}\ and\ \bibinfo {author} {\bibfnamefont {P.}~\bibnamefont
  {Coleman}},\ }\href {\doibase 10.1103/PhysRevB.79.014424} {\bibfield
  {journal} {\bibinfo  {journal} {Phys. Rev. B}\ }\textbf {\bibinfo {volume}
  {79}},\ \bibinfo {pages} {014424} (\bibinfo {year} {2009})}\BibitemShut
  {NoStop}%
\bibitem [{\citenamefont {Hirsch}\ and\ \citenamefont
  {Tang}(1989)}]{Hirsch1989}%
  \BibitemOpen
  \bibfield  {author} {\bibinfo {author} {\bibfnamefont {J.~E.}\ \bibnamefont
  {Hirsch}}\ and\ \bibinfo {author} {\bibfnamefont {S.}~\bibnamefont {Tang}},\
  }\href {\doibase 10.1103/PhysRevB.39.2850} {\bibfield  {journal} {\bibinfo
  {journal} {Phys. Rev. B}\ }\textbf {\bibinfo {volume} {39}},\ \bibinfo
  {pages} {2850} (\bibinfo {year} {1989})}\BibitemShut {NoStop}%
\bibitem [{\citenamefont {Sarker}\ \emph {et~al.}(1989)\citenamefont {Sarker},
  \citenamefont {Jayaprakash}, \citenamefont {Krishnamurthy},\ and\
  \citenamefont {Ma}}]{Sarker1989}%
  \BibitemOpen
  \bibfield  {author} {\bibinfo {author} {\bibfnamefont {S.}~\bibnamefont
  {Sarker}}, \bibinfo {author} {\bibfnamefont {C.}~\bibnamefont {Jayaprakash}},
  \bibinfo {author} {\bibfnamefont {H.~R.}\ \bibnamefont {Krishnamurthy}}, \
  and\ \bibinfo {author} {\bibfnamefont {M.}~\bibnamefont {Ma}},\ }\href
  {\doibase 10.1103/PhysRevB.40.5028} {\bibfield  {journal} {\bibinfo
  {journal} {Phys. Rev. B}\ }\textbf {\bibinfo {volume} {40}},\ \bibinfo
  {pages} {5028} (\bibinfo {year} {1989})}\BibitemShut {NoStop}%
\bibitem [{\citenamefont {Chandra}\ \emph {et~al.}(1990)\citenamefont
  {Chandra}, \citenamefont {Coleman},\ and\ \citenamefont
  {Larkin}}]{Chandra1990}%
  \BibitemOpen
  \bibfield  {author} {\bibinfo {author} {\bibfnamefont {P.}~\bibnamefont
  {Chandra}}, \bibinfo {author} {\bibfnamefont {P.}~\bibnamefont {Coleman}}, \
  and\ \bibinfo {author} {\bibfnamefont {A.~I.}\ \bibnamefont {Larkin}},\
  }\href {\doibase 10.1088/0953-8984/2/39/008} {\bibfield  {journal} {\bibinfo
  {journal} {Journal of Physics: Condensed Matter}\ }\textbf {\bibinfo {volume}
  {2}},\ \bibinfo {pages} {7933} (\bibinfo {year} {1990})}\BibitemShut
  {NoStop}%
\bibitem [{\citenamefont {Auerbach}\ and\ \citenamefont
  {Arovas}(1988)}]{Auerbach1988}%
  \BibitemOpen
  \bibfield  {author} {\bibinfo {author} {\bibfnamefont {A.}~\bibnamefont
  {Auerbach}}\ and\ \bibinfo {author} {\bibfnamefont {D.~P.}\ \bibnamefont
  {Arovas}},\ }\href {\doibase 10.1103/PhysRevLett.61.617} {\bibfield
  {journal} {\bibinfo  {journal} {Phys. Rev. Lett.}\ }\textbf {\bibinfo
  {volume} {61}},\ \bibinfo {pages} {617} (\bibinfo {year} {1988})}\BibitemShut
  {NoStop}%
\bibitem [{\citenamefont {Mila}\ \emph {et~al.}(1991)\citenamefont {Mila},
  \citenamefont {Poilblanc},\ and\ \citenamefont {Bruder}}]{Mila1991}%
  \BibitemOpen
  \bibfield  {author} {\bibinfo {author} {\bibfnamefont {F.}~\bibnamefont
  {Mila}}, \bibinfo {author} {\bibfnamefont {D.}~\bibnamefont {Poilblanc}}, \
  and\ \bibinfo {author} {\bibfnamefont {C.}~\bibnamefont {Bruder}},\ }\href
  {\doibase 10.1103/PhysRevB.43.7891} {\bibfield  {journal} {\bibinfo
  {journal} {Phys. Rev. B}\ }\textbf {\bibinfo {volume} {43}},\ \bibinfo
  {pages} {7891} (\bibinfo {year} {1991})}\BibitemShut {NoStop}%
\bibitem [{\citenamefont {Yoshioka}\ and\ \citenamefont
  {Miyazaki}(1991)}]{Yoshioka1991}%
  \BibitemOpen
  \bibfield  {author} {\bibinfo {author} {\bibfnamefont {D.}~\bibnamefont
  {Yoshioka}}\ and\ \bibinfo {author} {\bibfnamefont {J.}~\bibnamefont
  {Miyazaki}},\ }\href {\doibase 10.1143/JPSJ.60.614} {\bibfield  {journal}
  {\bibinfo  {journal} {Journal of the Physical Society of Japan}\ }\textbf
  {\bibinfo {volume} {60}},\ \bibinfo {pages} {614} (\bibinfo {year}
  {1991})}\BibitemShut {NoStop}%
\bibitem [{\citenamefont {Sachdev}(1992)}]{Sachdev1992}%
  \BibitemOpen
  \bibfield  {author} {\bibinfo {author} {\bibfnamefont {S.}~\bibnamefont
  {Sachdev}},\ }\href {\doibase 10.1103/PhysRevB.45.12377} {\bibfield
  {journal} {\bibinfo  {journal} {Phys. Rev. B}\ }\textbf {\bibinfo {volume}
  {45}},\ \bibinfo {pages} {12377} (\bibinfo {year} {1992})}\BibitemShut
  {NoStop}%
\bibitem [{\citenamefont {Lefmann}\ and\ \citenamefont
  {Hedeg\aa{}rd}(1994)}]{Lefmann1994}%
  \BibitemOpen
  \bibfield  {author} {\bibinfo {author} {\bibfnamefont {K.}~\bibnamefont
  {Lefmann}}\ and\ \bibinfo {author} {\bibfnamefont {P.}~\bibnamefont
  {Hedeg\aa{}rd}},\ }\href {\doibase 10.1103/PhysRevB.50.1074} {\bibfield
  {journal} {\bibinfo  {journal} {Phys. Rev. B}\ }\textbf {\bibinfo {volume}
  {50}},\ \bibinfo {pages} {1074} (\bibinfo {year} {1994})}\BibitemShut
  {NoStop}%
\bibitem [{\citenamefont {Mattsson}(1995)}]{Mattsson1995}%
  \BibitemOpen
  \bibfield  {author} {\bibinfo {author} {\bibfnamefont {A.}~\bibnamefont
  {Mattsson}},\ }\href {\doibase 10.1103/PhysRevB.51.11574} {\bibfield
  {journal} {\bibinfo  {journal} {Phys. Rev. B}\ }\textbf {\bibinfo {volume}
  {51}},\ \bibinfo {pages} {11574} (\bibinfo {year} {1995})}\BibitemShut
  {NoStop}%
\bibitem [{\citenamefont {Mezio}\ \emph {et~al.}(2011)\citenamefont {Mezio},
  \citenamefont {Sposetti}, \citenamefont {Manuel},\ and\ \citenamefont
  {Trumper}}]{Mezio2011}%
  \BibitemOpen
  \bibfield  {author} {\bibinfo {author} {\bibfnamefont {A.}~\bibnamefont
  {Mezio}}, \bibinfo {author} {\bibfnamefont {C.~N.}\ \bibnamefont {Sposetti}},
  \bibinfo {author} {\bibfnamefont {L.~O.}\ \bibnamefont {Manuel}}, \ and\
  \bibinfo {author} {\bibfnamefont {A.~E.}\ \bibnamefont {Trumper}},\ }\href
  {\doibase 10.1209/0295-5075/94/47001} {\bibfield  {journal} {\bibinfo
  {journal} {EPL (Europhysics Letters)}\ }\textbf {\bibinfo {volume} {94}},\
  \bibinfo {pages} {47001} (\bibinfo {year} {2011})}\BibitemShut {NoStop}%
\bibitem [{\citenamefont {Ceccatto}\ \emph {et~al.}(1993)\citenamefont
  {Ceccatto}, \citenamefont {Gazza},\ and\ \citenamefont
  {Trumper}}]{Ceccatto1993}%
  \BibitemOpen
  \bibfield  {author} {\bibinfo {author} {\bibfnamefont {H.~A.}\ \bibnamefont
  {Ceccatto}}, \bibinfo {author} {\bibfnamefont {C.~J.}\ \bibnamefont {Gazza}},
  \ and\ \bibinfo {author} {\bibfnamefont {A.~E.}\ \bibnamefont {Trumper}},\
  }\href {\doibase 10.1103/PhysRevB.47.12329} {\bibfield  {journal} {\bibinfo
  {journal} {Phys. Rev. B}\ }\textbf {\bibinfo {volume} {47}},\ \bibinfo
  {pages} {12329} (\bibinfo {year} {1993})}\BibitemShut {NoStop}%
\bibitem [{\citenamefont {Gazza}\ and\ \citenamefont
  {Ceccatto}(1993)}]{Gazza1993}%
  \BibitemOpen
  \bibfield  {author} {\bibinfo {author} {\bibfnamefont {C.~J.}\ \bibnamefont
  {Gazza}}\ and\ \bibinfo {author} {\bibfnamefont {H.~A.}\ \bibnamefont
  {Ceccatto}},\ }\href {http://stacks.iop.org/0953-8984/5/i=10/a=003}
  {\bibfield  {journal} {\bibinfo  {journal} {Journal of Physics: Condensed
  Matter}\ }\textbf {\bibinfo {volume} {5}},\ \bibinfo {pages} {L135} (\bibinfo
  {year} {1993})}\BibitemShut {NoStop}%
\bibitem [{\citenamefont {Manuel}\ \emph {et~al.}(1994)\citenamefont {Manuel},
  \citenamefont {Trumper}, \citenamefont {Gazza},\ and\ \citenamefont
  {Ceccatto}}]{Manuel1994}%
  \BibitemOpen
  \bibfield  {author} {\bibinfo {author} {\bibfnamefont {L.~O.}\ \bibnamefont
  {Manuel}}, \bibinfo {author} {\bibfnamefont {A.~E.}\ \bibnamefont {Trumper}},
  \bibinfo {author} {\bibfnamefont {C.~J.}\ \bibnamefont {Gazza}}, \ and\
  \bibinfo {author} {\bibfnamefont {H.~A.}\ \bibnamefont {Ceccatto}},\ }\href
  {\doibase 10.1103/PhysRevB.50.1313} {\bibfield  {journal} {\bibinfo
  {journal} {Phys. Rev. B}\ }\textbf {\bibinfo {volume} {50}},\ \bibinfo
  {pages} {1313} (\bibinfo {year} {1994})}\BibitemShut {NoStop}%
\bibitem [{\citenamefont {Mezio}\ \emph {et~al.}(2012)\citenamefont {Mezio},
  \citenamefont {Manuel}, \citenamefont {Singh},\ and\ \citenamefont
  {Trumper}}]{Mezio2012}%
  \BibitemOpen
  \bibfield  {author} {\bibinfo {author} {\bibfnamefont {A.}~\bibnamefont
  {Mezio}}, \bibinfo {author} {\bibfnamefont {L.~O.}\ \bibnamefont {Manuel}},
  \bibinfo {author} {\bibfnamefont {R.~R.~P.}\ \bibnamefont {Singh}}, \ and\
  \bibinfo {author} {\bibfnamefont {A.~E.}\ \bibnamefont {Trumper}},\ }\href
  {\doibase 10.1088/1367-2630/14/12/123033} {\bibfield  {journal} {\bibinfo
  {journal} {New Journal of Physics}\ }\textbf {\bibinfo {volume} {14}},\
  \bibinfo {pages} {123033} (\bibinfo {year} {2012})}\BibitemShut {NoStop}%
\bibitem [{\citenamefont {Wang}\ and\ \citenamefont
  {Vishwanath}(2006)}]{Wang2006}%
  \BibitemOpen
  \bibfield  {author} {\bibinfo {author} {\bibfnamefont {F.}~\bibnamefont
  {Wang}}\ and\ \bibinfo {author} {\bibfnamefont {A.}~\bibnamefont
  {Vishwanath}},\ }\href {\doibase 10.1103/PhysRevB.74.174423} {\bibfield
  {journal} {\bibinfo  {journal} {Phys. Rev. B}\ }\textbf {\bibinfo {volume}
  {74}},\ \bibinfo {pages} {174423} (\bibinfo {year} {2006})}\BibitemShut
  {NoStop}%
\bibitem [{\citenamefont {Messio}\ \emph {et~al.}(2013)\citenamefont {Messio},
  \citenamefont {Lhuillier},\ and\ \citenamefont {Misguich}}]{Messio2013}%
  \BibitemOpen
  \bibfield  {author} {\bibinfo {author} {\bibfnamefont {L.}~\bibnamefont
  {Messio}}, \bibinfo {author} {\bibfnamefont {C.}~\bibnamefont {Lhuillier}}, \
  and\ \bibinfo {author} {\bibfnamefont {G.}~\bibnamefont {Misguich}},\ }\href
  {\doibase 10.1103/PhysRevB.87.125127} {\bibfield  {journal} {\bibinfo
  {journal} {Phys. Rev. B}\ }\textbf {\bibinfo {volume} {87}},\ \bibinfo
  {pages} {125127} (\bibinfo {year} {2013})}\BibitemShut {NoStop}%
\bibitem [{\citenamefont {Auerbach}(1994)}]{Auerbach1994}%
  \BibitemOpen
  \bibfield  {author} {\bibinfo {author} {\bibfnamefont {A.}~\bibnamefont
  {Auerbach}},\ }\href@noop {} {\emph {\bibinfo {title} {Interacting electrons
  and quantum magnetism}}}\ (\bibinfo  {publisher} {Springer-Verlag},\ \bibinfo
  {address} {New York},\ \bibinfo {year} {1994})\BibitemShut {NoStop}%
\bibitem [{\citenamefont {Ghioldi}\ \emph {et~al.}(2015)\citenamefont
  {Ghioldi}, \citenamefont {Mezio}, \citenamefont {Manuel}, \citenamefont
  {Singh}, \citenamefont {Oitmaa},\ and\ \citenamefont
  {Trumper}}]{Ghioldi2015}%
  \BibitemOpen
  \bibfield  {author} {\bibinfo {author} {\bibfnamefont {E.~A.}\ \bibnamefont
  {Ghioldi}}, \bibinfo {author} {\bibfnamefont {A.}~\bibnamefont {Mezio}},
  \bibinfo {author} {\bibfnamefont {L.~O.}\ \bibnamefont {Manuel}}, \bibinfo
  {author} {\bibfnamefont {R.~R.~P.}\ \bibnamefont {Singh}}, \bibinfo {author}
  {\bibfnamefont {J.}~\bibnamefont {Oitmaa}}, \ and\ \bibinfo {author}
  {\bibfnamefont {A.~E.}\ \bibnamefont {Trumper}},\ }\href {\doibase
  10.1103/PhysRevB.91.134423} {\bibfield  {journal} {\bibinfo  {journal} {Phys.
  Rev. B}\ }\textbf {\bibinfo {volume} {91}},\ \bibinfo {pages} {134423}
  (\bibinfo {year} {2015})}\BibitemShut {NoStop}%
\bibitem [{\citenamefont {Trumper}\ \emph {et~al.}(1997)\citenamefont
  {Trumper}, \citenamefont {Manuel}, \citenamefont {Gazza},\ and\ \citenamefont
  {Ceccatto}}]{Trumper1997}%
  \BibitemOpen
  \bibfield  {author} {\bibinfo {author} {\bibfnamefont {A.~E.}\ \bibnamefont
  {Trumper}}, \bibinfo {author} {\bibfnamefont {L.~O.}\ \bibnamefont {Manuel}},
  \bibinfo {author} {\bibfnamefont {C.~J.}\ \bibnamefont {Gazza}}, \ and\
  \bibinfo {author} {\bibfnamefont {H.~A.}\ \bibnamefont {Ceccatto}},\ }\href
  {\doibase 10.1103/PhysRevLett.78.2216} {\bibfield  {journal} {\bibinfo
  {journal} {Phys. Rev. Lett.}\ }\textbf {\bibinfo {volume} {78}},\ \bibinfo
  {pages} {2216} (\bibinfo {year} {1997})}\BibitemShut {NoStop}%
\bibitem [{\citenamefont {Negele}\ and\ \citenamefont
  {Orland}(1998)}]{Negele1998}%
  \BibitemOpen
  \bibfield  {author} {\bibinfo {author} {\bibfnamefont {J.}~\bibnamefont
  {Negele}}\ and\ \bibinfo {author} {\bibfnamefont {H.}~\bibnamefont
  {Orland}},\ }\href@noop {} {\emph {\bibinfo {title} {Quantum many-particle
  systems}}}\ (\bibinfo  {publisher} {Perseus books},\ \bibinfo {address}
  {Massachusetts},\ \bibinfo {year} {1998})\BibitemShut {NoStop}%
\bibitem [{\citenamefont {Manuel}\ \emph {et~al.}(1998)\citenamefont {Manuel},
  \citenamefont {Trumper},\ and\ \citenamefont {Ceccatto}}]{Manuel1998}%
  \BibitemOpen
  \bibfield  {author} {\bibinfo {author} {\bibfnamefont {L.~O.}\ \bibnamefont
  {Manuel}}, \bibinfo {author} {\bibfnamefont {A.~E.}\ \bibnamefont {Trumper}},
  \ and\ \bibinfo {author} {\bibfnamefont {H.~A.}\ \bibnamefont {Ceccatto}},\
  }\href {\doibase 10.1103/PhysRevB.57.8348} {\bibfield  {journal} {\bibinfo
  {journal} {Phys. Rev. B}\ }\textbf {\bibinfo {volume} {57}},\ \bibinfo
  {pages} {8348} (\bibinfo {year} {1998})}\BibitemShut {NoStop}%
\bibitem [{\citenamefont {Gonzalez}\ \emph {et~al.}(2017)\citenamefont
  {Gonzalez}, \citenamefont {Ghioldi}, \citenamefont {Gazza}, \citenamefont
  {Manuel},\ and\ \citenamefont {Trumper}}]{Gonzalez2017}%
  \BibitemOpen
  \bibfield  {author} {\bibinfo {author} {\bibfnamefont {M.~G.}\ \bibnamefont
  {Gonzalez}}, \bibinfo {author} {\bibfnamefont {E.~A.}\ \bibnamefont
  {Ghioldi}}, \bibinfo {author} {\bibfnamefont {C.~J.}\ \bibnamefont {Gazza}},
  \bibinfo {author} {\bibfnamefont {L.~O.}\ \bibnamefont {Manuel}}, \ and\
  \bibinfo {author} {\bibfnamefont {A.~E.}\ \bibnamefont {Trumper}},\ }\href
  {\doibase 10.1103/PhysRevB.96.174423} {\bibfield  {journal} {\bibinfo
  {journal} {Phys. Rev. B}\ }\textbf {\bibinfo {volume} {96}},\ \bibinfo
  {pages} {174423} (\bibinfo {year} {2017})}\BibitemShut {NoStop}%
\bibitem [{\citenamefont {Shindou}\ \emph {et~al.}(2013)\citenamefont
  {Shindou}, \citenamefont {Yunoki},\ and\ \citenamefont
  {Momoi}}]{Shindou2013}%
  \BibitemOpen
  \bibfield  {author} {\bibinfo {author} {\bibfnamefont {R.}~\bibnamefont
  {Shindou}}, \bibinfo {author} {\bibfnamefont {S.}~\bibnamefont {Yunoki}}, \
  and\ \bibinfo {author} {\bibfnamefont {T.}~\bibnamefont {Momoi}},\ }\href
  {\doibase 10.1103/PhysRevB.87.054429} {\bibfield  {journal} {\bibinfo
  {journal} {Phys. Rev. B}\ }\textbf {\bibinfo {volume} {87}},\ \bibinfo
  {pages} {054429} (\bibinfo {year} {2013})}\BibitemShut {NoStop}%
\bibitem [{\citenamefont {Chubukov}\ and\ \citenamefont
  {Starykh}(1996)}]{Chubukov1996}%
  \BibitemOpen
  \bibfield  {author} {\bibinfo {author} {\bibfnamefont {A.~V.}\ \bibnamefont
  {Chubukov}}\ and\ \bibinfo {author} {\bibfnamefont {O.~A.}\ \bibnamefont
  {Starykh}},\ }\href {\doibase 10.1103/PhysRevB.53.R14729} {\bibfield
  {journal} {\bibinfo  {journal} {Phys. Rev. B}\ }\textbf {\bibinfo {volume}
  {53}},\ \bibinfo {pages} {R14729} (\bibinfo {year} {1996})}\BibitemShut
  {NoStop}%
\bibitem [{not()}]{notemd}%
  \BibitemOpen
  \href@noop {} {\ }\bibinfo {note} {We note that this decay process must be
  replaced by single-magnon to two-magnon decay in the long wavelength limit
  because low-energy spinons are removed from the spectrum via the Higgs
  mechanism. However, single to two-magnon decay arises from corrections to the
  RPA propagator (higher order diagrams) that have not been included in our
  calculation..}\BibitemShut {Stop}%
\bibitem [{\citenamefont {Coleman}\ and\ \citenamefont
  {Chandra}(1994)}]{Coleman1994}%
  \BibitemOpen
  \bibfield  {author} {\bibinfo {author} {\bibfnamefont {P.}~\bibnamefont
  {Coleman}}\ and\ \bibinfo {author} {\bibfnamefont {P.}~\bibnamefont
  {Chandra}},\ }\href {\doibase 10.1103/PhysRevLett.72.1944} {\bibfield
  {journal} {\bibinfo  {journal} {Phys. Rev. Lett.}\ }\textbf {\bibinfo
  {volume} {72}},\ \bibinfo {pages} {1944} (\bibinfo {year}
  {1994})}\BibitemShut {NoStop}%
\bibitem [{\citenamefont {Dombre}\ and\ \citenamefont {Read}(1989)}]{Dombre89}%
  \BibitemOpen
  \bibfield  {author} {\bibinfo {author} {\bibfnamefont {T.}~\bibnamefont
  {Dombre}}\ and\ \bibinfo {author} {\bibfnamefont {N.}~\bibnamefont {Read}},\
  }\href {\doibase 10.1103/PhysRevB.39.6797} {\bibfield  {journal} {\bibinfo
  {journal} {Phys. Rev. B}\ }\textbf {\bibinfo {volume} {39}},\ \bibinfo
  {pages} {6797} (\bibinfo {year} {1989})}\BibitemShut {NoStop}%
\bibitem [{\citenamefont {Horn}\ and\ \citenamefont
  {Johnson}(2012)}]{Horn2012}%
  \BibitemOpen
  \bibfield  {author} {\bibinfo {author} {\bibfnamefont {R.~A.}\ \bibnamefont
  {Horn}}\ and\ \bibinfo {author} {\bibfnamefont {C.~R.}\ \bibnamefont
  {Johnson}},\ }\href {\doibase 10.1017/9781139020411} {\emph {\bibinfo {title}
  {Matrix Analysis}}},\ \bibinfo {edition} {2nd}\ ed.\ (\bibinfo  {publisher}
  {Cambridge University Press},\ \bibinfo {year} {2012})\BibitemShut {NoStop}%
\bibitem [{\citenamefont {Kamenev}\ and\ \citenamefont
  {Levchenko}(2009)}]{Kamenev2009}%
  \BibitemOpen
  \bibfield  {author} {\bibinfo {author} {\bibfnamefont {A.}~\bibnamefont
  {Kamenev}}\ and\ \bibinfo {author} {\bibfnamefont {A.}~\bibnamefont
  {Levchenko}},\ }\href@noop {} {\bibfield  {journal} {\bibinfo  {journal}
  {Advances in Physics}\ }\textbf {\bibinfo {volume} {58}},\ \bibinfo {pages}
  {197} (\bibinfo {year} {2009})}\BibitemShut {NoStop}%
\end{thebibliography}%

\end{document}